\documentclass[aps,prb,floatfix,nopacs,superscriptaddress,twocolumn,reprint]{revtex4-2}
\usepackage[utf8]{inputenc}
\usepackage{amsmath}
\usepackage{amsfonts}
\usepackage{graphicx}
\usepackage{bm}
\usepackage[dvipsnames]{xcolor}
\usepackage{physics}

\newcommand{\bb}[1]{\mathbf{#1}}
\newcommand{\m}[1]{\mathcal{#1}}
\usepackage{url}
\usepackage{mathtools}

\usepackage[american,]{babel}
\usepackage[T1]{fontenc}
\usepackage{hyperref}
\usepackage{xcolor}
\hypersetup{colorlinks,bookmarksopen,bookmarksnumbered,
citecolor=[rgb]{0.255,0.412,0.882},
linkcolor=[rgb]{0.255,0.412,0.882},
pdfstartview=false,
urlcolor=[rgb]{0.255,0.412,0.882}}

\begin{document}

\title{Dissipative Kondo physics in the Anderson Impurity Model with two-body losses}

\author{Matthieu Vanhoecke}
\email{matthieu.vanhoecke@college-de-france.fr}
\affiliation{JEIP, UAR 3573 CNRS, Coll\`ege de France,   PSL  Research  University, 11,  place  Marcelin  Berthelot,75231 Paris Cedex 05, France}
\author{Naoto Tsuji}
\affiliation{Department of Physics, The University of Tokyo, Hongo, Tokyo 113-0033, Japan}
\affiliation{RIKEN Center for Emergent Matter Science (CEMS), Wako 351-0198, Japan}
\author{Marco Schir\`o}
\affiliation{JEIP, UAR 3573 CNRS, Coll\`ege de France,   PSL  Research  University, 11,  place  Marcelin  Berthelot,75231 Paris Cedex 05, France}

\begin{abstract}
We study a dissipative version of the Anderson Impurity model, where an interacting impurity is coupled to a fermionic reservoir and exposed to Markovian dissipation in the form of two-body losses. Using a self-consistent hybridization expansion based on the Non-Crossing Approximation (NCA) we compute the dynamics of the impurity, its steady-state and spectral function. We show that the interplay between strong Coulomb repulsion and correlated dissipation gives rise to robust signatures of Kondo physics both at weak and strong losses. These include a strongly suppressed spin relaxation rate, displaying a characteristic Kondo-Zeno crossover and a spectral function where doublon band is quickly destroyed by dissipation while the coherent Kondo peak remains visible for weak losses, then disappears at intermediate values and finally re-emerge as the system enters in the Kondo-Zeno regime. As compared to the case of single particle losses we show that two-body dissipation protects Kondo physics. The picture obtained with NCA is confirmed by numerical simulations of exact dynamics on finite-size chains.
We interpret these results using a dissipative Schrieffer-Wolff transformation, which leads to an effective Kondo model with residual impurity-bath losses which are suppressed by strong correlations or strong losses.
\end{abstract}

\maketitle

\section{Introduction}
\label{sec:intro}

The Anderson impurity model of a localized spinful electronic level coupled to a continuum is one of the simplest and most studied quantum many-body problem~\cite{hewson1993thekondo}.  It captures the physics of the Kondo effect~\cite{kondo1964resistance}, relevant for magnetic impurities in metals~\cite{anderson1961localized,nozieres1974fermiliquid,wilson1975therenormalization}, quantum dots~\cite{Pustilnik_2004} or single molecules~\cite{roch2009observation} coupled to metallic leads as well as atomic impurities in ultracold gases of alkaline-earth atoms~\cite{riegger2018localized,zhang2020controlling,nagy2018exploring,amaricci2025engineeringkondoimpurityproblem}. Nonequilibrium extensions of this and related quantum impurity models, for example due to an external bias voltage~\cite{cronenwett1998atunable,rosch2003nonequilibrium,mehta2006nonequilibrium,heidrich2009realtime,antipov2016voltage} or a local quantum quench~\cite{elzerman2004single,anders2005realtime,latta2011quantum,schiro2010realtime,schiro2012nonequilibrium}, have been studied extensively and played an important role in the development of quantum many-body physics out of equilibrium.

Recently, a new class of dissipative quantum impurity models has attracted interest: here the impurity is both coupled to a quantum bath, i.e. a structured frequency-dependent environment, and exposed to fast Markovian dissipation describing incoherent processes. The interest in this type of settings arise from recent developments in quantum simulations, where different types of dissipative environment can coexist and be controlled with high degree of tunability. Experiments with ultracold atoms, for example, have realized quantum transport through a dissipative quantum point contact~\cite{lebrat2019quantized,corman2019quantized,huang2023superfluid}, where the constriction between two quantum conductors is exposed to additional particle losses. 
Atomic impurities in ultracold gases~\cite{Hewitt_2024} are naturally exposed to correlated dissipative processes, such as dephasing due to spontaneous emission~\cite{gerbier2010heating,bouganne2020} or two-body losses due to inelastic scattering~\cite{garcia-ripoll2009,TomitaEtAlScienceAdv17,honda2022observation}. 
Quantum devices in the solid state and superconducting circuits for example are also emerging as platform to explore the role of local dissipation in a controlled way~\cite{google_dephasing_abanin,mi2023stable} and proposals to realize quantum impurity models on chip have appeared~\cite{sarma2025designbenchmarksemulatingkondo}.

Motivated by these developments there has been theoretical progress in understanding dissipative quantum impurity models. These include
non-interacting chains with localised single particle losses~\cite{Froml2019,damanet2019controlling,visuri2022symmetry,visuri2023nonlinear,visuri2023dc} or pumps~\cite{krapivsky2019free,krapivsky2020free} or local dephasing~\cite{Scarlatella2019,tonielli2019orthogonality,dolgirev2020nongaussian,ferreira2023exact}. Strongly interacting quantum impurity models in dissipative environments have been studied in presence of charge monitoring, dephasing or projective measurements~\cite{hasegawa2021kondo,vanhoecke2025kondozenocrossoverdynamicsmonitored}. 
A dissipative realization of the Kondo model due to two-body losses,  starting from a non-interacting quantum dot problem has been recently studied~\cite{stefanini2024dissipative,qu2025variational}. Finally, non-Hermitian generalization of Kondo or Anderson impurity models, corresponding to the no-click limit of suitable Lindblad dynamics, have been recently solved with Bethe Ansatz~\cite{Nakagawa2018,lourenco2018kondo,kattel2024dissipation,yamamoto2025correlation}. Despite this recent progress many questions remain open concerning the dynamics and steady-state properties of dissipative impurities and in particular the interplay between Kondo physics and local dissipation.

In this work we study the dynamics of an Anderson impurity model in the presence of two-body losses. We solve for the dynamics and steady-state impurity properties using a self-consistent dynamical map based on the Non-Crossing Approximation (NCA)~\cite{scarlatella2021dynamical,scarlatella2023selfconsistent,vanhoecke2025kondozenocrossoverdynamicsmonitored}. We first show how the charge dynamics on the impurity is affected by dissipation via the onset of a Zeno effect. Then we focus on spin dynamics and demonstrate the interplay between Kondo screening and dissipation, giving rise to a characteristic non-monotonic behavior in the spin-relaxation rate. We compute the steady-state spectral function of the impurity and show that while two-body losses destroy very effectively the upper Hubbard band, the Kondo peak remains clearly visible at weak dissipation and re-emerges at strong dissipation when the doublons are completely projected out. As opposed to single body losses, which we show to destroy all interesting many-body effects, the correlated nature of the dissipation allows to retain non-trivial features in our problem. We further compare our NCA results to exact finite-size simulations via quantum jump methods showing qualitative agreement, in particular for the non-monotonic behavior of the magnetization dynamics and the Kondo-Zeno crossover. Finally, using a Schrieffer-Wolff transformation on the Lindbladian of the dissipative Anderson impurity model we derive an effective Kondo model with residual impurity-bath losses, which we use to interpret qualitatively our results. Both Kondo coupling and residual dissipation depends on interaction and bare two-body losses, confirming how the interplay between these two mechanisms give rise to robust Kondo physics in the dissipative problem.

The paper is organized as follows. In Sec.~\ref{sec:Model}, we introduce the dissipative Anderson impurity model and discuss the methods used in this work to study its dynamics. In Sec.~\ref{sec:results}, we present our main results on the dynamics and steady-state properties of the model. In particular, we discuss charge and spin dynamics, the spectral function for increasing values of dissipation and compare our results to the case of single body losses and to exact dynamics on a finite chain. Finally in Sec.~\ref{sec:SW} we derive an effective Kondo model via a dissipative Schrieffer-Wolff transformation. In Sec.~\ref{sec:disc} we discuss our results in light of recent literature while in Sec.~\ref{sec:conclusions} is devoted to conclusions. 

\section{Model and Method}\label{sec:Model}
\begin{figure}[!t] 
    \centering
 \includegraphics[width=0.35\textwidth]{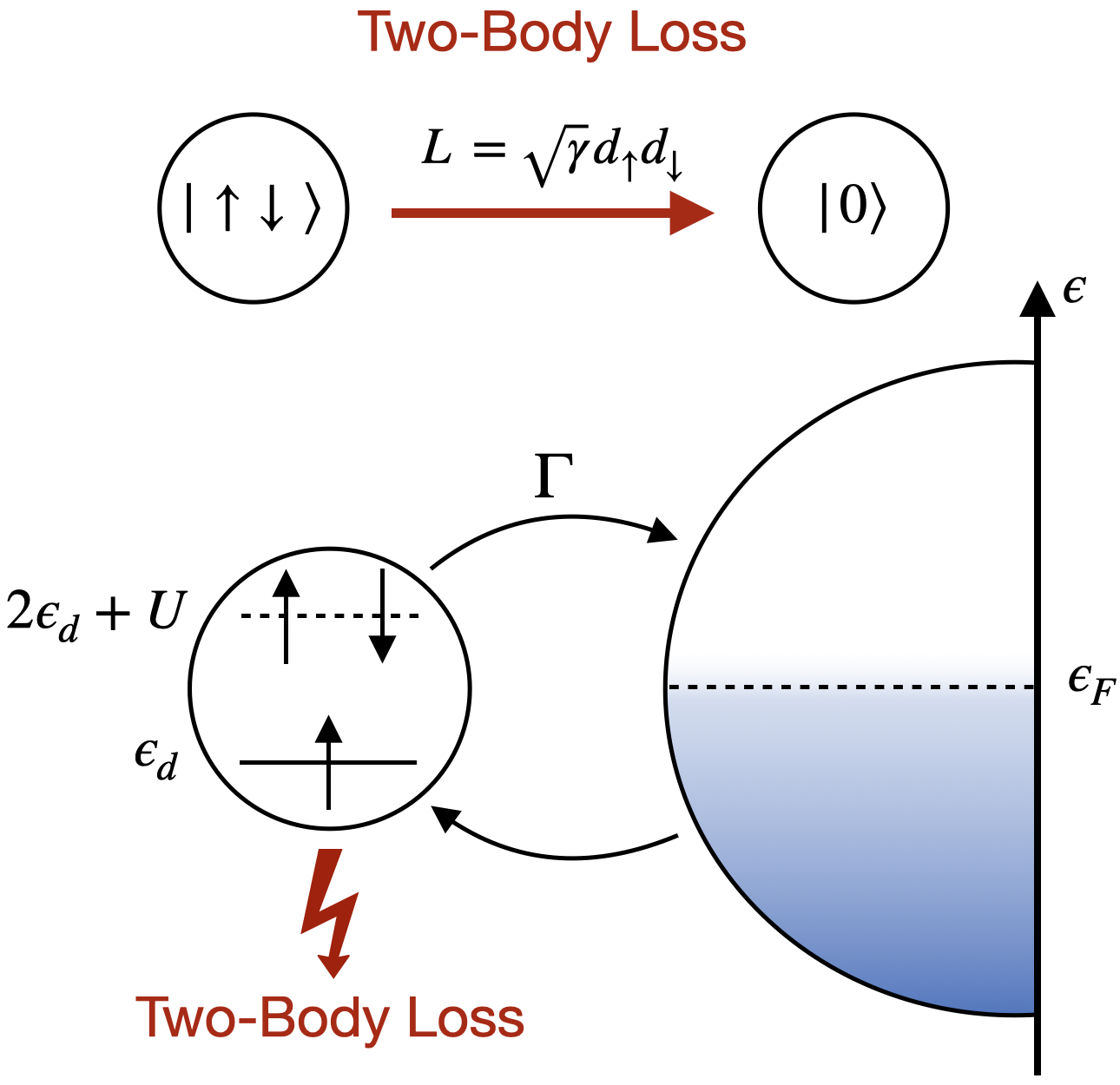}
    \caption{\label{fig:sketchtwobody} Sketch of the setup for the Anderson Impurity Model with Losses: an impurity subjected to two-body loss and coupled to a large metallic lead via a hybridization $\Gamma$. The dynamics of the combined impurity–lead system are governed by a Lindblad master equation, with a jump operator $L = \sqrt{\gamma}, d_{\uparrow} d_{\downarrow}$ that models the two-body loss process.}  
\end{figure}
We start describing the model and the setup, see also the sketch in Fig.~\ref{fig:sketchtwobody}. 
We consider a dissipative version of the celebrated Anderson Impurity model
(AIM), describing a spinfull fermionic level which is coupled to a metallic bath via a hybridization term. The AIM Hamiltonian reads~\cite{hewson1993thekondo}
\begin{align}\label{eq:H_aim}
H=\sum_{\bb{k},\sigma}\varepsilon_{\bb{k}}c^\dagger_{\bb{k},\sigma}c_{\bb{k},\sigma}+ \sum_{\bb{k},\sigma} \left(V_{\bb{k}} d_{\sigma}^\dagger c_{\bb{k},\sigma} + h.c \right)+H_{\rm imp}
\end{align}
Here the first term describes the Hamiltonian of the metallic bath with fermionic operators $c_{\bb{k},\sigma} ,c^\dagger_{\bb{k},\sigma} $, the second term describes the hybridization between impurity and bath with coupling $V_{\bb{k}}$ and  the last term describes the impurity Hamiltonian
\begin{align}
	H_{\rm imp}= \sum_\sigma \varepsilon_d n_{\sigma}+Un_{\uparrow}n_{\downarrow}
\end{align}
with impurity level $\varepsilon_d$ and Coulomb repulsion $U$ and $n_{\sigma}=d^\dagger_{\sigma}d_{\sigma}$. In the following we consider a half-filled metallic bath at zero temperature with a semicircular density of states of bandwidth $W$, giving rise to an  hybridization function  $\Gamma(\varepsilon)= 2 \pi \sum_{\bb{k}}V_{\bb{k}}^2\delta(\varepsilon-\varepsilon_{\bb{k}})= \Gamma\sqrt{1-(\varepsilon/W)^2} $. 

As shown in the sketch of Fig.~\ref{fig:sketchtwobody}, we consider the quantum impurity to be exposed to a Markovian bath characterized by set of local jump operators $L[d_{\sigma},d^{\dagger}_{\sigma}],L^{\dagger}[d_{\sigma},d^{\dagger}_{\sigma}]$. Because of this local dissipation the total system (impurity plus metallic bath) is in a mixed state described by the  density matrix $\rho_t$ which
evolves according to the many-body Lindblad master equation~\cite{fazio2025manybodyopenquantumsystems} 
\begin{align}\label{eqn:lindblad}
    \partial_t \rho_t = -i \left[ H, \rho_t \right] + \left( L \rho_t L^\dagger -\frac{1}{2} \{ L^\dagger L , \rho_t \}\right)
\end{align}
where $H$ is the AIM Hamiltonian while the jump operators that we will consider in the following describe 
two-body losses
\begin{align}
L=\sqrt{\gamma} d_\uparrow d_\downarrow
\label{eq:jump_operator}
\end{align}

It is worth to briefly comment on the general properties of this dissipator and the associated Lindblad master equation. First, due to losses the total number of particles in the system $N_{\rm tot}=\sum_\sigma n_\sigma + \sum_{\bb{k}\sigma} c_{\bb{k}\sigma}^\dagger c_{\bb{k}\sigma} $ is not conserved in time. Instead its time derivative, which we refer to as loss current, reads
\begin{align}\label{eqn:losscurrent}
\frac{d }{dt } \mbox{Tr}\left(\rho_t N_{\rm tot} \right)&=
-2\gamma \langle n_\uparrow n_\downarrow \rangle_t =-I_{\rm loss}
\end{align}
as one can obtain from the equations of motion. Losses here reflect the Markovian nature of the environment: particles once lost cannot go back into the system. This implies that a system with finite particle number would eventually deplete at long time, even if very slowly. In the context of dissipative quantum impurity models we are interested in the situation in which the thermodynamic limit of the fermionic bath is taken before the long-time limit.  In this case, the bath acts as an infinite reservoir and the system at long times sets into a non-equilibrium steady state that in general contains particle flow, i.e. a current~\cite{Froml2019}.

Finally, we comment on the symmetries of the Lindblad master equation. Two-body losses conserve the total spin which is therefore a strong symmetry of the Lindbladian. On the other hand this type of dissipation breaks particle-hole symmetry: indeed under the transformation $d_{\sigma} \rightarrow d^{\dagger}_{\sigma}$ the dissipative part of the Lindblad master equation is affected, as such we do not expect to be able to fix the energy level of the impurity $\varepsilon_d$ to guarantee half-filling during the time evolution. 
 
\subsection{Non-Crossing Approximation for Dissipative Impurity Models}

We solve the dynamics of the dissipative AIM described in the previous section using the recently developed self-consistent NCA dynamical maps~\cite{scarlatella2023selfconsistent}, which we briefly recall here for consistency. To this extent, we first reformulate the Lindblad many-body dynamics in Eq.~(\ref{eqn:lindblad}) using the purification/superfermion representation~\cite{Prosen_2008,dzhioev2011,HARBOLA2008191,dorda2014auxiliary,Arrigoni2018,werner2023configuration,takahashi96,ojima1981}. This amounts to represent density matrices as pure states $\vert\rho_t \rangle$ in an extended Hilbert space $\mathcal{H}\otimes \tilde{\mathcal{H}}$ that contains a copy of our degrees of freedom. In this formalism the Linbdlad master equation takes the form of a Schrodinger-like equation 
\begin{align}
\partial_t \vert\rho_t \rangle=\mathcal{L}\vert\rho_t\rangle
\end{align}
generated by a non-Hermitian operator $\mathcal{L}$, the Lindbladian, which reads
\begin{align}\label{eqn:Lvector}
    \m{L}= -i \left( H - \tilde{H}\right) + \gamma \left(L\tilde{L} - \frac{1}{2} L^{\dagger}L -\frac{1}{2} \tilde{L}^{\dagger}\tilde{L}\right)
\end{align}
where $H$ is given in Eq.~(\ref{eq:H_aim}) and $\tilde{H}$ takes the very same form in terms of fermions living in the $\tilde{\mathcal{H}}$ Hilbert space, $\tilde{c}_{\bb{k},\sigma} ,\tilde{c}^\dagger_{\bb{k},\sigma}$ and $\tilde{d}_{\sigma} ,\tilde{d}^\dagger_{\sigma}$ for the bath and impurity operators respectively, which satisfy the usual fermionic algebra. This representation bears similarities with the Keldysh formalism~\cite{mcdonald2023third}, where the role of the contours is played here by the duplicated Hilbert spaces 
$\mathcal{H}\otimes \tilde{\mathcal{H}}$ while the coupling between them is implemented by the quantum jump term in the master equation~\cite{fazio2025manybodyopenquantumsystems}.

Once the Lindbladian is written as in Eq. (\ref{eqn:Lvector}) we can then perform an exact hybridization expansion in the impurity-bath coupling, starting from the density matrix in the interaction picture~\cite{Scarlatella2019,scarlatella2023selfconsistent,vanhoecke2024diagrammatic}, and obtain an exact equation for the dressed impurity time-evolution operator $\m{V}(t,0)=\mbox{Tr}_{\rm bath}\left[\exp \left(\m{L}t \right)\right]$, which reads
\begin{align}
    \partial_t \m{V}(t,0) = \m{L}_{\rm imp} \m{V}(t,0) + \int_0^t d\tau \Sigma(t,\tau) \m{V}(\tau,0)
\end{align}
where $\m{L}_{\rm imp}$ is the impurity Lindbladian which reads in the present case
\begin{align}
\m{L}_{\rm imp}= - i \left( H_{\rm imp} - \tilde{H}_{\rm imp} \right) + \gamma d_\uparrow d_\downarrow \tilde{d}_\uparrow \tilde{d}_\downarrow - \frac{\gamma}{2} \left( n_\uparrow n_\downarrow + \tilde{n}_\uparrow \tilde{n}_\downarrow \right) 
\end{align}
while the self-energy $\Sigma(t,\tau)$ takes into account the effect of the metallic bath and it is a priori given by an infinite class of one-particle irreducible diagrams. To close the hierarchy we use the NCA dynamical map~\cite{scarlatella2023selfconsistent} corresponding to a self-consistent approximation on the series for $\m{V}$, by keeping only the compact diagrams in which hybridization lines do not cross~\cite{bickersBickers1987b,muller-hartmannMuller-Hartmann1984,meir1993low,nordlander1999how,ecksteinWerner2010a,hartle2013decoherence,erpenbeck2021resolving}
. This amounts to restrict the self-energy to the form $\Sigma(\tau,\bar{\tau})\equiv \Sigma_{NCA}(\tau,\bar{\tau})$ 
\begin{align}\label{eqn:sigmaNCA}
    \Sigma_{NCA}(\tau,\bar{\tau}) &= -i \sum_{\sigma\alpha,\bar{\alpha}}  \Psi^\alpha_\sigma \m{V}\left( \tau,\bar{\tau} \right) \Bar{\Psi}^{\bar{\alpha}}_\sigma \Delta_\sigma^{\alpha \bar{\alpha}}\left( \tau, \bar{\tau}\right) \nonumber\\
    &+i\sum_{\sigma\alpha,\bar{\alpha}} \Bar{\Psi}^{\bar{\alpha}}_\sigma \m{V}\left( \tau,\bar{\tau} \right) \Psi^\alpha_\sigma  \Delta_\sigma^{\alpha \bar{\alpha}}\left( \bar{\tau},\tau\right)
\end{align}
which is itself a function of $\m{V}(t,0)$, hence the self-consistent non-perturbative nature of this method. We note that in Eq.~(\ref{eqn:sigmaNCA}) we have collected the impurity operators in $\mathcal{H}$ and $\tilde{\mathcal{H}}$ into a field
$\bar{\Psi}_{\sigma}=(d^{\dagger}_{\sigma}\; \tilde{d}_{\sigma})$ whose components are labelled by the index $\alpha=0,1$ and introduced the real-time hybridization function $\Delta_{\sigma}^{\alpha\bar{\alpha}}(\tau,\bar{\tau})$, that keeps into account the full non-Markovian nature of the metallic bath 
(see Appendix~\ref{sec:hyb_exp} for definition.)
From the knowledge of the NCA dynamical map $\m{V}(t)$ we can compute the dynamics of the impurity degrees of freedom, the steady state properties as well as the impurity Green's functions describing excitations on top of the stationary state, using a result analog to the quantum regression theorem for Lindblad evolution~\cite{scarlatella2021dynamical}. 

\begin{figure*}[t!]
 	\centering
 \includegraphics[width=\textwidth]{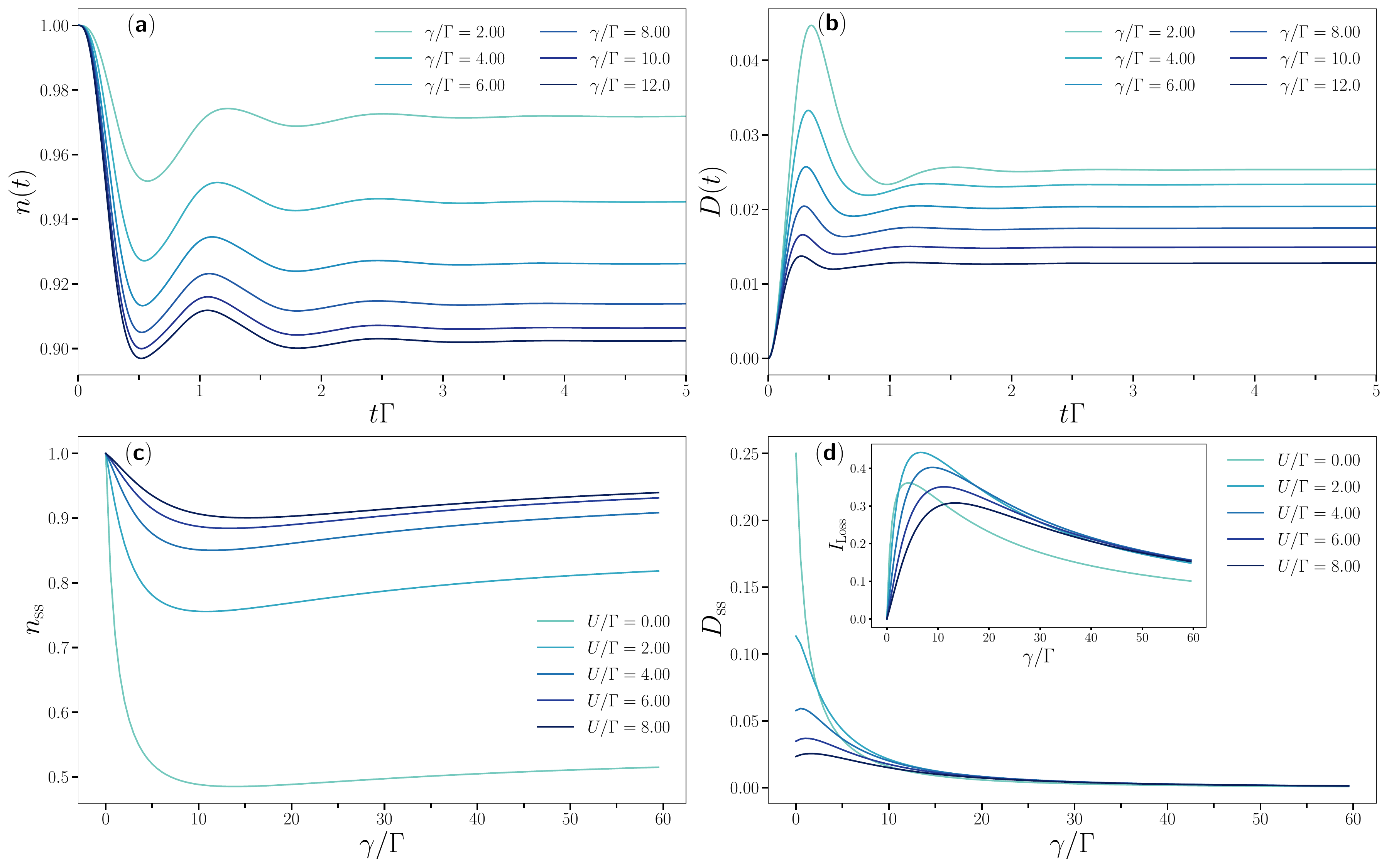}    \caption{\label{fig:TwoBodylosses_charge} Anderson Impurity Model with two-body losses -  Charge Dynamics in presence of two-body losses. (a-b) Dynamics of the impurity density and its steady-state value, for different value of dissipation $\gamma$ and $U = 8\Gamma$. (c-d) Dynamics of double occupation and steady-state value as a function of the dissipation strength $\gamma$ and interaction $U$. Inset: steady-state of the lossy current $I_{\rm loss} = - \frac{dN_{\rm tot}}{dt} = 2 \gamma D_{\rm ss}$. Here the impurity level is fixed to $\epsilon_d = -\frac{U}{2}$.}
\end{figure*}

\subsection{Exact Dynamics with Quantum Trajectories}\label{sec:exact}

We benchmark our numerical results with exact dynamics obtained by unravelling the Lindblad dynamics of the dissipative Anderson impurity model in quantum jump trajectories~\cite{Dalibard1992, Carmichael_book, Daley2014, nakagawa2020dynamical, fazio2025manybodyopenquantumsystems}. In the case of a dissipative quantum impurity, as discussed earlier, there is only one jump channel with the associated jump operator $L=\sqrt{\gamma}d_{\uparrow}d_{\downarrow}$. Specifically, we introduce a pure state $\vert \psi(\xi_t,t)\rangle$ which evolves according to the stochastic Schr\"odinger equation,
\begin{align}
d\ket{\psi_{\xi_t}(t )} 
&= -i dt \left[\mathcal{H}-\frac{i}{2} \left(L^\dagger L-\langle L^\dagger L\rangle_t \right)\right]\ket{\psi_{\xi_t}(t )} 
+ \nonumber\\
&+\left(\frac{L}{\sqrt{\langle L^\dagger L\rangle}}-1\right) d\xi_{t}\ket{\psi(\xi_t,t )}, 
\label{eq:qjump}
\end{align}
where $\langle\circ\rangle_t\equiv \langle \psi(\xi_t,t) \vert\circ\vert \psi(\xi_t,t)\rangle$, and $\xi_t=\left\{\xi_{t}\right\}$ represents independent Poisson processes ${d\xi_{t}\in\{0,1\}}$ characterized by the mean $\overline{d\xi_{t}} = dt \langle L^\dagger L\rangle_t$. In practical implementation, the dynamics is described as a series of quantum jumps, causing sudden changes in the wave function, alongside a deterministic, non-unitary evolution governed by a non-Hermitian Hamiltonian
\begin{equation}
    \mathcal{H}_{\rm nH}=\mathcal{H}-\frac{i}{2} L^\dagger L,
\end{equation}
which in the present case takes the form of an Anderson impurity model with a complex interaction. Upon averaging over the quantum jump trajectories, the resulting density matrix takes the form of
\begin{equation}
    \rho_t=\overline{\vert \psi(\xi_t,t)\rangle\langle  \psi(\xi_t,t)\vert},
\end{equation}
which evolves according to the Lindblad master equation written in Eq.~(\ref{eqn:lindblad}). In practice to implement the method it is convenient to choose a one dimensional geometry for the Hamiltonian of the Anderson impurity model. Here we take a one dimensional chain of non-interacting spinful fermions with the periodic boundary condition as a bath, and place the lossy Anderson impurity in the middle of the chain.

\section{Results}\label{sec:results}

In the following we present our numerical results obtained with the NCA dynamical map. Unless stated otherwise we consider the fermionic bath to be initially at zero temperature, we fix the energy level to $\varepsilon_d=-U/2$ and start from an initial state $\rho_0=\rho_{\rm 0,imp}\otimes \rho_{\rm 0,bath}$ with the impurity prepared in a polarised (up) state $\rho_{\rm 0,imp}=\vert \uparrow\rangle\langle\uparrow\vert$.

\begin{figure*}[t!]
 	\centering
 \includegraphics[width=\textwidth]{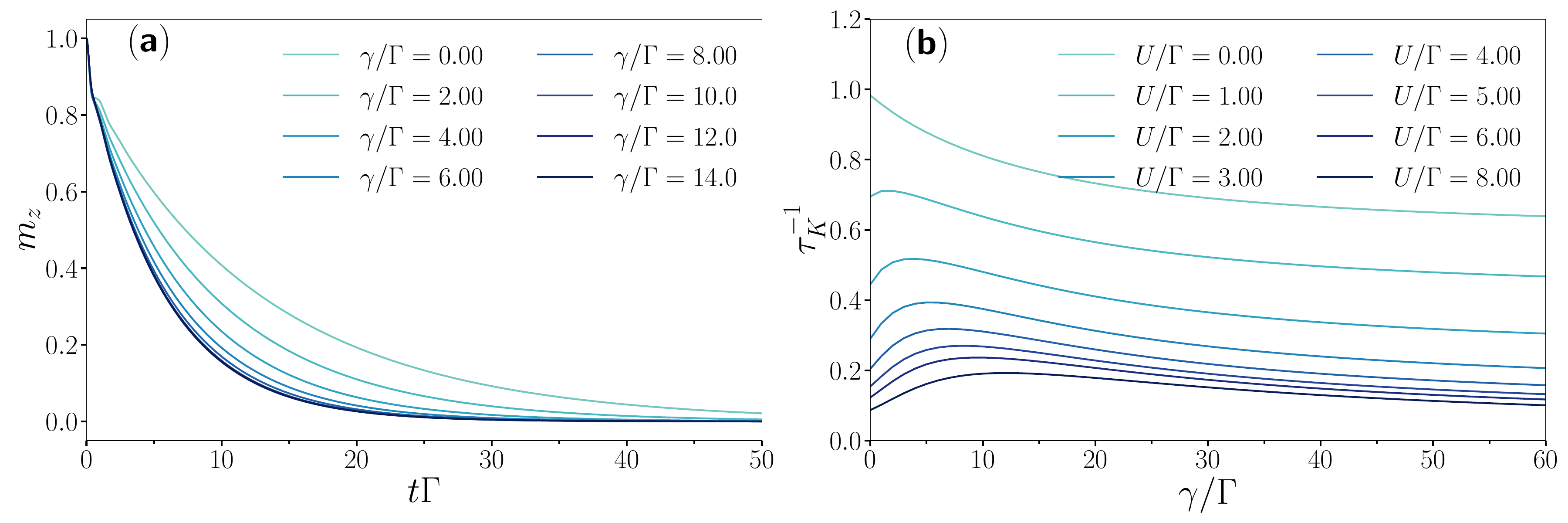}
    \caption{\label{fig:TwoBodylosses_Spindynamics} Anderson Impurity Model with two-body losses - The decay rate $\tau_K^{-1}$ is obtained by performing an exponential fit to the time evolution of the magnetization $m_z \sim \exp\left( -t/\tau_K \right)$, starting from a polarized initial state $\rho_{0,\rm imp}= \vert \uparrow \rangle \langle \uparrow \vert$. (a) the decay rate $\tau_K^{-1}$ depends non-monotonously on the dissipation rate $\gamma$. Contrary to the charge dephasing case, the spin relaxation time displays a strong $U$-dependence even for large $\gamma$. (b) The $U$-dependence is clearly visible in the dependence of $\tau_K^{-1}$ from the interaction. Here the impurity level is fixed to $\epsilon_d = -\frac{U}{2} = - 4 \Gamma $.}
\end{figure*} 
\subsection{Charge Dynamics and Steady-state population}

We start our analysis by discussing the dynamics of the impurity charge degrees of freedom. In Fig.~\ref{fig:TwoBodylosses_charge}(a)  we plot the impurity density as a function of time, 
\begin{align}
n(t)=\sum_{\sigma}\mbox{Tr} \left[ \rho_{\rm t,imp}n_{\sigma} \right]
\end{align} 
for $U=8\Gamma$ and different values of the loss rate $\gamma$. We first note that for $\gamma=0$, when the system is at particle-hole symmetry, the density remains constant to $n=1$. On the other hand for $\gamma\neq0$ the impurity density decreases in time and rapidly reaches a stationary value. This dynamics is purely induced by the two-body losses and contains some interesting physics. While $\gamma$ does not affect significantly the single particle charge dynamics, which is essentially controlled by the hybridization $\Gamma$,  it controls the steady-state value (see panel c). The steady-state impurity density decreases rapidly with $\gamma$ for $U=0$, where it approaches $n\simeq 0.5$ for large $\gamma$, while for larger value of the interaction we see that the impurity remains closer to half-filling.
Furthermore we can notice a characteristic non-monotonic behavior of particle density with the loss rate, signature of the Zeno effect~\cite{garcia-ripoll2009,scarlatella2021dynamical}.
To understand this effect we note that in presence of two-body losses the singly occupied impurity site can decay only via Zeno-like hybridization-induced processes, in which an electron from the metallic bath hops on the impurity site, creates a doublon which then decay. For $U=0$ the probability of creating a doublon is high and so losses are effective in depleting the impurity site, while for strong interactions doublon creation is energetically blocked, leading to an impurity which essentially remains close to half-filling.  We will come back to this point when discussing the effective model in Sec.~\ref{sec:SW}.

We can also look at the dynamics of double occupation,
\begin{align}
D(t)=\mbox{Tr} \left[ \rho_{\rm t,imp}n_{\uparrow}n_{\downarrow}  \right]
\end{align} 
that we plot in Fig.~\ref{fig:TwoBodylosses_charge}(b) for $U=8\Gamma$ and different values of $\gamma$. Starting from an initially singly occupied site, doublons are created by hybridization events with the metallic bath. For strong interactions, $U=8\Gamma$, this process is inefficient due to the large energy cost of a doublon and the two-body losses contribute further to reduce doublon injection. In panel (d) we plot the steady-state doublon population for different values of the loss rates and different interaction strengths.  The doublon number decreases monotonically with the loss rate for $U=0$, while for $U\neq0$ we observe a weak non-monotonic behavior for small $\gamma$, suggesting that in this regime adding two-body losses can increase the doublon fraction. For strong dissipation on the other hand the doublon fraction decays to zero, almost independently from the interaction. Overall, we see that the combination of two-body losses and strong interactions lead to a strong suppression of double occupations on the impurity site. The fraction of doublons at the impurity site is directly relate to the loss current, see Eq.~(\ref{eqn:losscurrent}). In the inset of Fig.~\ref{fig:TwoBodylosses_charge}(d) we plot the loss current in the steady-state as a function of the dissipation $\gamma/\Gamma$ which displays a clear signature of Zeno effect. The loss current grows at small $\gamma$ as expected, since dissipation results in a flow of pair of particles, then reaches a maximum around a scale $\gamma_{\rm max}\sim U$ and starts decreasing at large $\gamma$. Indeed, as doublon population decreases in the steady-state at strong dissipation the effective channel for particle losses closes and the system becomes effectively closed in the strong monitoring regime.   To summarize the analysis of charge dynamics we can say that interactions and two-body losses conspire together to create strong correlations at the impurity site, with a low doublon fraction and almost unit filling. In Appendix~\ref{app:steadystatepop} we present additional results on the steady-state population of the impurity.
\begin{figure*}[t!]
 	\centering
 \includegraphics[width=\textwidth]{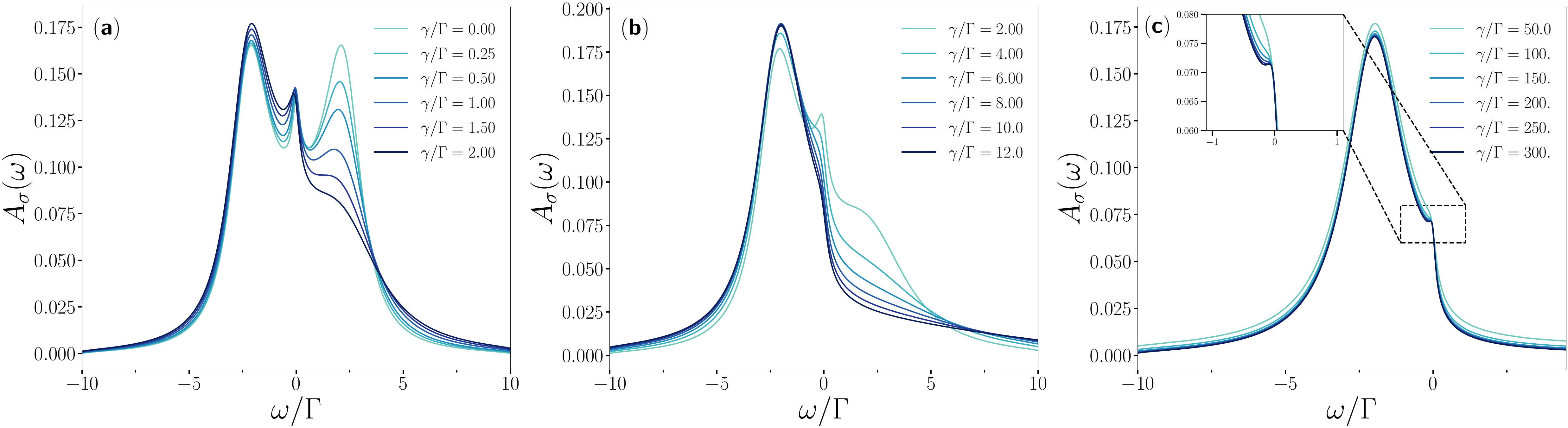}\caption{\label{fig:TwoBodylosses_SpectralFunct} Anderson Impurity Model with two-body losses - Impurity spectral function for $U = -2 \epsilon_d = 4\Gamma$ (Half-filled point) and for increasing values of $\gamma$. (a) We see that the low-frequency structure is robust to losses.(b) Upon increasing further $\gamma$ the upper Hubbard band merges with the Kondo resonance and the lower one.}
\end{figure*} 
\subsection{Spin relaxation rate and Kondo-Zeno crossover}

We then move to discuss the dynamics of the impurity spin degree of freedom, which is sensitive to the Kondo effect, and how it is affected by the two-body losses. We consider a protocol similar to the one discussed in the previous section, namely an initial decoupled spin, polarised along the $z$ direction, which is suddenly coupled to the bath and exposed to the dissipative process. We focus on the dynamics of the spin magnetization along the $z-$axis, defined as
\begin{align}
m_z(t)=\mbox{Tr} \left[ \rho_{\rm t,imp}(n_{\uparrow}-n_{\downarrow})  \right]
\end{align} 
where $\rho_{\rm t,imp}$ is the reduced density matrix of the impurity at time $t$ that we obtain directly from the NCA dynamical map. In Fig.~\ref{fig:TwoBodylosses_Spindynamics}(a) we plot the spin dynamics for different values of $\gamma$ and we show that the magnetization decays to zero at long-times with a behavior compatible with an exponential relaxation. 
We extract the spin-relaxation rate, $\tau_K^{-1}$, by fitting $m_z(t)\sim \exp(-t/\tau_K)$ and plot it in Fig.~\ref{fig:TwoBodylosses_Spindynamics}(b) as a function of the loss rate and for different values of $U/\Gamma$. In absence of two-body losses, $\gamma=0$, we see that the spin relaxation rate is strongly suppressed increasing $U/\Gamma$, a signature of the Kondo effect~\cite{anders2005realtime,wauters2023simulations,vanhoecke2025kondozenocrossoverdynamicsmonitored}. In this regime charge fluctuations are quenched and the only decay channel of the initially polarised spin is via the Kondo exchange $J_K\sim V^2/U$, a process that takes a time of order $1/T_K$, with $T_K$ the Kondo temperature. 

When losses are present but the impurity is non-interacting, i.e. $U=0$, we see that the spin decay rate  $\tau_K^{-1}$ decreases only weakly with $\gamma$ and remains large and order $o(1)$ even in the strongly dissipative regime $\gamma\gg \Gamma$. At finite dissipation $\gamma$ and finite interaction $U$ the relaxation rate displays a characteristic non-monotonic effect, a signature of Kondo-Zeno crossover~\cite{vanhoecke2025kondozenocrossoverdynamicsmonitored}. First the spin inverse relaxation time grows at small $\gamma$, since dissipation  helps the spin to decay: a doublon is formed on the impurity via a hybridization event, which is then lost leading a zero spin state. The spin decay rate  $\tau_K^{-1}$ reaches a maximum and then decreases again as the system enters the Zeno regime for $\gamma/\Gamma\gg 1$. Here doublons have all decayed and so the effective decay channel of the spin is suppressed. As for the loss current, the maximum as a function of $\gamma$ is for values $\gamma\sim U$. 
As compared to the non-interacting case $U=0$, we note that the spin-relaxation rate is suppressed by interactions across all values of $\gamma$. This is a first hint to the fact that Kondo physics in presence of a finite $U$ is robust to two-body losses.

\begin{figure*}[t!]
 	\centering
 \includegraphics[width=\textwidth]{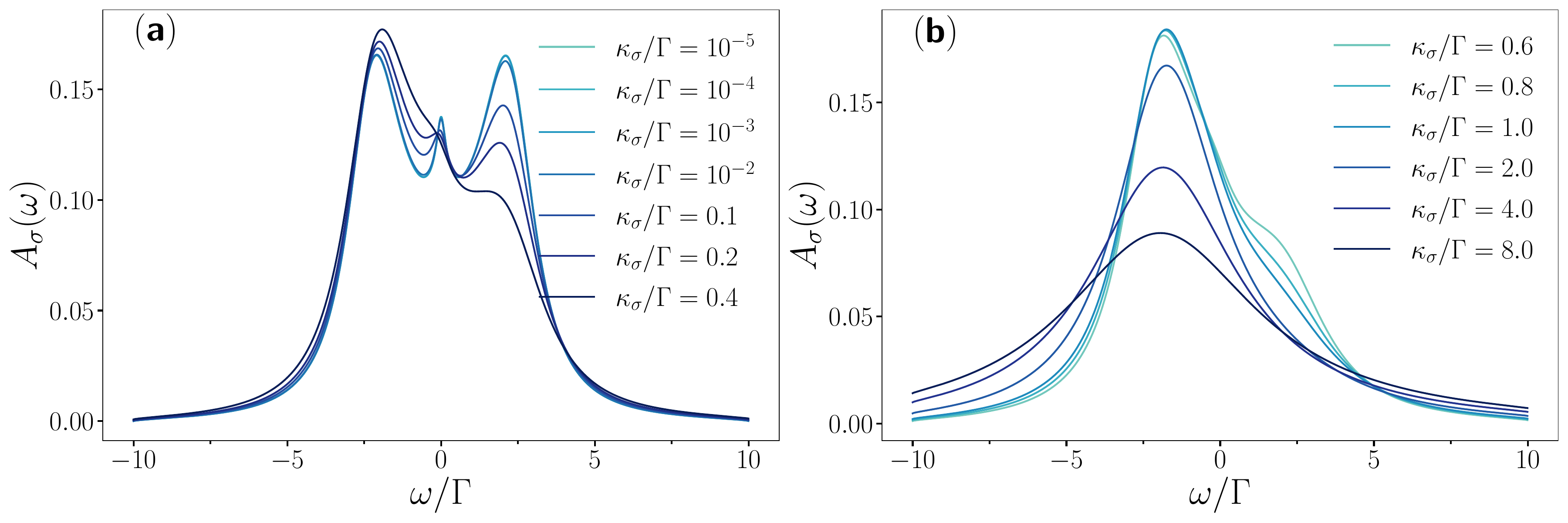}\caption{\label{fig:lossesSpectralFunct} Anderson Impurity Model with single-body losses - (a) Impurity spectral function for the half-filled lossy AIM and for $U = -2 \epsilon_d = 4 \Gamma $ and increasing values of $\kappa_\sigma$.(b) Upon increasing further $\kappa_\sigma$ the central peak and the upper Hubbard band are destroyed and ends up merging into a single-peaked structure. }
\end{figure*} 

\subsection{Impurity Spectral Function}

To obtain a better understanding of the steady-state properties of the system, and in particular of the interplay between Kondo physics and two-body losses, it is crucial to study the spectral function of the impurity, $A_{\sigma}(\omega)$, defined as the imaginary part of the impurity retarded Green's function
\begin{align}
    A_{\sigma  }(\omega) = -\frac{1}{\pi} \Im \left[ G^{R}_{\sigma }(\omega)\right]\,.
\end{align}
where  $G^{R}_{\sigma} (t,t^\prime) = -i \theta (t-t^\prime) \langle \{ d_\sigma(t) , d^\dagger_{\sigma} (t^\prime) \} \rangle$.  We plot this quantity in Fig.~\ref{fig:TwoBodylosses_SpectralFunct}(a-c)  for increasing values of the loss rate and at fixed interaction $U= -2 \epsilon_d =4 \Gamma$. Starting from panel (a) we see that two-body losses are very effective in collapsing the upper Hubbard band associated to doublons, whose height decreases rapidly already when $\gamma <U/2$. This indicates that virtual transitions where a doublon excitation is created have reduced probability to occur. On the other hand, the lower Hubbard band containing holon excitations is almost unchanged, its height weakly increases with $\gamma$. As a result and as previously anticipated, the steady-state is not particle-hole symmetric. Interestingly, we see that the coherent Kondo peak at zero frequency is not immediately washed out by weak two-body losses, but acquires a very asymmetric line shape. It is partially filled for negative frequencies while remaining sharply defined for positive ones, where it connects smoothly with the residual Hubbard-like excitation at higher frequencies. The strongly asymmetric lineshape of the Kondo resonance for weak two-body losses is one of the important result of this work. When the strength of two-body losses is further increased (panel b), the central coherent peak is also collapsed and merged with the lower-band. In this regime, which corresponds to $\gamma\sim U$ when the impurity population is the lowest, the impurity spectral function only describes one band of holon excitations. Finally, in the strongly dissipative regime $\gamma\gg U$ where doublons are effectively projected out and the system approaches half-filling (see Fig.~\ref{fig:TwoBodylosses_charge}) we see the emergence of a small coherent peak in the spectral function, an indication of a resurgence of the Kondo effect in this limit~\cite{stefanini2024dissipative}. From the spectral function we conclude that Kondo physics depends non-monotonously from two-body losses which at first protect the coherent peak, then collapse it and finally make it reappear. As we will discuss in Sec.~\ref{sec:SW}
we can interpret this behavior using an effective dissipative Kondo model.

\subsection{Comparison with Single Particle Losses}

\label{sec:onebody}

To appreciate the role of two-body losses with respect to Kondo physics it is useful to compare the results obtained so far with those from a different type of dissipative process,  which is single particle losses. This process,  more ubiquitous, is described by a jump operator acting on the impurity site of the form
\begin{align}
L=\sqrt{\kappa_\sigma}d_{\sigma}
\end{align}
We repeat the same analysis done in the previous section, namely we consider a dynamical protocol starting from an uncoupled spin-up which is suddenly coupled to a zero-temperature bath and to Markovian single-particle losses with strength $\kappa_\sigma$. We discuss the spin dynamics in Appendix~\ref{app:spin_singlebody} and focus here on the impurity spectral function in the steady-state, $A_{\sigma}(\omega)$, that we plot in Fig.~\ref{fig:lossesSpectralFunct}(a-b) for increasing values of $\kappa_\sigma$ at fixed $U=4\Gamma=-2\varepsilon_d$. We see that single particle losses also collapse the upper Hubbard band and increase the strength of the lower one, as it was the case for the two-body losses. However their impact on the Kondo peak is much stronger. Indeed we see that only for $\kappa_\sigma \gg \Gamma$ the spectral function is not affected and the Kondo peak remain well visible, while for $\gamma\sim 0.4\Gamma$ the coherent peak is already destroyed. Eventually, the peak is merged by the lower Hubbard band into an incoherent featureless band. This single peak structure eventually broadens up and acquires a Lorentzian like lineshape for large $\kappa_\sigma$ (panel b). Overall this result can be understood intuitively since single particle losses directly act on the single particle sector, destroying both charge and spin excitations and thus affecting directly the Kondo singlet. On the other hand this dissipative process lack dark states and thus the associated Zeno regime leading to a Kondo effect at strong dissipation.

\subsection{Comparison With Exact Dynamics}


\begin{figure*}[t!]
 	\centering
 \includegraphics[width=\textwidth]{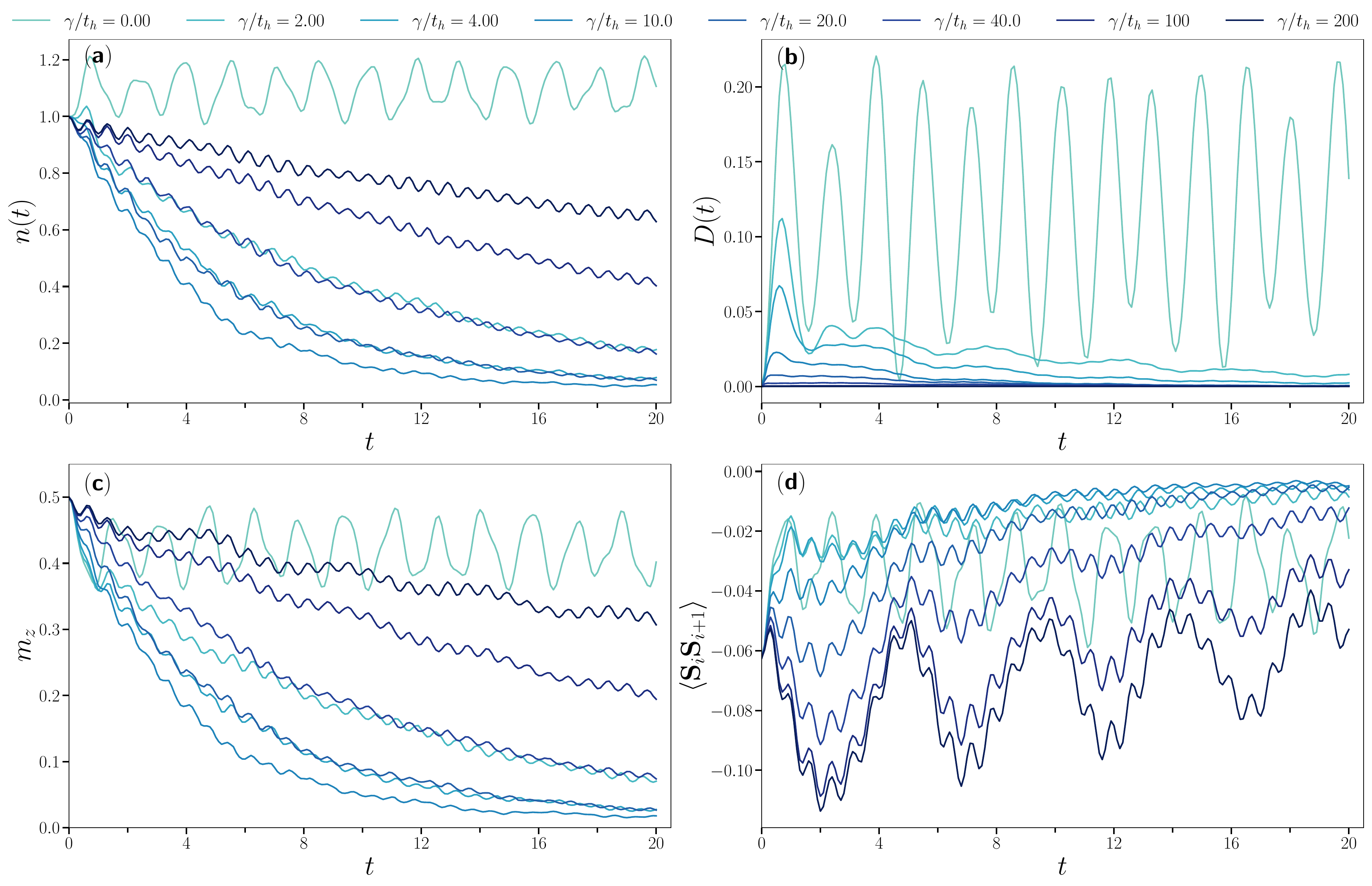}\caption{\label{fig:exactdynamics} Exact dynamics of the Anderson impurity model with two-body losses on a finite chain ($L=8, N_\uparrow=N_\downarrow=4, U=4$) for (a) the particle density, (b) double occupation, (c) magnetization at the impurity site, and (d) the spin-spin correlation between the impurity and neighboring sites. Here the hopping amplitude is set to $t_h=1$.}
\end{figure*} 

In this section, we present numerically exact results obtained with the quantum trajectory method \cite{Dalibard1992, Carmichael_book, Daley2014, nakagawa2020dynamical, fazio2025manybodyopenquantumsystems} combined with the exact diagonalization for the dissipative Anderson impurity model with finite bath sites (see Sec.~\ref{sec:exact}). The goal is to show that the predictions obtained with NCA are qualitatively captured by the exact dynamics at finite size. We emphasize that a direct quantitative comparison is less instructive, since the exact results are obtained for a finite-size system with a one-dimensional geometry, while the NCA works directly in the thermodynamic limit (with the semi-circular density of states).
We consider a chain of $L=8$ sites with the periodic boundary condition, and the impurity is located at site $i=0$. Using the quantum trajectory method, we solve the Lindblad master equation (\ref{eqn:lindblad}), where the Hamiltonian corresponds to that of the Anderson impurity model (\ref{eq:H_aim}) and the jump operator is given by Eq.~(\ref{eq:jump_operator}). The initial state is set to be $\rho_0=\rho_{0,{\rm imp}}\otimes \rho_{0,{\rm bath}}$ with $\rho_{0,{\rm imp}}=|\uparrow\rangle \langle\uparrow|$ and $\rho_{0,{\rm bath}}$ being the Gibbs state at zero temperature ($T=0$).
We use the hopping amplitude $t_h=1$ along the chain as the unit of energy, and fix the parameters to $U=4$ and $\varepsilon_d=-\frac{U}{2}$. The results are averaged over 1000 trajectories.

In Fig.~\ref{fig:exactdynamics}(a), we plot the particle density at the impurity site, which decays due to the two-body losses at $\gamma>0$. In contrast to the NCA results (Fig.~\ref{fig:TwoBodylosses_charge}(a)), we do not clearly observe a quasi-stationary state with a constant density, since the bath has a finite number of particles. Nevertheless, if we look at the dynamics of the double occupancy (Fig.~\ref{fig:exactdynamics}(b)) and magnetization (Fig.~\ref{fig:exactdynamics}(c)), we can see a similar behavior with those of the NCA calculations (see Fig.~\ref{fig:TwoBodylosses_charge}(b) and Fig.~\ref{fig:TwoBodylosses_Spindynamics}(a)). The double occupancy is quickly suppressed after an overshoot due to the two-body loss, and the suppression is monotonically enhanced as we increase $\gamma$. The dynamics of the magnetization, on the other hand, shows a non-monotonic dependence on $\gamma$, which is characteristic of the Kondo-Zeno crossover.
Namely, the spin relaxation is first accelerated as we increase $\gamma$ up to $\gamma\lesssim U$, and then it slows down for $\gamma\gg U$.
We emphasize that for such a small system the decay of the magnetization due to Kondo effect is not really visible, but the non-monotonic behavior confirms the prediction of our NCA results.
We also notice a similar non-monotonic behavior for the particle density, which is again consistent with the NCA results (see Fig.~\ref{fig:TwoBodylosses_charge}(c)).

In Fig.~\ref{fig:exactdynamics}(d), we plot the spin-spin correlation function $\langle \bm S_i \cdot \bm S_{i+1}\rangle$
between the impurity site $i=0$ and the neighboring bath site $i+1$ as a function of time. When $\gamma$ is small ($\gamma\lesssim U$), the spin correlation gradually decays from the initial antiferromagnetic correlation.
As we increase $\gamma$, we find that the antiferromagnetic correlation is enhanced and stays around a nonzero value for a long time, which is reminiscent of a partial recovery of the Kondo resonance at large $\gamma$ regime (see Fig.~\ref{fig:TwoBodylosses_SpectralFunct}(c)).
This behavior is in sharp contrast to the case of the Hubbard model with two-body losses \cite{nakagawa2020dynamical}, where the spin correlation is flipped from the antiferromagnetic to ferromagnetic one due to the effect of dissipation.

We can remove the effect of holes produced by quantum jumps and focus on the contribution of spins remaining at the impurity site by looking at the conditional projected correlator \cite{nakagawa2020dynamical}. The results (not shown) are similar to those of the original spin-spin correlator (Fig.~\ref{fig:exactdynamics}(d)), indicating that
the enhancement of the antiferromagnetic correlation at large $\gamma$ is essentially due to the consequence of the effective non-Hermitian dynamics.
We also confirm that when the system is noninteracting (i.e., $U=0$) there is neither clear enhancement of the antiferromagnetic correlation nor characteristic non-monotonic behavior of the magnetization, suggesting that these signatures arise as a result of the interplay between the Coulomb interaction and two-body losses.

\section{Effective Model via Schrieffer-Wolff Transformation}\label{sec:SW}

In this section we derive an effective Lindbladian for our model, using a generalized Schrieffer-Wolff (SW) transformation adapted to open quantum systems~\cite{kessler2012generalized,rosso2020dissipative,vanhoecke2025kondozenocrossoverdynamicsmonitored}. We begin with the vectorized form of the Lindbladian expressed in the superfermion representation, i.e. Eq.~(\ref{eqn:Lvector}), which we rewrite as:
\begin{align}
    \m{L} = \m{L}_0 + \m{L}_{\rm hyb}
\end{align}
Here, $\mathcal{L}_0$ denotes the diagonal part of the Lindbladian, meaning that it does not couple the high- and low-energy sectors, 
\begin{align}
    \m{L}_0 = \m{L}_{\rm imp} + \m{L}_{\rm bath}
\end{align}
where $\m{L}_{\rm imp}$ and $\m{L}_{\rm bath}$ include the decoupled impurity and
bath terms as well as the two-body losses. The term $\m{L}_{\rm hyb} = -i \left( H_{\rm hyb} - \tilde{H}_{\rm hyb}\right)$ captures the off-diagonal part. The SW transformation integrates out the bath–impurity coupling, via a non-unitary transformation generated by a generator S, such that the new Lindbladian
\begin{align}
    \m{L}_{\rm eff} = e^{S} \m{L} e^{-S} = \m{L} + \left[S,\m{L} \right] + \frac{1}{2} \left[ S ,\left[ S, \m{L} \right] \right] + \cdots 
\end{align}
is diagonal order by order in a perturbative expansion. In particular this can be achieved by choosing the generator $S$ to be
fully off-diagonal with respect to the hybridization and requiring it to satisfy the condition
\begin{align}
    \m{L}_{\rm hyb} + \left[ S, \m{L}_0\right] = 0
\end{align}
which ensures the cancellation of first-order hybridization contributions. Under this choice, the effective Lindbladian is obtained perturbatively to second order in $V_{\bb{k}}$ as,
\begin{align}
    \m{L}_{\rm eff} = \m{L}_0 + \frac{1}{2} \left[ S, \m{L}_{\rm hyb} \right] + O(V_{\bb{k}}^2)
\end{align}
In the following, we first derive the form of the generator S and then write down the effective Lindbladian. After some algebra (see App.~\ref{app:SW}) we obtain the following form of effective Lindbladian. The effective Lindbladian is composed of different contributions, which we now discuss in detail
\begin{align}
    \m{L}_{\rm eff} = \m{L}_{0} + \m{L}_{\rm imp}^\prime+ \m{L}_{\rm pair} + \m{L}_{\rm Kondo} + \m{L}_{\rm scatt} + \m{L}_{\rm diss}
\end{align}
The first contribution is the diagonal part of the Lindbladian. $\m{L}_{\rm Kondo} = -i \left( H_{\rm Kondo} - \tilde{H}_{\rm Kondo}\right) $ describes a Kondo coupling between the impurity spin and the spin of the bath,
\begin{align}
    H_{\rm Kondo} =  - \sum_{\bb{q} \bb{k}} J_{\bb{q}\bb{k}} \vec{S}_d \cdot \vec{s}_{\bb{q}\bb{k}}
\end{align}
with a Kondo coupling $J_{\bb{q}\bb{k}} =\mbox{Re}\left( V_\bb{k} B_{\bb{k}} + V_\bb{q} B_\bb{k}\right)$ which reads
\begin{align}
J_{\bb{q}\bb{k}}=\mbox{Re}\left(\frac{V_{\bb{k}} V_{\bb{q}} \left( U -i \gamma/2 \right)}{\left(\epsilon_\bb{k} - \epsilon_d \right) \left( \epsilon_\bb{k} - \epsilon_d - U + i \gamma/2\right)}\right) + \left( \bb{q}  \leftrightarrow \bb{k} \right)
\end{align}
Disregarding the momentum dependence of the coupling, which is usually a reasonable approximation at long-times/low energy, this gives rise to a Kondo coupling of the form
\begin{align}\label{eq:JKondo}
J= - \frac{8V^2 \left[U^2 + \gamma^2 /2 \right]}{U \left[ U^2 + \gamma^2 \right]} 
\end{align}  
In addition to the renormalized Kondo coupling, two body losses also give rise to a number of other terms (see App.~\ref{app:SW}). A particularly relevant one describes effective non-local two-body losses, where one particle is destroyed on the impurity and one in a bath-mode. This is described by a term of the form
\begin{align}
    \m{L}_{\rm TbL,eff} \left[ \bullet \right]  = \sum_{\bb{k}\bb{q}} \kappa_{\bb{k}\bb{q},\rm eff} L_{\bb{k},\rm eff} \bullet L_{\bb{q},\rm eff}^\dagger - \frac{\kappa_{\bb{k}\bb{q},\rm eff} }{2} \{L_{\bb{k}, \rm eff}^\dagger L_{\bb{q}, \rm eff}, \bullet \}
\end{align}
where $L_{\bb{k}, \rm eff}$ is the effective jump operator,
\begin{align}
    L_{\bb{k},\rm eff} = \sum_\sigma \sigma c_{\bb{k}\sigma} d_{\bar{\sigma}}
\end{align}
and $\kappa_{\bb{k}\bb{q},\rm eff}$ the dissipation strength,
\begin{align}
    \kappa_{\bb{k}\bb{q},\rm eff} =  \frac{V_{\bb{k}} V_{\bb{q}} \gamma }{2  \left[  (\epsilon_\bb{k} - \epsilon_d - U)^2 + \gamma^2 /4 \right]}  + \left( \bb{q}  \leftrightarrow \bb{k} \right)
\end{align}
At the Fermi energy ($\epsilon_\bb{k} = 0$) this dissipation strength takes the form
\begin{align}\label{eq:kappa_eff}
    \kappa_{\rm eff} = \frac{ 4 V^2 \gamma}{U^2 + \gamma^2} 
\end{align}
The non-local dissipation plays a crucial role in determining the stability of the Kondo state. Indeed, the Kondo effect relies on the formation of long-lived quantum correlations between the localized impurity spin and the surrounding conduction electrons. This process is develops over the characteristic Kondo time scale $\tau_k \sim 1/T_K$, where $T_K$ is the Kondo temperature. So if the timescale $\kappa_{\rm eff}^{-1}$ associated with the non-local loss mechanism is shorter than $\tau_k$, these many-body correlations are disrupted before the Kondo singlet can form, effectively destroying the Kondo state.
Remarkably, our results show that this effective non-local loss can be suppressed by increasing either the coherent interaction strength 
U ($\kappa_{\rm eff} \sim 4V^2 \gamma/ U^2 $ for $U\gg \gamma$) or the dissipation rate $\gamma$ ($\kappa_{\rm eff} \sim 4V^2 / \gamma $ for $U\ll \gamma$).


\section{Discussion}\label{sec:disc}


The results presented in the previous sections demonstrate that the interplay between strong Coulomb repulsion and two-body losses gives rise to new regimes for Kondo physics. In particular we show that a strongly correlated dissipative Kondo state emerges both at weak and at strong dissipation. In the first regime one can naively expect that the two-body losses are not effective on a well formed Kondo state since they conserve the spin and only affect doublons.  The key point for this protection is that the residual, Zeno mediated, single particle loss rate is suppressed at small $\gamma$ by strong Coulomb repulsion and at large dissipation by the combination of Kondo and Zeno effects.

It is interesting to compare and discuss our results in light of recent literature. In the non-interacting dissipative case, corresponding to $U=0$ and large $\gamma$, the problem was studied in Refs.~\cite{stefanini2024dissipative,qu2025variational}. There it was shown that Kondo physics associated to an infinite-$U$ Anderson model emerge in the strongly dissipative regime $\gamma\rightarrow\infty$, with residual losses at finite $\gamma$. Our results complete this picture by showing the impact of strong Coulomb correlations. These are crucial to give rise to Kondo physics, see Eq.~(\ref{eq:JKondo}) controlling the dynamics of the effective model. Importantly, as we see in Eq.~(\ref{eq:JKondo}), the Kondo coupling remains finite also for large dissipation. Furthermore, interaction is responsible for renormalizing the residual rate of losses, see Eq.~(\ref{eq:kappa_eff}), which can be made parametrically smaller than Kondo coupling by suitably increasing $U$.

Our results share some similarities but also key differences with respect to the Kondo-Zeno crossover obtained in the case of charge dephasing~\cite{vanhoecke2025kondozenocrossoverdynamicsmonitored}. In both cases the Kondo effect appears to be robust in the regime of weak dissipation, as seen from the spin dynamics and the spectral function. At the level of the effective model both problems give rise to a Kondo model with residual dissipation. In the charge dephasing case particle-hole symmetry is preserved, which together with strong correlations suppressing heating allows to project away dissipative terms in the effective Lindbladian. In the case of losses on the other hand we have shown that weak two-body losses between impurity and bath remains and are likely to be relevant, even if they can be suppressed for $U\gg \gamma$. Major differences between the two settings appear in the strongly dissipative regime, where losses do not give rise to the extensive heating and doublon production obtained with dephasing. As such, for example the spin relaxation rate or the spectral function display signatures of Kondo correlations also for $\gamma/\Gamma\gg 1$, while in the dephasing case heating due to doublon production was controlling the physics. This difference appears clearly in the form of the effective Kondo coupling in Eq.~(\ref{eq:JKondo}). In the case of charge dephasing it was shown in Ref.~\cite{vanhoecke2025kondozenocrossoverdynamicsmonitored} that the effective Kondo coupling scale as $J\sim -8V^2U/\gamma^2-i8V^2/\gamma$, i.e. vanishing for $\gamma\gg U$. On the other hand in the case of two-body losses the real-part of the effective Kondo coupling remains finite even for $\gamma\gg U$.

Finally, it is worth mentioning that our results also share important insights on the role of quantum jumps in the physics of dissipative Anderson impurity models. Indeed, the no-click non-Hermitian Hamiltonian associated to the two-body losses and charge dephasing are directly related and given by an Anderson impurity model with complex interaction $U-i\gamma/2$ (plus a trivial shift). As discussed in the weak dissipation regime the two models behave similarly which is tempting to attribute to the physics of the non-Hermitian impurity model. Yet, for large dissipation the role of quantum jumps is crucial and control the differences observed between two body losses and charge dephasing.

\section{Conclusions}\label{sec:conclusions}

In this work, we have studied the physics of the Anderson impurity model with local two-body losses as an incarnation of a strongly correlated dissipative quantum impurity model. Using a self-consistent hybridization expansion and a non-crossing approximation for the many-body Lindbladian we have solved the dynamics of impurity operators and calculated as well the correlation functions above the steady-state. Our results show that the interplay between strong correlation and losses give rise to rich and robust Kondo physics both at weak dissipation, where strong interactions suppress charge fluctuations and the associated effective single particle losses, as well as at strong dissipation where the Zeno effect push the impurity towards a regime of half-filling with small residual charge fluctuations. This is clearly seen in the behavior of the spin relaxation rate, where strong suppression due to interactions (a hallmark of Kondo physics) remains visible throughout the Kondo-Zeno crossover. Furthermore, it appears clearly in the spectral features which show rapid destruction of the upper Hubbard band but a low frequency peak which is resilient to weak dissipation and re-emerge for strong one. We have compared the results on dynamics against exact numerical simulations on finite-size chains, confirming the picture obtained with NCA. Finally, we have shown that the above results are due to the correlated nature of two-body losses: specifically, we have shown that single particle losses are strongly detrimental for any interesting many-body effect. In particular their effect on the dynamics of the spin or the impurity spectral function opens up a new channel of dissipation and leads to an incoherent mixed state which decays faster and has broad spectral features.

\begin{acknowledgements}
We acknowledge computational resources on the Coll\'ege de France IPH cluster. This project has received funding from the European Research Council (ERC) under the European Union’s Horizon 2020 research and innovation programme (Grant agreement No. 101002955 — CONQUER).
N.T. acknowledges support by JST FOREST (Grant No.~JPMJFR2131) and JSPS KAKENHI (Grant Nos.~JP24H00191, JP25H01246, and JP25H01251).
\end{acknowledgements}

\appendix

\section{Hybridization Expansion in Superfermion Representation}\label{sec:hyb_exp}
In this section, we detail for completeness the hybridization expansion in the superfermion representation of the lossy Anderson impurity model. This expansion corresponds to a perturbative series of the evolution superoperator in powers of the impurity-bath coupling, taken to all orders 
\begin{align}
    \vert \rho_t \rangle = \m{V}_{0}(t) \m{T}_t \exp \left( \int_0^t ds \m{L}_{\rm hyb}(s) \right) \vert \rho_0 \rangle
\end{align}
where $\m{V}_0 (t)= \exp \left( \m{L}_0 t\right)$ is the time evolution operator in the decoupeld limit, i.e $\m{L}_0 = \m{L}_{\rm imp} + \m{L}_{\rm bath}$. The impurity-bath coupling term is expressed in the interaction picture as $\m{L}_{\rm hyb}(t) = e^{-\m{L}_0 t}\m{L}_{\rm hyb} e^{\m{L}_0 t}$, evolving under the free dynamics generated by $\m{L}_0$. We now proceed to perform a formal expansion in the hybridization term between impurity and bath. Using the fact that the hybridization is a bilinear of fermionic operators and that the bath is made of non-interacting (gaussian) degrees of freedom we can further simplify the expansion. In particular, we can perform the trace over the bath degrees of freedom, $\Tr_{\rm bath} \left[ \cdots \right] = \langle I_{\rm bath} \vert \cdots \vert \rho_0 \rangle $ assuming that the initial state $\vert \rho_0 \rangle$ s factorized, i.e. $\rho_0 = \rho_{\rm 0,imp} \otimes \rho_{\rm 0,bath}$. Given that the bath is non-interacting, Wick's theorem can be applied to evaluate the time-ordered bath correlations. This leads to the hybridization expansion of the reduced impurity dynamics, captured by the dressed propagator $\langle I_{\rm bath} \vert \rho_t \rangle = \m{V}(t,0) \vert \rho_{\rm 0,imp} \rangle$, which takes the form:
\begin{widetext}
    \begin{align*}
    \vert \rho_t \rangle = \sum_n \sum_{ \{ \sigma, \bar{\sigma} \} }  \frac{1}{2n!}& \int_0^t d\tau_1 \cdots d\tau_{2n}  \m{V}_{0}(t) \m{T}_t \left[ \bar{\Psi}_\sigma(\tau_1)  \cdots  \Psi_\sigma (\tau_{2n})\right]\vert \rho_0 \rangle \times \rm{Det}_\sigma \left[ \{ \bar{\Phi}_\sigma, \Phi_\sigma \} \right] \rm{Det}_{\bar{\sigma}} \left[ \{ \bar{\Phi}_\sigma, \Phi_\sigma \} \right]
\end{align*}
\end{widetext}
where $\rm{Det}_\sigma \left[ \{ \bar{\Phi}_\sigma, \Phi_\sigma \} \right]$ are determinants built out of the real-time hybridization functions of the bath defined as
\begin{align}
    \Delta_\sigma^{ \bar{\alpha} \alpha} (\tau, \bar{\tau})= - i \langle I_{\rm bath} \vert \m{T}_t \left[ \Phi_\sigma^{\bar{\alpha}} (\bar{\tau}) \bar{\Phi}_\sigma^{\alpha}(\tau) \right] \vert \rho_{\rm 0,bath} \rangle
\end{align}
which is basically the two-times correlation function of the bath. For the non-interacting
bath considered here, a simple calculation yields
\begin{align}
    & \Delta^{01}_\sigma (\tau ,\bar{\tau}) = \int d\epsilon n_{\rm F}(\epsilon) \Gamma_\sigma (\epsilon) e^{-i\epsilon (\tau -\bar{\tau})} \notag \\ & 
    \Delta^{10}_\sigma (\tau ,\bar{\tau}) = -\int d\epsilon (1-n_{\rm F}(\epsilon) ) \Gamma_\sigma(\epsilon) e^{-i\epsilon (\tau -\bar{\tau})}
\end{align}
where $n_F(\epsilon)$ is the Fermi distribution and $\Gamma(\epsilon)$ the energy-dependent hybridization for the spin channel $\sigma$, defined in the main text.

The hybridization expansion we have derived above is a bare expansion in the impurity-bath coupling. It can be reorganized in a self-consistent diagrammatic expansion by identifying one-particle irreducible diagrams and introducing a self-energy $\Sigma(t,t')$ to obtain a Dyson equation for $\m{V}(t,0)$ of the form~\cite{scarlatella2023selfconsistent}\begin{align}\label{eqn:dynamicalmap}
    \m{V}(t,0) = \m{V}_{\rm imp}(t,0) + \int_0^t d\tau \int_0^\tau d \bar{\tau} \m{V}_{\rm imp} \left( t, \tau \right) \Sigma \left( \tau,\bar{\tau} \right) \m{V} \left( \bar{\tau}, 0\right)
\end{align}
which can be rewritten as in the main text upon taking a time-derivative on both sides of the equation and using $\partial_t \m{V}_{\rm imp}(t,0)=\m{L}_{\rm imp}\m{V}_{\rm imp}(t,0)$. As in any interacting diagrammatic theory the self-energy is not known in closed form. It can however reorganised as an expansion in diagrams with increasing number of crossing hybridization lines~\cite{scarlatella2021dynamical,scarlatella2023selfconsistent}. The lowest order self-consistent level of this hierarchy is the so called non-crossing approximation (NCA) which corresponds to keeping only the compact diagrams in which hybridization lines do not cross. The formal expression for the NCA self-energy is therefore the one given in the main text.
which describes a rainbow diagram with a dot vertex 
$\bar{\Psi}_{\sigma}^{\bar{\alpha}}$ at time $\bar{\tau}$, a dressed impurity propagation from $\bar{\tau}$ to $\tau$ with propagator $\m{V}(\tau,\bar{\tau})$ and a hybridization line $\Delta_\sigma^{\alpha \bar{\alpha}}\left( \tau, \bar{\tau}\right)$ and finally another impurity vertex operator $\Psi_{\sigma}^{\alpha}$ at time $\tau$, with the second term in Eq.~(\ref{eqn:sigmaNCA}) describing the diagram with $\tau,\bar{\tau}$ exchanged.

\begin{figure*}[t!]
 	\centering
 \includegraphics[width=\textwidth]{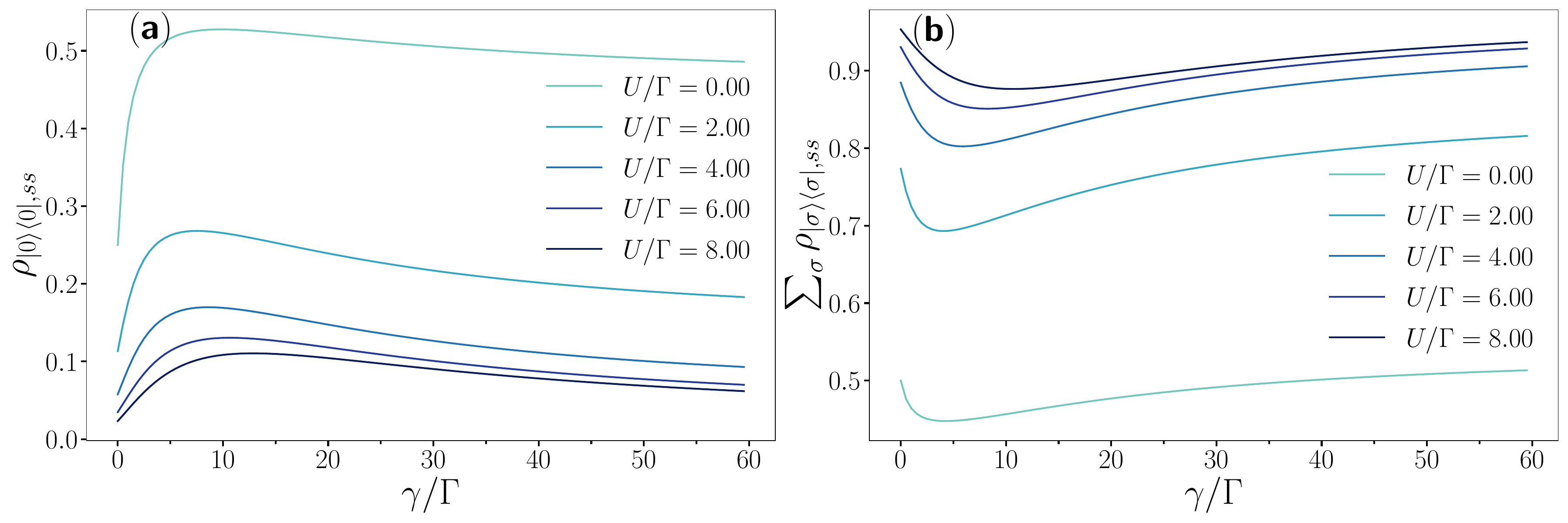}
    \caption{\label{fig:HolonSingly} Half-Filled Anderson Impurity Model with Two-Body Loss - (a) Steady-state fraction of holon occupancy as a function of dissipation rate $\gamma$ and interaction strength $U$. (b) Steady-state fraction of singly occupied states as a function of dissipation rate $\gamma$ and interaction strength $U$. }
\end{figure*} 

\section{Impurity Steady-State Population}\label{app:steadystatepop}

To gain insight into the steady-state properties of the system, we analyze the effects of the interaction $U$ and the two-body loss rate $\gamma$ on the holon state ($\vert \alpha \rangle = \vert 0 \rangle$)  and the singly occupied states ($\vert \alpha \rangle = \vert \sigma \rangle$). These occupations are defined as
\begin{align}
    \rho_{\vert \alpha \rangle \langle \alpha \vert , ss} = \Tr \left( P_\alpha \rho_{\rm ss,imp}  \right)
\end{align}
where $P_\alpha = \vert \alpha \rangle \langle \alpha \vert$ is the projector into the state $\vert \alpha \rangle$ and $\rho_{\rm ss,imp}$ denotes the steady-state density matrix. In Fig.~\ref{fig:HolonSingly}, we show the steady-state occupations of the holon state (panel a) and the singly occupied states (panel b) as functions of the dissipation $\gamma$, for several values of the interaction $U$. From Fig.~\ref{fig:HolonSingly}(a), we observe that the holon population increases with the dissipation $\gamma$, especially in the weakly interacting regime. This reflects the fact that two-body losses primarily act on doublon states, effectively converting them into empty states.
Conversely, the singly occupied state population shows a non-monotonic dependence on $\gamma$.

\section{Anderson Impurity Model with Two-body losses away from Half-Filling }
\begin{figure}[t!]
 \includegraphics[width=0.45\textwidth]{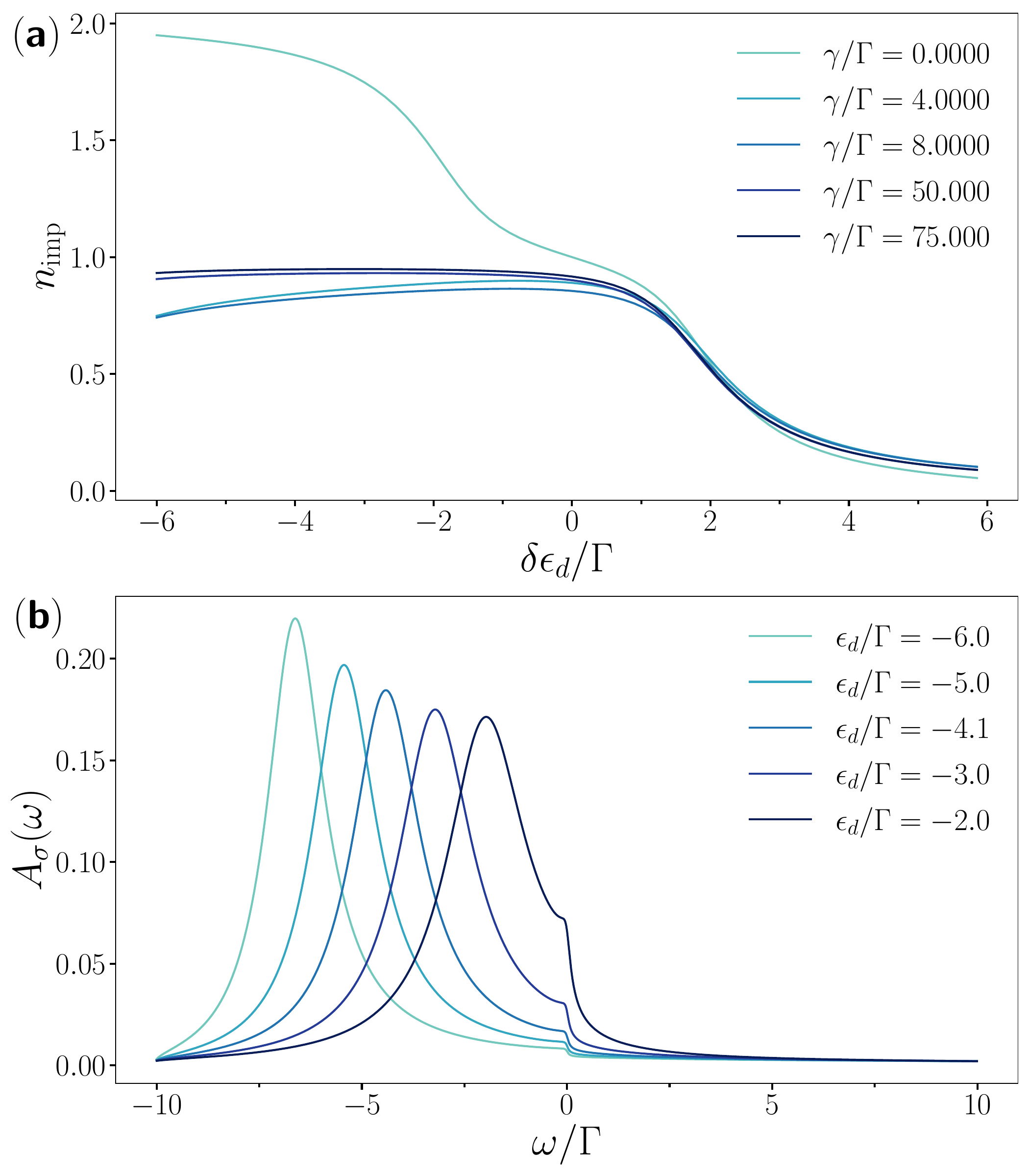}
    \caption{\label{fig:Function_epsilond} Anderson Impurity Model with two-body losses - (a) Impurity occupancy, as a function of the shifted impurity level $\delta \epsilon_d = \epsilon_d + U/2$ and dissipation $\gamma$. (b) Impurity spectral function $A_{\sigma}(\omega)$ for the asymmetric Anderson impurity model. For fixed interaction $U=  4 \Gamma$ and dissipation $\gamma = 150 \Gamma$. }
\end{figure}
Here, we present additional results for the spectral function and the impurity occupancy away from the particle-hole symmetric point. In Fig.~\ref{fig:Function_epsilond}(a), we plot the impurity density as a function of the shifted impurity level $\delta \epsilon_d =\epsilon_d + U/2 $ for increasing values of the loss rate $\gamma$, at fixed interaction strength $U = 4 \Gamma$. In the absence of loss,  the impurity density exhibits the characteristic "staircase" profile: it approaches the doubly occupied value  $n_{\rm imp} \approx 2$ for $ \delta \epsilon_d \ll 0$, the empty density $n_{\rm imp} \approx 0$ for $ \delta \epsilon_d \gg 0$ and the local moment regime with the impurity density $n_{\rm imp} \approx 1$ near $\delta \epsilon_d = 0$. As the loss rate $\gamma$ increases, this staircase structure is progressively washed out. Notably, the doubly occupied regime is strongly suppressed, with the impurity occupancy dropping below half-filling. For sufficiently large dissipation, the occupancy flattens and approaches a constant value of $n_{\rm imp}$ throughout the regime $\delta \epsilon_d < 0$. 
This behavior is further confirmed by examining the spectral function in the strong dissipation regime $\gamma \gg U$, shown in Fig.~\ref{fig:Function_epsilond}(b). There, we observe that for all values of $\delta \epsilon_d$ , the spectral function exhibits qualitatively similar features, including a persistent Kondo resonance, with only a shift in the position of the lower Hubbard band.

\section{Spin Decay Rate for Single Particle Losses}\label{app:spin_singlebody}
\begin{figure}[t!]
 	\centering
    \includegraphics[width=0.47\textwidth]{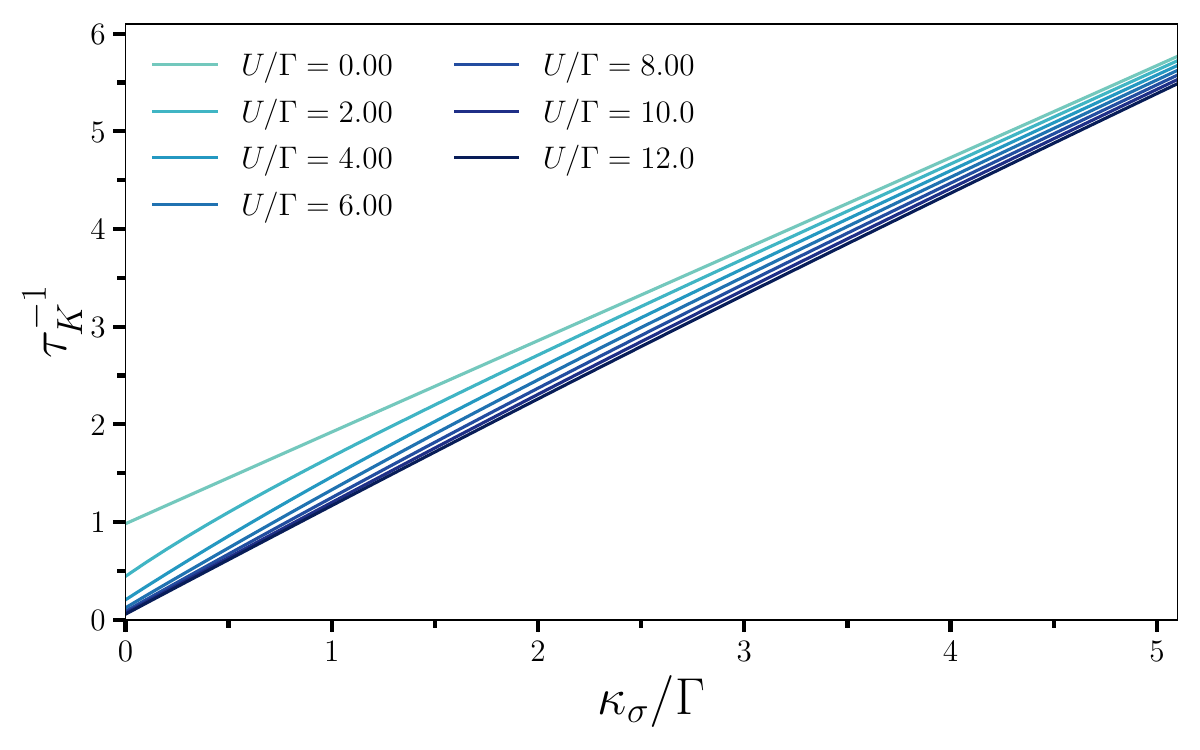}\caption{\label{fig:lossesTauK} Anderson Impurity Model with One-Body Loss - Decay rate $\tau_K^{-1}$ as a function of the loss rate $\kappa_\sigma$. Here the dissipation considered is the single particle loss, described by the following quantum jump operators $L_\sigma = \sqrt{\kappa_\sigma} d_\sigma$. Here the impurity level is fixed to $\epsilon_d = -\frac{U}{2}$.  }
\end{figure} 

We first focus on the spin dynamics and in particular on the spin relaxation time. By fitting the dynamics of the impurity magnetization $m_z(t)$ with an exponential decay, we can extract the inverse Kondo time $\tau_K^{-1}$, that we plot in Fig.~\ref{fig:lossesTauK} as a function of $\kappa_\sigma$ and different values of the interaction $U$. We see that the lifetime of the spin excitation increases almost linearly with $\kappa_\sigma$, as expected when adding a new dissipative channel. This implies that the impurity spin decays faster, which is consistent with the fact that single particle losses are already enough to make decay the isolated spin, without the need of the coupling to the bath. Furthermore, the inverse Kondo lifetime depends very weakly from the interaction and only at weak $\kappa_\sigma$,
again pointing to the fact that single particle losses are detrimental for any interesting many-body effect.

\section{Schrieffer-Wolff Transformation}\label{app:SW}
In this section, we present a detailed derivation of the effective Lindbladian, by using a generalized Schrieffer-Wolff (SW) transformation adapted to open quantum systems. We begin with the vectorized form of the Lindbladian expressed in the superfermion representation, which we rewrite as:
\begin{align}
    \m{L} = \m{L}_0 + \m{L}_{\rm hyb}
\end{align}
Here, $\mathcal{L}_0$ denotes the diagonal part of the Lindbladian, meaning that it does not couple the high- and low-energy sectors, 
\begin{align}
    \m{L}_0 = \m{L}_{\rm imp} + \m{L}_{\rm bath}
\end{align}
where $\m{L}_{\rm imp}$ and $\m{L}_{\rm bath}$ include the decoupled impurity and
bath terms as well as the two-body losses. The term $\m{L}_{\rm hyb} = -i \left( H_{\rm hyb} - \tilde{H}_{\rm hyb}\right)$ captures the off-diagonal part. The SW transformation integrates out the bath–impurity coupling, via a non-unitary transformation generated by a generator S, such that the new Lindbladian
\begin{align}
    \m{L}_{\rm eff} = e^{S} \m{L} e^{-S} = \m{L} + \left[S,\m{L} \right] + \frac{1}{2} \left[ S ,\left[ S, \m{L} \right] \right] + \cdots 
\end{align}
is diagonal order by order in a perturbative expansion. In particular this can be achieved by choosing the generator $S$ to be
fully off-diagonal with respect to the hybridization and requiring it to satisfy the condition
\begin{align}
    \m{L}_{\rm hyb} + \left[ S, \m{L}_0\right] = 0
\end{align}
which ensures the cancellation of first-order hybridization contributions. Under this choice, the effective Lindbladian is obtained perturbatively to second order in $V_{\bb{k}}$ as,
\begin{align}
    \m{L}_{\rm eff} = \m{L}_0 + \frac{1}{2} \left[ S, \m{L}_{\rm hyb} \right] + O(V_{\bb{k}}^2)
\end{align}
In the following, we first derive the form of the generator S and then write down the effective Lindbladian.
\subsection{The generator S}
The constraint equation for $S$ indicates that its structure must include a diagonal component, acting separately within each Hilbert space and arising from the coherent Hamiltonian, as well as an off-diagonal part that couples the original and tilde Hilbert spaces, generated by the dissipative quantum jump term. We therefore parametrize it as following
\begin{align}
    S = S_0 + S_\gamma 
\end{align}
where $S_0$ has the standard structure of the SW generator for the Anderson impurity model
\begin{align}
    S_0 = \sum_{\bb{k}\sigma} X_{\bb{k}\sigma} \left[c_{\bb{k}\sigma}^\dagger d_\sigma - h.c \right] + \sum_{\bb{k}\sigma} \tilde{X}_{\bb{k}\sigma} \left[\tilde{c}_{\bb{k}\sigma}^\dagger \tilde{d}_\sigma - h.c \right]
\end{align}
Here the operators $X_{\bb{k}\sigma}, \tilde{X}_{\bb{k}\sigma}$ read,
\begin{align}
    &X_{\bb{k}\sigma} = A_{\bb{k}} + B_{\bb{k}} n_{\bar{\sigma}} \notag \\ & 
    \tilde{X}_{\bb{k}\sigma} = \tilde{A}_{\bb{k}} + \tilde{B}_{\bb{k}} \tilde{n}_{\bar{\sigma}}
\end{align}
The generator $S_\gamma$ accounts for the effects of dissipation by introducing terms that couple the tilde and non-tilde Hilbert space
\begin{align}
    S_\gamma = \sum_{\bb{k} \sigma} Y_{k\sigma}  c_{\bb{k}\sigma} d_{\bar{\sigma}} + \tilde{Y}_{\bb{k}\sigma} \tilde{c}_{\bb{k}\sigma} \tilde{d}_{\bar{\sigma}}
\end{align}
with the operators $Y_{\bb{k}\sigma} ,\tilde{Y}_{\bb{k}\sigma} $ defined as,
\begin{align}
     &Y_{\bb{k}\sigma} = C_{\bb{k}\sigma}  + D_{\bb{k}\sigma} n_{\sigma} \notag \\ & \tilde{Y}_{\bb{k}\sigma} = \tilde{C}_{\bb{k}\sigma}  + \tilde{D}_{\bb{k}\sigma} \tilde{n}_{\sigma} \notag 
\end{align}
By evaluating the commutator between $S$ and $\m{L}_0$, we obtain the conditions that fix the generator,
\begin{align}
    &A_{\bb{k}} = \frac{V_{\bb{k}}}{\epsilon_\bb{k} - \epsilon_d} \notag \\ & B_{\bb{k}} = \frac{V_{\bb{k}} \left( U -i \gamma/2 \right)}{\left(\epsilon_\bb{k} - \epsilon_d \right) \left( \epsilon_\bb{k} - \epsilon_d - U + i \gamma/2\right)} 
    \notag \\ & 
    C_{\bb{k}\sigma} = \frac{-i \gamma \sigma \left(  A_\bb{k} + B_\bb{k} \right)}{\epsilon_\bb{k} - \epsilon_d - U - i \gamma /2} \notag \\ &
    D_{\bb{k}\sigma} = \frac{i U C_{\bb{k}\sigma}}{ i \epsilon_d - i \epsilon_\bb{k} - \sigma \gamma}
\end{align}
Similarly, the tilde-space counterparts of these coefficients can be derived by observing that
\begin{align}
    &\tilde{A}_{\bb{k}} = A_{\bb{k}}^\ast \quad \tilde{B}_{\bb{k}} = B_{\bb{k}}^\ast \notag \\ &
    \tilde{C}_{\bb{k}} = C_{\bb{k}}^\ast \quad \tilde{D}_{\bb{k}} = D_{\bb{k}}^\ast
\end{align}
\subsection{Derivation of the Effective Lindbladian}
The effective Lindbladian is composed of different contributions, which we now discuss in detail
\begin{align}
    \m{L}_{\rm eff} = \m{L}_{0} + \m{L}_{\rm imp}^\prime+ \m{L}_{\rm pair} + \m{L}_{\rm Kondo} + \m{L}_{\rm scatt} + \m{L}_{\rm diss}
\end{align}
The first contribution is the diagonal part of the Lindbladian. $\m{L}_{\rm Kondo} = -i \left( H_{\rm Kondo} - \tilde{H}_{\rm Kondo}\right) $ describes a Kondo coupling between the impurity spin and the spin of the bath,
\begin{align}
    H_{\rm Kondo} =  - \sum_{\bb{q} \bb{k}} J_{\bb{q}\bb{k}} \vec{S}_d \cdot \vec{s}_{\bb{q}\bb{k}}
\end{align}
with a complex coupling $J_{\bb{q}\bb{k}} = V_\bb{k} B_{\bb{k}} + V_\bb{q} B_\bb{k}$. Unlike in the charge dephasing case, here the Kondo coupling remains sensitive to the interaction strength
U, even in the presence of strong dissipation
\begin{align}
    J_{\bb{q}\bb{k}}  \underset{\gamma \gg U}{\sim} -\frac{4 V_{\bb{k}} V_{\bb{q}}}{U} - i \frac{4 V_{\bb{k}} V_{\bb{q}}}{\gamma }
\end{align}
in the limit of $U \gg \gamma$,
\begin{align}
    J_{\bb{q}\bb{k}}  \underset{U  \gg \gamma }{\sim} -  \frac{8 V_{\bb{k}} V_{\bb{q}}}{U} - i \frac{4 \gamma V_{\bb{k}} V_{\bb{q}} }{ U^2}
\end{align}
Then $\m{L}_{\rm scatt} = -i \left( H_{\rm scatt}  - \tilde{H}_{\rm scatt}\right)$describes a scattering potential for the conduction electrons,
\begin{align}
    H_{\rm scatt} = \sum_{qk} \left[ W_{qk} + \frac{1}{4} J_{qk} \left(\Psi_d^\dagger \Psi_d \right)  \right] \Phi_q^\dagger \Phi_k
\end{align}
with coupling constant $W_{\bb{q}\bb{k}} = V_{\bb{q}} A_{\bb{k}} + V_{\bb{k}} A_{\bb{q}}$.
$\m{L}_{\rm imp}^\prime = -i \left( H_{\rm imp}^\prime  - \tilde{H}_{\rm imp}^\prime \right) $ describes a local Lindblad contribution
\begin{align}
    H_{\rm imp}^\prime = - \sum_{\bb{k}\sigma} V_\bb{k} \left( A_\bb{k} n_\sigma + B_\bb{k} n_{\bar{\sigma}} n_\sigma  \right) 
\end{align}
$\m{L}_{\rm pair} = -i \left( H_{\rm pair}  - \tilde{H}_{\rm pair}\right) $ describes a pair tunneling term
\begin{align}
    H_{\rm pair} = - \frac{1}{2}\sum_{\bb{q} \bb{k} \sigma} V_{\bb{k}} B_{\bb{q} } \left( d_\sigma^\dagger d_{\bar{\sigma}}^\dagger c_{\bb{q} \bar{\sigma}} c_{\bb{k}\sigma} + c_{\bb{k}\sigma}^\dagger d_\sigma c_{\bb{q}\bar{\sigma}}^\dagger d_{\bar{\sigma}} \right) 
\end{align}
Finally $\m{L}_{\rm diss}$ describes dissipative process arising from the two-body losses,
\begin{align}
    \m{L}_{\rm diss} = \m{L}_{\rm diss,High} + \m{L}_{\rm diss,Low}
\end{align}
As in the charge dephasing case, two-body losses are responsible for generating intrinsically new terms such that $\m{L}_{\rm diss}$, which couples operators in the
two Hilbert spaces $\m{H},\tilde{\m{H}}$. In particular $\m{L}_{\rm diss,High}$ describes dissipative processes acting on high-energy doublon states, involving transitions that either create or annihilate a doublon from an empty state. These terms couple only doublon and holon configurations, and do not directly affect the singly occupied sector.
\begin{widetext}
\begin{align}
    \m{L}_{\rm diss,High} &= i \frac{1}{2} \sum_{\bb{k}} V_\bb{k}\left(  C_{\bb{k} \downarrow} - C_{\bb{k} \uparrow} \right)  d_\uparrow d_\downarrow \tilde{d}_\uparrow \tilde{d}_{\downarrow} - i \frac{1}{2} \sum_{\bb{k} \bb{q} \sigma}V_{\bb{k}} D_{\bb{q}\sigma} c_{\bb{k}\sigma}^\dagger c_{\bb{q}\sigma} d_\sigma d_{\bar{\sigma}} \tilde{d}_\uparrow \tilde{d}_\downarrow  \notag \\ & -i \frac{1}{2} \sum_{\bb{k}\bb{q} \sigma} V_\bb{k} \left( C_{\bb{q}\bar{\sigma}} + D_{\bb{q}\bar{\sigma}} n_{\bar{\sigma}} \right) c_{\bb{q}\sigma} c_{\bb{k}\sigma} \tilde{d}_\uparrow \tilde{d}_\downarrow  
    -i \sum_{\bb{q}\bb{k}\sigma} V_\bb{k} D_{\bb{q}\sigma} d_\sigma c_{\bb{q}\sigma} d_{\bar{\sigma}} c_{k\sigma} \tilde{d}_\uparrow \tilde{d}_\downarrow   + i \frac{1}{2} \sum_{\bb{k}\bb{q}\sigma \mu} \mu   V_{\bb{k}} D_{\bb{q}\sigma} n_{\sigma} c_{q\sigma} d_{\bar{\sigma}} \tilde{d}_\mu  \tilde{c}_{\bb{k} \bar{\mu}} 
    \notag \\ &  + \rm{TildeVersion}
    \end{align}
\end{widetext}
More precisely, the structure of $\m{L}_{\rm diss,High}$ involves operator products of the form $d_\uparrow d_\downarrow$, $\tilde{d}_\uparrow \tilde{d}_\downarrow$ and $n_\sigma d_{\bar{\sigma}}$, all of which act non-trivially only on doublon or empty states, and thus yield vanishing contributions when applied to singly occupied configurations.
In contrast, $\m{L}_{\rm diss,Low}$ governs dissipation within the singly occupied spin sector to the charge fluctuations and so allow the system to go out from the singly occupied manifold. As a result, this term allows the system to deviate from the particle-hole symmetric point, and plays a crucial role in low-energy relaxation dynamics and charge redistribution. To clarify this structure, we can express $\m{L}_{\rm diss,Low}$ explicitly as,
\begin{align}
        \m{L}_{\rm diss,Low} &=  \frac{i}{2} \sum_{\bb{k} \bb{q}\sigma \mu} \mu V_\bb{k}C_{\bb{q}\sigma} c_{\bb{q}\sigma} d_{\bar{\sigma}} \tilde{d}_\mu \tilde{c}_{\bb{k}\bar{\mu }} -  \frac{i}{2} \sum_{\bb{k} \bb{q}\sigma \mu} \mu V_\bb{k} \tilde{C}_{\bb{q}\sigma} \tilde{c}_{\bb{q}\sigma} \tilde{d}_{\bar{\sigma}} d_\mu c_{\bb{k}\bar{\mu }} \notag \\ & = \sum_{\bb{q}\bb{k}} \kappa_{\bb{q}\bb{k},\rm eff} L_{\bb{q},\rm eff} 
        \tilde{L}_{\bb{k},\rm eff}    
\end{align}
where we have introduced the effective quantum jump operator
\begin{align}
    L_{\bb{k},\rm eff} = \sum_\sigma \sigma c_{\bb{k}\sigma} d_{\bar{\sigma}} 
\end{align}
which describes a non-local two-body loss process involving the impurity and a conduction electron of momentum $\bb{k}$ form the bath. Recasting this expression back into the non-vectorized framework, the term $\m{L}_{\rm diss,Low}$ takes a Lindblad-type structure of the form, 
\begin{align} \label{eqn:QJ_twobodyNonlocal}
    \m{L}_{\rm diss,Low} \left[ \bullet \right] = \sum_{\bb{q}\bb{k}} \kappa_{\bb{q}\bb{k},\rm eff} L_{\bb{q},\rm eff} \bullet  L_{\bb{k},\rm eff}^\dagger
\end{align}
explicitly showing that $\m{L}_{\rm diss,Low}$ arises from effective quantum jump operators and reflects dissipative spin–charge coupling processes.

\subsection{Effective Non-local Two-body Losses}
Having derived the effective dissipative terms induced by two-body losses, we now explore how to recast this non-local two-body loss term into a Lindblad form, explicitly identifying both the quantum jump operators and the non-Hermitian contribution. To proceed, we begin by expressing the dissipative part arising from the Kondo Hamiltonian and the potential scattering term,
\begin{align}
    \rm{Re} \left[  -i H_{\rm Kondo}\right] = \rm{Re} \left[  \sum_{\bb{k}\bb{q}} i J_{\bb{q}\bb{k}} \left( S_d^z s_{\bb{q}\bb{k}}^z + S_d^x s_{\bb{q}\bb{k}}^x + S_d^y s_{\bb{q}\bb{k}}^y \right)  \right] 
\end{align}
Each of these spin components can be expressed in terms of fermionic creation and annihilation operators for both the impurity and the bath. For the longitudinal z-component, the Kondo interaction becomes:
\begin{align}
    \rm{Re} \left[  \sum_{\bb{k}\bb{q}} i J_{\bb{q}\bb{k}}  S_d^z s_{\bb{q}\bb{k}}^z \right] = -  \sum_{\bb{k}\bb{q}} \sum_{\sigma \mu } \sigma \mu  \frac{\rm{Im}\left[ J_{\bb{q}\bb{k}} \right]}{4}   d^\dagger_\sigma d_\sigma c_{\bb{q}\mu}^\dagger c_{\bb{k}\mu }
\end{align}
where $\sigma,\mu=\pm 1$ encode spin indices.
Similarly, for the transverse (x and y) components , we find
\begin{align}
    \rm{Re} \left[  \sum_{\bb{k}\bb{q}} i J_{\bb{q}\bb{k}} ( S_d^x s_{\bb{q}\bb{k}}^x + S_d^y s_{\bb{q}\bb{k}}^y)   \right] =   \sum_{\bb{k}\bb{q} \sigma }  \sigma \mu  \frac{\rm{Im}\left[ J_{\bb{q}\bb{k}} \right]}{2}   d^\dagger_\sigma d_{\bar{\sigma}} c_{\bb{q}\bar{\sigma}}^\dagger c_{\bb{k}\sigma }
\end{align}
The dissipative contribution from the potential scattering term can likewise be expressed in fermionic form,
\begin{align}
    \rm{Re} \left[ -i \sum_{\bb{q}\bb{k}} \frac{J_{\bb{q}\bb{k}}}{4}  \Psi^\dagger_d \Psi_d \Phi_\bb{q}^\dagger \Phi_\bb{k}  \right] = \sum_{\bb{k}\bb{q}} \sum_{\sigma \mu }  \frac{\rm{Im}\left[ J_{\bb{q}\bb{k}} \right]}{4}   d^\dagger_\sigma d_\sigma c_{\bb{q}\mu}^\dagger c_{\bb{k}\mu }
\end{align}
By summing over all spin components of the Kondo interaction and including the dissipative part of the potential scattering term, we recover the non-Hermitian contribution associated with non-local two-body losses,
\begin{align}\label{eqn:Nh_twobodyNonlocal}
    &\rm{Re}\left[ -i H_{\rm Kondo} -i \sum_{\bb{q}\bb{k}} \frac{J_{\bb{q}\bb{k}}}{4}  \Psi^\dagger_d \Psi_d \Phi_\bb{q}^\dagger \Phi_\bb{k}   \right] = -\frac{1}{2} \sum_{\bb{q}\bb{k}}\kappa_{\bb{q}\bb{k},\rm eff} L_\bb{q}^\dagger L_{\bb{k}}
\end{align}
where the two-body losses rate $\kappa_{\bb{q}\bb{k},\rm eff}$ are defined via the relation $\rm{Im} \left[ J_{\bb{q}\bb{k}} \right] = - \kappa_{\bb{q}\bb{k},\rm eff}$. An analogous expression holds for the "tilde" non-Hermitian contribution, derived from the tilde Kondo Hamiltonian $\tilde{H}_{\rm Kondo}$ and the corresponding tilde potential scattering term $\tilde{H}_{\rm scatt}$.
Finally, by combining the non-Hermitian contribution obtained in Eq.\eqref{eqn:Nh_twobodyNonlocal} with the quantum jump term derived in Eq.\eqref{eqn:QJ_twobodyNonlocal}, we obtain the effective Lindbladian describing non-local two-body losses as introduced in the main text:
\begin{align}
    \m{L}_{\rm TbL,eff} \left[ \bullet \right] = \sum_{\bb{k}\bb{q}} \kappa_{\bb{k}\bb{q},\rm eff} L_{\bb{k},\rm eff} \bullet L_{\bb{q},\rm eff}^\dagger - \frac{\kappa_{\bb{k}\bb{q},\rm eff} }{2} \{L_{\bb{k}, \rm eff}^\dagger L_{\bb{q}, \rm eff}, \bullet \} 
\end{align}

\begin{thebibliography}{79}%
\makeatletter
\providecommand \@ifxundefined [1]{%
 \@ifx{#1\undefined}
}%
\providecommand \@ifnum [1]{%
 \ifnum #1\expandafter \@firstoftwo
 \else \expandafter \@secondoftwo
 \fi
}%
\providecommand \@ifx [1]{%
 \ifx #1\expandafter \@firstoftwo
 \else \expandafter \@secondoftwo
 \fi
}%
\providecommand \natexlab [1]{#1}%
\providecommand \enquote  [1]{``#1''}%
\providecommand \bibnamefont  [1]{#1}%
\providecommand \bibfnamefont [1]{#1}%
\providecommand \citenamefont [1]{#1}%
\providecommand \href@noop [0]{\@secondoftwo}%
\providecommand \href [0]{\begingroup \@sanitize@url \@href}%
\providecommand \@href[1]{\@@startlink{#1}\@@href}%
\providecommand \@@href[1]{\endgroup#1\@@endlink}%
\providecommand \@sanitize@url [0]{\catcode `\\12\catcode `\$12\catcode
  `\&12\catcode `\#12\catcode `\^12\catcode `\_12\catcode `\%12\relax}%
\providecommand \@@startlink[1]{}%
\providecommand \@@endlink[0]{}%
\providecommand \url  [0]{\begingroup\@sanitize@url \@url }%
\providecommand \@url [1]{\endgroup\@href {#1}{\urlprefix }}%
\providecommand \urlprefix  [0]{URL }%
\providecommand \Eprint [0]{\href }%
\providecommand \doibase [0]{https://doi.org/}%
\providecommand \selectlanguage [0]{\@gobble}%
\providecommand \bibinfo  [0]{\@secondoftwo}%
\providecommand \bibfield  [0]{\@secondoftwo}%
\providecommand \translation [1]{[#1]}%
\providecommand \BibitemOpen [0]{}%
\providecommand \bibitemStop [0]{}%
\providecommand \bibitemNoStop [0]{.\EOS\space}%
\providecommand \EOS [0]{\spacefactor3000\relax}%
\providecommand \BibitemShut  [1]{\csname bibitem#1\endcsname}%
\let\auto@bib@innerbib\@empty
\bibitem [{\citenamefont {Hewson}(1993)}]{hewson1993thekondo}%
  \BibitemOpen
  \bibfield  {author} {\bibinfo {author} {\bibfnamefont {A.~C.}\ \bibnamefont
  {Hewson}},\ }\href {https://doi.org/DOI: 10.1017/CBO9780511470752} {\emph
  {\bibinfo {title} {Cambridge Studies in Magnetism}}}\ (\bibinfo  {publisher}
  {Cambridge University Press},\ \bibinfo {address} {Cambridge},\ \bibinfo
  {year} {1993})\BibitemShut {NoStop}%
\bibitem [{\citenamefont {Kondo}(1964)}]{kondo1964resistance}%
  \BibitemOpen
  \bibfield  {author} {\bibinfo {author} {\bibfnamefont {J.}~\bibnamefont
  {Kondo}},\ }\bibfield  {title} {\bibinfo {title} {{Resistance Minimum in
  Dilute Magnetic Alloys}},\ }\href {https://doi.org/10.1143/PTP.32.37}
  {\bibfield  {journal} {\bibinfo  {journal} {Progress of Theoretical Physics}\
  }\textbf {\bibinfo {volume} {32}},\ \bibinfo {pages} {37} (\bibinfo {year}
  {1964})},\ \Eprint
  {https://arxiv.org/abs/https://academic.oup.com/ptp/article-pdf/32/1/37/5193092/32-1-37.pdf}
  {https://academic.oup.com/ptp/article-pdf/32/1/37/5193092/32-1-37.pdf}
  \BibitemShut {NoStop}%
\bibitem [{\citenamefont {Anderson}(1961)}]{anderson1961localized}%
  \BibitemOpen
  \bibfield  {author} {\bibinfo {author} {\bibfnamefont {P.~W.}\ \bibnamefont
  {Anderson}},\ }\bibfield  {title} {\bibinfo {title} {Localized magnetic
  states in metals},\ }\href {https://doi.org/10.1103/PhysRev.124.41}
  {\bibfield  {journal} {\bibinfo  {journal} {Phys. Rev.}\ }\textbf {\bibinfo
  {volume} {124}},\ \bibinfo {pages} {41} (\bibinfo {year} {1961})}\BibitemShut
  {NoStop}%
\bibitem [{\citenamefont {Nozi{\`e}res}(1974)}]{nozieres1974fermiliquid}%
  \BibitemOpen
  \bibfield  {author} {\bibinfo {author} {\bibfnamefont {P.}~\bibnamefont
  {Nozi{\`e}res}},\ }\bibfield  {title} {\bibinfo {title} {A
  ``fermi-liquid''description of the kondo problem at low temperatures},\
  }\href {https://doi.org/10.1007/BF00654541} {\bibfield  {journal} {\bibinfo
  {journal} {Journal of Low Temperature Physics}\ }\textbf {\bibinfo {volume}
  {17}},\ \bibinfo {pages} {31} (\bibinfo {year} {1974})}\BibitemShut {NoStop}%
\bibitem [{\citenamefont {Wilson}(1975)}]{wilson1975therenormalization}%
  \BibitemOpen
  \bibfield  {author} {\bibinfo {author} {\bibfnamefont {K.~G.}\ \bibnamefont
  {Wilson}},\ }\bibfield  {title} {\bibinfo {title} {The renormalization group:
  Critical phenomena and the kondo problem},\ }\href
  {https://doi.org/10.1103/RevModPhys.47.773} {\bibfield  {journal} {\bibinfo
  {journal} {Rev. Mod. Phys.}\ }\textbf {\bibinfo {volume} {47}},\ \bibinfo
  {pages} {773} (\bibinfo {year} {1975})}\BibitemShut {NoStop}%
\bibitem [{\citenamefont {Pustilnik}\ and\ \citenamefont
  {Glazman}(2004)}]{Pustilnik_2004}%
  \BibitemOpen
  \bibfield  {author} {\bibinfo {author} {\bibfnamefont {M.}~\bibnamefont
  {Pustilnik}}\ and\ \bibinfo {author} {\bibfnamefont {L.}~\bibnamefont
  {Glazman}},\ }\bibfield  {title} {\bibinfo {title} {Kondo effect in quantum
  dots},\ }\href {https://doi.org/10.1088/0953-8984/16/16/R01} {\bibfield
  {journal} {\bibinfo  {journal} {Journal of Physics: Condensed Matter}\
  }\textbf {\bibinfo {volume} {16}},\ \bibinfo {pages} {R513} (\bibinfo {year}
  {2004})}\BibitemShut {NoStop}%
\bibitem [{\citenamefont {Roch}\ \emph {et~al.}(2009)\citenamefont {Roch},
  \citenamefont {Florens}, \citenamefont {Costi}, \citenamefont {Wernsdorfer},\
  and\ \citenamefont {Balestro}}]{roch2009observation}%
  \BibitemOpen
  \bibfield  {author} {\bibinfo {author} {\bibfnamefont {N.}~\bibnamefont
  {Roch}}, \bibinfo {author} {\bibfnamefont {S.}~\bibnamefont {Florens}},
  \bibinfo {author} {\bibfnamefont {T.~A.}\ \bibnamefont {Costi}}, \bibinfo
  {author} {\bibfnamefont {W.}~\bibnamefont {Wernsdorfer}},\ and\ \bibinfo
  {author} {\bibfnamefont {F.}~\bibnamefont {Balestro}},\ }\bibfield  {title}
  {\bibinfo {title} {Observation of the underscreened kondo effect in a
  molecular transistor},\ }\href
  {https://doi.org/10.1103/PhysRevLett.103.197202} {\bibfield  {journal}
  {\bibinfo  {journal} {Phys. Rev. Lett.}\ }\textbf {\bibinfo {volume} {103}},\
  \bibinfo {pages} {197202} (\bibinfo {year} {2009})}\BibitemShut {NoStop}%
\bibitem [{\citenamefont {Riegger}\ \emph {et~al.}(2018)\citenamefont
  {Riegger}, \citenamefont {Darkwah~Oppong}, \citenamefont {H\"ofer},
  \citenamefont {Fernandes}, \citenamefont {Bloch},\ and\ \citenamefont
  {F\"olling}}]{riegger2018localized}%
  \BibitemOpen
  \bibfield  {author} {\bibinfo {author} {\bibfnamefont {L.}~\bibnamefont
  {Riegger}}, \bibinfo {author} {\bibfnamefont {N.}~\bibnamefont
  {Darkwah~Oppong}}, \bibinfo {author} {\bibfnamefont {M.}~\bibnamefont
  {H\"ofer}}, \bibinfo {author} {\bibfnamefont {D.~R.}\ \bibnamefont
  {Fernandes}}, \bibinfo {author} {\bibfnamefont {I.}~\bibnamefont {Bloch}},\
  and\ \bibinfo {author} {\bibfnamefont {S.}~\bibnamefont {F\"olling}},\
  }\bibfield  {title} {\bibinfo {title} {Localized magnetic moments with
  tunable spin exchange in a gas of ultracold fermions},\ }\href
  {https://doi.org/10.1103/PhysRevLett.120.143601} {\bibfield  {journal}
  {\bibinfo  {journal} {Phys. Rev. Lett.}\ }\textbf {\bibinfo {volume} {120}},\
  \bibinfo {pages} {143601} (\bibinfo {year} {2018})}\BibitemShut {NoStop}%
\bibitem [{\citenamefont {Zhang}\ \emph {et~al.}(2020)\citenamefont {Zhang},
  \citenamefont {Cheng}, \citenamefont {Zhang},\ and\ \citenamefont
  {Zhai}}]{zhang2020controlling}%
  \BibitemOpen
  \bibfield  {author} {\bibinfo {author} {\bibfnamefont {R.}~\bibnamefont
  {Zhang}}, \bibinfo {author} {\bibfnamefont {Y.}~\bibnamefont {Cheng}},
  \bibinfo {author} {\bibfnamefont {P.}~\bibnamefont {Zhang}},\ and\ \bibinfo
  {author} {\bibfnamefont {H.}~\bibnamefont {Zhai}},\ }\bibfield  {title}
  {\bibinfo {title} {Controlling the interaction of ultracold alkaline-earth
  atoms},\ }\href {https://doi.org/10.1038/s42254-020-0157-9} {\bibfield
  {journal} {\bibinfo  {journal} {Nature Reviews Physics}\ }\textbf {\bibinfo
  {volume} {2}},\ \bibinfo {pages} {213} (\bibinfo {year} {2020})}\BibitemShut
  {NoStop}%
\bibitem [{\citenamefont {Kan\'asz-Nagy}\ \emph {et~al.}(2018)\citenamefont
  {Kan\'asz-Nagy}, \citenamefont {Ashida}, \citenamefont {Shi}, \citenamefont
  {Moca}, \citenamefont {Ikeda}, \citenamefont {F\"olling}, \citenamefont
  {Cirac}, \citenamefont {Zar\'and},\ and\ \citenamefont
  {Demler}}]{nagy2018exploring}%
  \BibitemOpen
  \bibfield  {author} {\bibinfo {author} {\bibfnamefont {M.}~\bibnamefont
  {Kan\'asz-Nagy}}, \bibinfo {author} {\bibfnamefont {Y.}~\bibnamefont
  {Ashida}}, \bibinfo {author} {\bibfnamefont {T.}~\bibnamefont {Shi}},
  \bibinfo {author} {\bibfnamefont {C.~u. u. u. u. P. m.~c.}\ \bibnamefont
  {Moca}}, \bibinfo {author} {\bibfnamefont {T.~N.}\ \bibnamefont {Ikeda}},
  \bibinfo {author} {\bibfnamefont {S.}~\bibnamefont {F\"olling}}, \bibinfo
  {author} {\bibfnamefont {J.~I.}\ \bibnamefont {Cirac}}, \bibinfo {author}
  {\bibfnamefont {G.}~\bibnamefont {Zar\'and}},\ and\ \bibinfo {author}
  {\bibfnamefont {E.~A.}\ \bibnamefont {Demler}},\ }\bibfield  {title}
  {\bibinfo {title} {Exploring the anisotropic kondo model in and out of
  equilibrium with alkaline-earth atoms},\ }\href
  {https://doi.org/10.1103/PhysRevB.97.155156} {\bibfield  {journal} {\bibinfo
  {journal} {Phys. Rev. B}\ }\textbf {\bibinfo {volume} {97}},\ \bibinfo
  {pages} {155156} (\bibinfo {year} {2018})}\BibitemShut {NoStop}%
\bibitem [{\citenamefont {Amaricci}\ \emph {et~al.}(2025)\citenamefont
  {Amaricci}, \citenamefont {Richaud}, \citenamefont {Capone}, \citenamefont
  {Oppong},\ and\ \citenamefont
  {Scazza}}]{amaricci2025engineeringkondoimpurityproblem}%
  \BibitemOpen
  \bibfield  {author} {\bibinfo {author} {\bibfnamefont {A.}~\bibnamefont
  {Amaricci}}, \bibinfo {author} {\bibfnamefont {A.}~\bibnamefont {Richaud}},
  \bibinfo {author} {\bibfnamefont {M.}~\bibnamefont {Capone}}, \bibinfo
  {author} {\bibfnamefont {N.~D.}\ \bibnamefont {Oppong}},\ and\ \bibinfo
  {author} {\bibfnamefont {F.}~\bibnamefont {Scazza}},\ }\href
  {https://arxiv.org/abs/2505.14630} {\bibinfo {title} {Engineering the kondo
  impurity problem with alkaline-earth atom arrays}} (\bibinfo {year} {2025}),\
  \Eprint {https://arxiv.org/abs/2505.14630} {arXiv:2505.14630
  [cond-mat.quant-gas]} \BibitemShut {NoStop}%
\bibitem [{\citenamefont {Cronenwett}\ \emph {et~al.}(1998)\citenamefont
  {Cronenwett}, \citenamefont {Oosterkamp},\ and\ \citenamefont
  {Kouwenhoven}}]{cronenwett1998atunable}%
  \BibitemOpen
  \bibfield  {author} {\bibinfo {author} {\bibfnamefont {S.~M.}\ \bibnamefont
  {Cronenwett}}, \bibinfo {author} {\bibfnamefont {T.~H.}\ \bibnamefont
  {Oosterkamp}},\ and\ \bibinfo {author} {\bibfnamefont {L.~P.}\ \bibnamefont
  {Kouwenhoven}},\ }\bibfield  {title} {\bibinfo {title} {A tunable kondo
  effect in quantum dots},\ }\href
  {https://doi.org/10.1126/science.281.5376.540} {\bibfield  {journal}
  {\bibinfo  {journal} {Science}\ }\textbf {\bibinfo {volume} {281}},\ \bibinfo
  {pages} {540} (\bibinfo {year} {1998})},\ \Eprint
  {https://arxiv.org/abs/https://www.science.org/doi/pdf/10.1126/science.281.5376.540}
  {https://www.science.org/doi/pdf/10.1126/science.281.5376.540} \BibitemShut
  {NoStop}%
\bibitem [{\citenamefont {Rosch}\ \emph {et~al.}(2003)\citenamefont {Rosch},
  \citenamefont {Paaske}, \citenamefont {Kroha},\ and\ \citenamefont
  {W\"olfle}}]{rosch2003nonequilibrium}%
  \BibitemOpen
  \bibfield  {author} {\bibinfo {author} {\bibfnamefont {A.}~\bibnamefont
  {Rosch}}, \bibinfo {author} {\bibfnamefont {J.}~\bibnamefont {Paaske}},
  \bibinfo {author} {\bibfnamefont {J.}~\bibnamefont {Kroha}},\ and\ \bibinfo
  {author} {\bibfnamefont {P.}~\bibnamefont {W\"olfle}},\ }\bibfield  {title}
  {\bibinfo {title} {Nonequilibrium transport through a kondo dot in a magnetic
  field: Perturbation theory and poor man's scaling},\ }\href
  {https://doi.org/10.1103/PhysRevLett.90.076804} {\bibfield  {journal}
  {\bibinfo  {journal} {Phys. Rev. Lett.}\ }\textbf {\bibinfo {volume} {90}},\
  \bibinfo {pages} {076804} (\bibinfo {year} {2003})}\BibitemShut {NoStop}%
\bibitem [{\citenamefont {Mehta}\ and\ \citenamefont
  {Andrei}(2006)}]{mehta2006nonequilibrium}%
  \BibitemOpen
  \bibfield  {author} {\bibinfo {author} {\bibfnamefont {P.}~\bibnamefont
  {Mehta}}\ and\ \bibinfo {author} {\bibfnamefont {N.}~\bibnamefont {Andrei}},\
  }\bibfield  {title} {\bibinfo {title} {Nonequilibrium transport in quantum
  impurity models: The bethe ansatz for open systems},\ }\href
  {https://doi.org/10.1103/PhysRevLett.96.216802} {\bibfield  {journal}
  {\bibinfo  {journal} {Phys. Rev. Lett.}\ }\textbf {\bibinfo {volume} {96}},\
  \bibinfo {pages} {216802} (\bibinfo {year} {2006})}\BibitemShut {NoStop}%
\bibitem [{\citenamefont {Heidrich-Meisner}\ \emph {et~al.}(2009)\citenamefont
  {Heidrich-Meisner}, \citenamefont {Feiguin},\ and\ \citenamefont
  {Dagotto}}]{heidrich2009realtime}%
  \BibitemOpen
  \bibfield  {author} {\bibinfo {author} {\bibfnamefont {F.}~\bibnamefont
  {Heidrich-Meisner}}, \bibinfo {author} {\bibfnamefont {A.~E.}\ \bibnamefont
  {Feiguin}},\ and\ \bibinfo {author} {\bibfnamefont {E.}~\bibnamefont
  {Dagotto}},\ }\bibfield  {title} {\bibinfo {title} {Real-time simulations of
  nonequilibrium transport in the single-impurity anderson model},\ }\href
  {https://doi.org/10.1103/PhysRevB.79.235336} {\bibfield  {journal} {\bibinfo
  {journal} {Phys. Rev. B}\ }\textbf {\bibinfo {volume} {79}},\ \bibinfo
  {pages} {235336} (\bibinfo {year} {2009})}\BibitemShut {NoStop}%
\bibitem [{\citenamefont {Antipov}\ \emph {et~al.}(2016)\citenamefont
  {Antipov}, \citenamefont {Dong},\ and\ \citenamefont
  {Gull}}]{antipov2016voltage}%
  \BibitemOpen
  \bibfield  {author} {\bibinfo {author} {\bibfnamefont {A.~E.}\ \bibnamefont
  {Antipov}}, \bibinfo {author} {\bibfnamefont {Q.}~\bibnamefont {Dong}},\ and\
  \bibinfo {author} {\bibfnamefont {E.}~\bibnamefont {Gull}},\ }\bibfield
  {title} {\bibinfo {title} {Voltage quench dynamics of a kondo system},\
  }\href {https://doi.org/10.1103/PhysRevLett.116.036801} {\bibfield  {journal}
  {\bibinfo  {journal} {Phys. Rev. Lett.}\ }\textbf {\bibinfo {volume} {116}},\
  \bibinfo {pages} {036801} (\bibinfo {year} {2016})}\BibitemShut {NoStop}%
\bibitem [{\citenamefont {Elzerman}\ \emph {et~al.}(2004)\citenamefont
  {Elzerman}, \citenamefont {Hanson}, \citenamefont {Willems~van Beveren},
  \citenamefont {Witkamp}, \citenamefont {Vandersypen},\ and\ \citenamefont
  {Kouwenhoven}}]{elzerman2004single}%
  \BibitemOpen
  \bibfield  {author} {\bibinfo {author} {\bibfnamefont {J.~M.}\ \bibnamefont
  {Elzerman}}, \bibinfo {author} {\bibfnamefont {R.}~\bibnamefont {Hanson}},
  \bibinfo {author} {\bibfnamefont {L.~H.}\ \bibnamefont {Willems~van
  Beveren}}, \bibinfo {author} {\bibfnamefont {B.}~\bibnamefont {Witkamp}},
  \bibinfo {author} {\bibfnamefont {L.~M.~K.}\ \bibnamefont {Vandersypen}},\
  and\ \bibinfo {author} {\bibfnamefont {L.~P.}\ \bibnamefont {Kouwenhoven}},\
  }\bibfield  {title} {\bibinfo {title} {Single-shot read-out of an individual
  electron spin in a quantum dot},\ }\href
  {https://doi.org/10.1038/nature02693} {\bibfield  {journal} {\bibinfo
  {journal} {Nature}\ }\textbf {\bibinfo {volume} {430}},\ \bibinfo {pages}
  {431} (\bibinfo {year} {2004})}\BibitemShut {NoStop}%
\bibitem [{\citenamefont {Anders}\ and\ \citenamefont
  {Schiller}(2005)}]{anders2005realtime}%
  \BibitemOpen
  \bibfield  {author} {\bibinfo {author} {\bibfnamefont {F.~B.}\ \bibnamefont
  {Anders}}\ and\ \bibinfo {author} {\bibfnamefont {A.}~\bibnamefont
  {Schiller}},\ }\bibfield  {title} {\bibinfo {title} {Real-time dynamics in
  quantum-impurity systems: A time-dependent numerical renormalization-group
  approach},\ }\href {https://doi.org/10.1103/PhysRevLett.95.196801} {\bibfield
   {journal} {\bibinfo  {journal} {Phys. Rev. Lett.}\ }\textbf {\bibinfo
  {volume} {95}},\ \bibinfo {pages} {196801} (\bibinfo {year}
  {2005})}\BibitemShut {NoStop}%
\bibitem [{\citenamefont {Latta}\ \emph {et~al.}(2011)\citenamefont {Latta},
  \citenamefont {Haupt}, \citenamefont {Hanl}, \citenamefont {Weichselbaum},
  \citenamefont {Claassen}, \citenamefont {Wuester}, \citenamefont {Fallahi},
  \citenamefont {Faelt}, \citenamefont {Glazman}, \citenamefont {von Delft},
  \citenamefont {T{\"u}reci},\ and\ \citenamefont
  {Imamoglu}}]{latta2011quantum}%
  \BibitemOpen
  \bibfield  {author} {\bibinfo {author} {\bibfnamefont {C.}~\bibnamefont
  {Latta}}, \bibinfo {author} {\bibfnamefont {F.}~\bibnamefont {Haupt}},
  \bibinfo {author} {\bibfnamefont {M.}~\bibnamefont {Hanl}}, \bibinfo {author}
  {\bibfnamefont {A.}~\bibnamefont {Weichselbaum}}, \bibinfo {author}
  {\bibfnamefont {M.}~\bibnamefont {Claassen}}, \bibinfo {author}
  {\bibfnamefont {W.}~\bibnamefont {Wuester}}, \bibinfo {author} {\bibfnamefont
  {P.}~\bibnamefont {Fallahi}}, \bibinfo {author} {\bibfnamefont
  {S.}~\bibnamefont {Faelt}}, \bibinfo {author} {\bibfnamefont
  {L.}~\bibnamefont {Glazman}}, \bibinfo {author} {\bibfnamefont
  {J.}~\bibnamefont {von Delft}}, \bibinfo {author} {\bibfnamefont {H.~E.}\
  \bibnamefont {T{\"u}reci}},\ and\ \bibinfo {author} {\bibfnamefont
  {A.}~\bibnamefont {Imamoglu}},\ }\bibfield  {title} {\bibinfo {title}
  {Quantum quench of kondo correlations in optical absorption},\ }\href
  {https://doi.org/10.1038/nature10204} {\bibfield  {journal} {\bibinfo
  {journal} {Nature}\ }\textbf {\bibinfo {volume} {474}},\ \bibinfo {pages}
  {627} (\bibinfo {year} {2011})}\BibitemShut {NoStop}%
\bibitem [{\citenamefont {Schir\'o}(2010)}]{schiro2010realtime}%
  \BibitemOpen
  \bibfield  {author} {\bibinfo {author} {\bibfnamefont {M.}~\bibnamefont
  {Schir\'o}},\ }\bibfield  {title} {\bibinfo {title} {Real-time dynamics in
  quantum impurity models with diagrammatic monte carlo},\ }\href
  {https://doi.org/10.1103/PhysRevB.81.085126} {\bibfield  {journal} {\bibinfo
  {journal} {Phys. Rev. B}\ }\textbf {\bibinfo {volume} {81}},\ \bibinfo
  {pages} {085126} (\bibinfo {year} {2010})}\BibitemShut {NoStop}%
\bibitem [{\citenamefont {Schir\'o}(2012)}]{schiro2012nonequilibrium}%
  \BibitemOpen
  \bibfield  {author} {\bibinfo {author} {\bibfnamefont {M.}~\bibnamefont
  {Schir\'o}},\ }\bibfield  {title} {\bibinfo {title} {Nonequilibrium dynamics
  across an impurity quantum critical point due to quantum quenches},\ }\href
  {https://doi.org/10.1103/PhysRevB.86.161101} {\bibfield  {journal} {\bibinfo
  {journal} {Phys. Rev. B}\ }\textbf {\bibinfo {volume} {86}},\ \bibinfo
  {pages} {161101} (\bibinfo {year} {2012})}\BibitemShut {NoStop}%
\bibitem [{\citenamefont {Lebrat}\ \emph {et~al.}(2019)\citenamefont {Lebrat},
  \citenamefont {H\"ausler}, \citenamefont {Fabritius}, \citenamefont
  {Husmann}, \citenamefont {Corman},\ and\ \citenamefont
  {Esslinger}}]{lebrat2019quantized}%
  \BibitemOpen
  \bibfield  {author} {\bibinfo {author} {\bibfnamefont {M.}~\bibnamefont
  {Lebrat}}, \bibinfo {author} {\bibfnamefont {S.}~\bibnamefont {H\"ausler}},
  \bibinfo {author} {\bibfnamefont {P.}~\bibnamefont {Fabritius}}, \bibinfo
  {author} {\bibfnamefont {D.}~\bibnamefont {Husmann}}, \bibinfo {author}
  {\bibfnamefont {L.}~\bibnamefont {Corman}},\ and\ \bibinfo {author}
  {\bibfnamefont {T.}~\bibnamefont {Esslinger}},\ }\bibfield  {title} {\bibinfo
  {title} {Quantized conductance through a spin-selective atomic point
  contact},\ }\href {https://doi.org/10.1103/PhysRevLett.123.193605} {\bibfield
   {journal} {\bibinfo  {journal} {Phys. Rev. Lett.}\ }\textbf {\bibinfo
  {volume} {123}},\ \bibinfo {pages} {193605} (\bibinfo {year}
  {2019})}\BibitemShut {NoStop}%
\bibitem [{\citenamefont {Corman}\ \emph {et~al.}(2019)\citenamefont {Corman},
  \citenamefont {Fabritius}, \citenamefont {H\"ausler}, \citenamefont {Mohan},
  \citenamefont {Dogra}, \citenamefont {Husmann}, \citenamefont {Lebrat},\ and\
  \citenamefont {Esslinger}}]{corman2019quantized}%
  \BibitemOpen
  \bibfield  {author} {\bibinfo {author} {\bibfnamefont {L.}~\bibnamefont
  {Corman}}, \bibinfo {author} {\bibfnamefont {P.}~\bibnamefont {Fabritius}},
  \bibinfo {author} {\bibfnamefont {S.}~\bibnamefont {H\"ausler}}, \bibinfo
  {author} {\bibfnamefont {J.}~\bibnamefont {Mohan}}, \bibinfo {author}
  {\bibfnamefont {L.~H.}\ \bibnamefont {Dogra}}, \bibinfo {author}
  {\bibfnamefont {D.}~\bibnamefont {Husmann}}, \bibinfo {author} {\bibfnamefont
  {M.}~\bibnamefont {Lebrat}},\ and\ \bibinfo {author} {\bibfnamefont
  {T.}~\bibnamefont {Esslinger}},\ }\bibfield  {title} {\bibinfo {title}
  {Quantized conductance through a dissipative atomic point contact},\ }\href
  {https://doi.org/10.1103/PhysRevA.100.053605} {\bibfield  {journal} {\bibinfo
   {journal} {Phys. Rev. A}\ }\textbf {\bibinfo {volume} {100}},\ \bibinfo
  {pages} {053605} (\bibinfo {year} {2019})}\BibitemShut {NoStop}%
\bibitem [{\citenamefont {Huang}\ \emph {et~al.}(2023)\citenamefont {Huang},
  \citenamefont {Mohan}, \citenamefont {Visuri}, \citenamefont {Fabritius},
  \citenamefont {Talebi}, \citenamefont {Wili}, \citenamefont {Uchino},
  \citenamefont {Giamarchi},\ and\ \citenamefont
  {Esslinger}}]{huang2023superfluid}%
  \BibitemOpen
  \bibfield  {author} {\bibinfo {author} {\bibfnamefont {M.-Z.}\ \bibnamefont
  {Huang}}, \bibinfo {author} {\bibfnamefont {J.}~\bibnamefont {Mohan}},
  \bibinfo {author} {\bibfnamefont {A.-M.}\ \bibnamefont {Visuri}}, \bibinfo
  {author} {\bibfnamefont {P.}~\bibnamefont {Fabritius}}, \bibinfo {author}
  {\bibfnamefont {M.}~\bibnamefont {Talebi}}, \bibinfo {author} {\bibfnamefont
  {S.}~\bibnamefont {Wili}}, \bibinfo {author} {\bibfnamefont {S.}~\bibnamefont
  {Uchino}}, \bibinfo {author} {\bibfnamefont {T.}~\bibnamefont {Giamarchi}},\
  and\ \bibinfo {author} {\bibfnamefont {T.}~\bibnamefont {Esslinger}},\
  }\bibfield  {title} {\bibinfo {title} {Superfluid signatures in a dissipative
  quantum point contact},\ }\href
  {https://doi.org/10.1103/PhysRevLett.130.200404} {\bibfield  {journal}
  {\bibinfo  {journal} {Phys. Rev. Lett.}\ }\textbf {\bibinfo {volume} {130}},\
  \bibinfo {pages} {200404} (\bibinfo {year} {2023})}\BibitemShut {NoStop}%
\bibitem [{\citenamefont {Hewitt}\ \emph {et~al.}(2024)\citenamefont {Hewitt},
  \citenamefont {Bertheas}, \citenamefont {Jain}, \citenamefont {Nishida},\
  and\ \citenamefont {Barontini}}]{Hewitt_2024}%
  \BibitemOpen
  \bibfield  {author} {\bibinfo {author} {\bibfnamefont {T.}~\bibnamefont
  {Hewitt}}, \bibinfo {author} {\bibfnamefont {T.}~\bibnamefont {Bertheas}},
  \bibinfo {author} {\bibfnamefont {M.}~\bibnamefont {Jain}}, \bibinfo {author}
  {\bibfnamefont {Y.}~\bibnamefont {Nishida}},\ and\ \bibinfo {author}
  {\bibfnamefont {G.}~\bibnamefont {Barontini}},\ }\bibfield  {title} {\bibinfo
  {title} {Controlling the interactions in a cold atom quantum impurity
  system},\ }\href {https://doi.org/10.1088/2058-9565/ad4c91} {\bibfield
  {journal} {\bibinfo  {journal} {Quantum Science and Technology}\ }\textbf
  {\bibinfo {volume} {9}},\ \bibinfo {pages} {035039} (\bibinfo {year}
  {2024})}\BibitemShut {NoStop}%
\bibitem [{\citenamefont {Gerbier}\ and\ \citenamefont
  {Castin}(2010)}]{gerbier2010heating}%
  \BibitemOpen
  \bibfield  {author} {\bibinfo {author} {\bibfnamefont {F.}~\bibnamefont
  {Gerbier}}\ and\ \bibinfo {author} {\bibfnamefont {Y.}~\bibnamefont
  {Castin}},\ }\bibfield  {title} {\bibinfo {title} {Heating rates for an atom
  in a far-detuned optical lattice},\ }\href
  {https://doi.org/10.1103/PhysRevA.82.013615} {\bibfield  {journal} {\bibinfo
  {journal} {Phys. Rev. A}\ }\textbf {\bibinfo {volume} {82}},\ \bibinfo
  {pages} {013615} (\bibinfo {year} {2010})}\BibitemShut {NoStop}%
\bibitem [{\citenamefont {Bouganne}\ \emph {et~al.}(2020)\citenamefont
  {Bouganne}, \citenamefont {Bosch~Aguilera}, \citenamefont {Ghermaoui},
  \citenamefont {Beugnon},\ and\ \citenamefont {Gerbier}}]{bouganne2020}%
  \BibitemOpen
  \bibfield  {author} {\bibinfo {author} {\bibfnamefont {R.}~\bibnamefont
  {Bouganne}}, \bibinfo {author} {\bibfnamefont {M.}~\bibnamefont
  {Bosch~Aguilera}}, \bibinfo {author} {\bibfnamefont {A.}~\bibnamefont
  {Ghermaoui}}, \bibinfo {author} {\bibfnamefont {J.}~\bibnamefont {Beugnon}},\
  and\ \bibinfo {author} {\bibfnamefont {F.}~\bibnamefont {Gerbier}},\
  }\bibfield  {title} {\bibinfo {title} {Anomalous decay of coherence in a
  dissipative many-body system},\ }\href
  {https://doi.org/10.1038/s41567-019-0678-2} {\bibfield  {journal} {\bibinfo
  {journal} {Nature Physics}\ }\textbf {\bibinfo {volume} {16}},\ \bibinfo
  {pages} {21} (\bibinfo {year} {2020})}\BibitemShut {NoStop}%
\bibitem [{\citenamefont {{Garc{\'i}a-Ripoll}}\ \emph
  {et~al.}(2009)\citenamefont {{Garc{\'i}a-Ripoll}}, \citenamefont {D{\"u}rr},
  \citenamefont {Syassen}, \citenamefont {Bauer}, \citenamefont {Lettner},
  \citenamefont {Rempe},\ and\ \citenamefont {Cirac}}]{garcia-ripoll2009}%
  \BibitemOpen
  \bibfield  {author} {\bibinfo {author} {\bibfnamefont {J.~J.}\ \bibnamefont
  {{Garc{\'i}a-Ripoll}}}, \bibinfo {author} {\bibfnamefont {S.}~\bibnamefont
  {D{\"u}rr}}, \bibinfo {author} {\bibfnamefont {N.}~\bibnamefont {Syassen}},
  \bibinfo {author} {\bibfnamefont {D.~M.}\ \bibnamefont {Bauer}}, \bibinfo
  {author} {\bibfnamefont {M.}~\bibnamefont {Lettner}}, \bibinfo {author}
  {\bibfnamefont {G.}~\bibnamefont {Rempe}},\ and\ \bibinfo {author}
  {\bibfnamefont {J.~I.}\ \bibnamefont {Cirac}},\ }\bibfield  {title} {\bibinfo
  {title} {Dissipation-induced hard-core boson gas in an optical lattice},\
  }\href@noop {} {\bibfield  {journal} {\bibinfo  {journal} {New Journal of
  Physics}\ }\textbf {\bibinfo {volume} {11}},\ \bibinfo {pages} {013053}
  (\bibinfo {year} {2009})}\BibitemShut {NoStop}%
\bibitem [{\citenamefont {Tomita}\ \emph {et~al.}(2017)\citenamefont {Tomita},
  \citenamefont {Nakajima}, \citenamefont {Danshita}, \citenamefont {Takasu},\
  and\ \citenamefont {Takahashi}}]{TomitaEtAlScienceAdv17}%
  \BibitemOpen
  \bibfield  {author} {\bibinfo {author} {\bibfnamefont {T.}~\bibnamefont
  {Tomita}}, \bibinfo {author} {\bibfnamefont {S.}~\bibnamefont {Nakajima}},
  \bibinfo {author} {\bibfnamefont {I.}~\bibnamefont {Danshita}}, \bibinfo
  {author} {\bibfnamefont {Y.}~\bibnamefont {Takasu}},\ and\ \bibinfo {author}
  {\bibfnamefont {Y.}~\bibnamefont {Takahashi}},\ }\bibfield  {title} {\bibinfo
  {title} {Observation of the mott insulator to superfluid crossover of a
  driven-dissipative bose-hubbard system},\ }\bibfield  {journal} {\bibinfo
  {journal} {Science Advances}\ }\textbf {\bibinfo {volume} {3}},\ \href
  {https://doi.org/10.1126/sciadv.1701513} {10.1126/sciadv.1701513} (\bibinfo
  {year} {2017})\BibitemShut {NoStop}%
\bibitem [{\citenamefont {Honda}\ \emph {et~al.}(2023)\citenamefont {Honda},
  \citenamefont {Taie}, \citenamefont {Takasu}, \citenamefont {Nishizawa},
  \citenamefont {Nakagawa},\ and\ \citenamefont
  {Takahashi}}]{honda2022observation}%
  \BibitemOpen
  \bibfield  {author} {\bibinfo {author} {\bibfnamefont {K.}~\bibnamefont
  {Honda}}, \bibinfo {author} {\bibfnamefont {S.}~\bibnamefont {Taie}},
  \bibinfo {author} {\bibfnamefont {Y.}~\bibnamefont {Takasu}}, \bibinfo
  {author} {\bibfnamefont {N.}~\bibnamefont {Nishizawa}}, \bibinfo {author}
  {\bibfnamefont {M.}~\bibnamefont {Nakagawa}},\ and\ \bibinfo {author}
  {\bibfnamefont {Y.}~\bibnamefont {Takahashi}},\ }\bibfield  {title} {\bibinfo
  {title} {Observation of the sign reversal of the magnetic correlation in a
  driven-dissipative fermi gas in double wells},\ }\href
  {https://doi.org/10.1103/PhysRevLett.130.063001} {\bibfield  {journal}
  {\bibinfo  {journal} {Phys. Rev. Lett.}\ }\textbf {\bibinfo {volume} {130}},\
  \bibinfo {pages} {063001} (\bibinfo {year} {2023})}\BibitemShut {NoStop}%
\bibitem [{\citenamefont {Mi}\ \emph {et~al.}(2022)\citenamefont {Mi},
  \citenamefont {Sonner}, \citenamefont {Niu}, \citenamefont {Lee},
  \citenamefont {Foxen}, \citenamefont {Acharya}, \citenamefont {Aleiner},
  \citenamefont {Andersen}, \citenamefont {Arute}, \citenamefont {Arya},
  \citenamefont {Asfaw}, \citenamefont {Atalaya}, \citenamefont {Bardin},
  \citenamefont {Basso}, \citenamefont {Bengtsson}, \citenamefont {Bortoli},
  \citenamefont {Bourassa}, \citenamefont {Brill}, \citenamefont {Broughton},
  \citenamefont {Buckley}, \citenamefont {Buell}, \citenamefont {Burkett},
  \citenamefont {Bushnell}, \citenamefont {Chen}, \citenamefont {Chiaro},
  \citenamefont {Collins}, \citenamefont {Conner}, \citenamefont {Courtney},
  \citenamefont {Crook}, \citenamefont {Debroy}, \citenamefont {Demura},
  \citenamefont {Dunsworth}, \citenamefont {Eppens}, \citenamefont {Erickson},
  \citenamefont {Faoro}, \citenamefont {Farhi}, \citenamefont {Fatemi},
  \citenamefont {Flores}, \citenamefont {Forati}, \citenamefont {Fowler},
  \citenamefont {Giang}, \citenamefont {Gidney}, \citenamefont {Gilboa},
  \citenamefont {Giustina}, \citenamefont {Dau}, \citenamefont {Gross},
  \citenamefont {Habegger}, \citenamefont {Harrigan}, \citenamefont {Hoffmann},
  \citenamefont {Hong}, \citenamefont {Huang}, \citenamefont {Huff},
  \citenamefont {Huggins}, \citenamefont {Ioffe}, \citenamefont {Isakov},
  \citenamefont {Iveland}, \citenamefont {Jeffrey}, \citenamefont {Jiang},
  \citenamefont {Jones}, \citenamefont {Kafri}, \citenamefont {Kechedzhi},
  \citenamefont {Khattar}, \citenamefont {Kim}, \citenamefont {Kitaev},
  \citenamefont {Klimov}, \citenamefont {Klots}, \citenamefont {Korotkov},
  \citenamefont {Kostritsa}, \citenamefont {Kreikebaum}, \citenamefont
  {Landhuis}, \citenamefont {Laptev}, \citenamefont {Lau}, \citenamefont {Lee},
  \citenamefont {Laws}, \citenamefont {Liu}, \citenamefont {Locharla},
  \citenamefont {Martin}, \citenamefont {McClean}, \citenamefont {McEwen},
  \citenamefont {Costa}, \citenamefont {Miao}, \citenamefont {Mohseni},
  \citenamefont {Montazeri}, \citenamefont {Morvan}, \citenamefont {Mount},
  \citenamefont {Mruczkiewicz}, \citenamefont {Naaman}, \citenamefont {Neeley},
  \citenamefont {Neill}, \citenamefont {Newman}, \citenamefont {O’Brien},
  \citenamefont {Opremcak}, \citenamefont {Petukhov}, \citenamefont {Potter},
  \citenamefont {Quintana}, \citenamefont {Rubin}, \citenamefont {Saei},
  \citenamefont {Sank}, \citenamefont {Sankaragomathi}, \citenamefont
  {Satzinger}, \citenamefont {Schuster}, \citenamefont {Shearn}, \citenamefont
  {Shvarts}, \citenamefont {Strain}, \citenamefont {Su}, \citenamefont
  {Szalay}, \citenamefont {Vidal}, \citenamefont {Villalonga}, \citenamefont
  {Vollgraff-Heidweiller}, \citenamefont {White}, \citenamefont {Yao},
  \citenamefont {Yeh}, \citenamefont {Yoo}, \citenamefont {Zalcman},
  \citenamefont {Zhang}, \citenamefont {Zhu}, \citenamefont {Neven},
  \citenamefont {Bacon}, \citenamefont {Hilton}, \citenamefont {Lucero},
  \citenamefont {Babbush}, \citenamefont {Boixo}, \citenamefont {Megrant},
  \citenamefont {Chen}, \citenamefont {Kelly}, \citenamefont {Smelyanskiy},
  \citenamefont {Abanin},\ and\ \citenamefont
  {Roushan}}]{google_dephasing_abanin}%
  \BibitemOpen
  \bibfield  {author} {\bibinfo {author} {\bibfnamefont {X.}~\bibnamefont
  {Mi}}, \bibinfo {author} {\bibfnamefont {M.}~\bibnamefont {Sonner}}, \bibinfo
  {author} {\bibfnamefont {M.~Y.}\ \bibnamefont {Niu}}, \bibinfo {author}
  {\bibfnamefont {K.~W.}\ \bibnamefont {Lee}}, \bibinfo {author} {\bibfnamefont
  {B.}~\bibnamefont {Foxen}}, \bibinfo {author} {\bibfnamefont
  {R.}~\bibnamefont {Acharya}}, \bibinfo {author} {\bibfnamefont
  {I.}~\bibnamefont {Aleiner}}, \bibinfo {author} {\bibfnamefont {T.~I.}\
  \bibnamefont {Andersen}}, \bibinfo {author} {\bibfnamefont {F.}~\bibnamefont
  {Arute}}, \bibinfo {author} {\bibfnamefont {K.}~\bibnamefont {Arya}},
  \bibinfo {author} {\bibfnamefont {A.}~\bibnamefont {Asfaw}}, \bibinfo
  {author} {\bibfnamefont {J.}~\bibnamefont {Atalaya}}, \bibinfo {author}
  {\bibfnamefont {J.~C.}\ \bibnamefont {Bardin}}, \bibinfo {author}
  {\bibfnamefont {J.}~\bibnamefont {Basso}}, \bibinfo {author} {\bibfnamefont
  {A.}~\bibnamefont {Bengtsson}}, \bibinfo {author} {\bibfnamefont
  {G.}~\bibnamefont {Bortoli}}, \bibinfo {author} {\bibfnamefont
  {A.}~\bibnamefont {Bourassa}}, \bibinfo {author} {\bibfnamefont
  {L.}~\bibnamefont {Brill}}, \bibinfo {author} {\bibfnamefont
  {M.}~\bibnamefont {Broughton}}, \bibinfo {author} {\bibfnamefont {B.~B.}\
  \bibnamefont {Buckley}}, \bibinfo {author} {\bibfnamefont {D.~A.}\
  \bibnamefont {Buell}}, \bibinfo {author} {\bibfnamefont {B.}~\bibnamefont
  {Burkett}}, \bibinfo {author} {\bibfnamefont {N.}~\bibnamefont {Bushnell}},
  \bibinfo {author} {\bibfnamefont {Z.}~\bibnamefont {Chen}}, \bibinfo {author}
  {\bibfnamefont {B.}~\bibnamefont {Chiaro}}, \bibinfo {author} {\bibfnamefont
  {R.}~\bibnamefont {Collins}}, \bibinfo {author} {\bibfnamefont
  {P.}~\bibnamefont {Conner}}, \bibinfo {author} {\bibfnamefont
  {W.}~\bibnamefont {Courtney}}, \bibinfo {author} {\bibfnamefont {A.~L.}\
  \bibnamefont {Crook}}, \bibinfo {author} {\bibfnamefont {D.~M.}\ \bibnamefont
  {Debroy}}, \bibinfo {author} {\bibfnamefont {S.}~\bibnamefont {Demura}},
  \bibinfo {author} {\bibfnamefont {A.}~\bibnamefont {Dunsworth}}, \bibinfo
  {author} {\bibfnamefont {D.}~\bibnamefont {Eppens}}, \bibinfo {author}
  {\bibfnamefont {C.}~\bibnamefont {Erickson}}, \bibinfo {author}
  {\bibfnamefont {L.}~\bibnamefont {Faoro}}, \bibinfo {author} {\bibfnamefont
  {E.}~\bibnamefont {Farhi}}, \bibinfo {author} {\bibfnamefont
  {R.}~\bibnamefont {Fatemi}}, \bibinfo {author} {\bibfnamefont
  {L.}~\bibnamefont {Flores}}, \bibinfo {author} {\bibfnamefont
  {E.}~\bibnamefont {Forati}}, \bibinfo {author} {\bibfnamefont {A.~G.}\
  \bibnamefont {Fowler}}, \bibinfo {author} {\bibfnamefont {W.}~\bibnamefont
  {Giang}}, \bibinfo {author} {\bibfnamefont {C.}~\bibnamefont {Gidney}},
  \bibinfo {author} {\bibfnamefont {D.}~\bibnamefont {Gilboa}}, \bibinfo
  {author} {\bibfnamefont {M.}~\bibnamefont {Giustina}}, \bibinfo {author}
  {\bibfnamefont {A.~G.}\ \bibnamefont {Dau}}, \bibinfo {author} {\bibfnamefont
  {J.~A.}\ \bibnamefont {Gross}}, \bibinfo {author} {\bibfnamefont
  {S.}~\bibnamefont {Habegger}}, \bibinfo {author} {\bibfnamefont {M.~P.}\
  \bibnamefont {Harrigan}}, \bibinfo {author} {\bibfnamefont {M.}~\bibnamefont
  {Hoffmann}}, \bibinfo {author} {\bibfnamefont {S.}~\bibnamefont {Hong}},
  \bibinfo {author} {\bibfnamefont {T.}~\bibnamefont {Huang}}, \bibinfo
  {author} {\bibfnamefont {A.}~\bibnamefont {Huff}}, \bibinfo {author}
  {\bibfnamefont {W.~J.}\ \bibnamefont {Huggins}}, \bibinfo {author}
  {\bibfnamefont {L.~B.}\ \bibnamefont {Ioffe}}, \bibinfo {author}
  {\bibfnamefont {S.~V.}\ \bibnamefont {Isakov}}, \bibinfo {author}
  {\bibfnamefont {J.}~\bibnamefont {Iveland}}, \bibinfo {author} {\bibfnamefont
  {E.}~\bibnamefont {Jeffrey}}, \bibinfo {author} {\bibfnamefont
  {Z.}~\bibnamefont {Jiang}}, \bibinfo {author} {\bibfnamefont
  {C.}~\bibnamefont {Jones}}, \bibinfo {author} {\bibfnamefont
  {D.}~\bibnamefont {Kafri}}, \bibinfo {author} {\bibfnamefont
  {K.}~\bibnamefont {Kechedzhi}}, \bibinfo {author} {\bibfnamefont
  {T.}~\bibnamefont {Khattar}}, \bibinfo {author} {\bibfnamefont
  {S.}~\bibnamefont {Kim}}, \bibinfo {author} {\bibfnamefont {A.~Y.}\
  \bibnamefont {Kitaev}}, \bibinfo {author} {\bibfnamefont {P.~V.}\
  \bibnamefont {Klimov}}, \bibinfo {author} {\bibfnamefont {A.~R.}\
  \bibnamefont {Klots}}, \bibinfo {author} {\bibfnamefont {A.~N.}\ \bibnamefont
  {Korotkov}}, \bibinfo {author} {\bibfnamefont {F.}~\bibnamefont {Kostritsa}},
  \bibinfo {author} {\bibfnamefont {J.~M.}\ \bibnamefont {Kreikebaum}},
  \bibinfo {author} {\bibfnamefont {D.}~\bibnamefont {Landhuis}}, \bibinfo
  {author} {\bibfnamefont {P.}~\bibnamefont {Laptev}}, \bibinfo {author}
  {\bibfnamefont {K.-M.}\ \bibnamefont {Lau}}, \bibinfo {author} {\bibfnamefont
  {J.}~\bibnamefont {Lee}}, \bibinfo {author} {\bibfnamefont {L.}~\bibnamefont
  {Laws}}, \bibinfo {author} {\bibfnamefont {W.}~\bibnamefont {Liu}}, \bibinfo
  {author} {\bibfnamefont {A.}~\bibnamefont {Locharla}}, \bibinfo {author}
  {\bibfnamefont {O.}~\bibnamefont {Martin}}, \bibinfo {author} {\bibfnamefont
  {J.~R.}\ \bibnamefont {McClean}}, \bibinfo {author} {\bibfnamefont
  {M.}~\bibnamefont {McEwen}}, \bibinfo {author} {\bibfnamefont {B.~M.}\
  \bibnamefont {Costa}}, \bibinfo {author} {\bibfnamefont {K.~C.}\ \bibnamefont
  {Miao}}, \bibinfo {author} {\bibfnamefont {M.}~\bibnamefont {Mohseni}},
  \bibinfo {author} {\bibfnamefont {S.}~\bibnamefont {Montazeri}}, \bibinfo
  {author} {\bibfnamefont {A.}~\bibnamefont {Morvan}}, \bibinfo {author}
  {\bibfnamefont {E.}~\bibnamefont {Mount}}, \bibinfo {author} {\bibfnamefont
  {W.}~\bibnamefont {Mruczkiewicz}}, \bibinfo {author} {\bibfnamefont
  {O.}~\bibnamefont {Naaman}}, \bibinfo {author} {\bibfnamefont
  {M.}~\bibnamefont {Neeley}}, \bibinfo {author} {\bibfnamefont
  {C.}~\bibnamefont {Neill}}, \bibinfo {author} {\bibfnamefont
  {M.}~\bibnamefont {Newman}}, \bibinfo {author} {\bibfnamefont {T.~E.}\
  \bibnamefont {O’Brien}}, \bibinfo {author} {\bibfnamefont {A.}~\bibnamefont
  {Opremcak}}, \bibinfo {author} {\bibfnamefont {A.}~\bibnamefont {Petukhov}},
  \bibinfo {author} {\bibfnamefont {R.}~\bibnamefont {Potter}}, \bibinfo
  {author} {\bibfnamefont {C.}~\bibnamefont {Quintana}}, \bibinfo {author}
  {\bibfnamefont {N.~C.}\ \bibnamefont {Rubin}}, \bibinfo {author}
  {\bibfnamefont {N.}~\bibnamefont {Saei}}, \bibinfo {author} {\bibfnamefont
  {D.}~\bibnamefont {Sank}}, \bibinfo {author} {\bibfnamefont {K.}~\bibnamefont
  {Sankaragomathi}}, \bibinfo {author} {\bibfnamefont {K.~J.}\ \bibnamefont
  {Satzinger}}, \bibinfo {author} {\bibfnamefont {C.}~\bibnamefont {Schuster}},
  \bibinfo {author} {\bibfnamefont {M.~J.}\ \bibnamefont {Shearn}}, \bibinfo
  {author} {\bibfnamefont {V.}~\bibnamefont {Shvarts}}, \bibinfo {author}
  {\bibfnamefont {D.}~\bibnamefont {Strain}}, \bibinfo {author} {\bibfnamefont
  {Y.}~\bibnamefont {Su}}, \bibinfo {author} {\bibfnamefont {M.}~\bibnamefont
  {Szalay}}, \bibinfo {author} {\bibfnamefont {G.}~\bibnamefont {Vidal}},
  \bibinfo {author} {\bibfnamefont {B.}~\bibnamefont {Villalonga}}, \bibinfo
  {author} {\bibfnamefont {C.}~\bibnamefont {Vollgraff-Heidweiller}}, \bibinfo
  {author} {\bibfnamefont {T.}~\bibnamefont {White}}, \bibinfo {author}
  {\bibfnamefont {Z.}~\bibnamefont {Yao}}, \bibinfo {author} {\bibfnamefont
  {P.}~\bibnamefont {Yeh}}, \bibinfo {author} {\bibfnamefont {J.}~\bibnamefont
  {Yoo}}, \bibinfo {author} {\bibfnamefont {A.}~\bibnamefont {Zalcman}},
  \bibinfo {author} {\bibfnamefont {Y.}~\bibnamefont {Zhang}}, \bibinfo
  {author} {\bibfnamefont {N.}~\bibnamefont {Zhu}}, \bibinfo {author}
  {\bibfnamefont {H.}~\bibnamefont {Neven}}, \bibinfo {author} {\bibfnamefont
  {D.}~\bibnamefont {Bacon}}, \bibinfo {author} {\bibfnamefont
  {J.}~\bibnamefont {Hilton}}, \bibinfo {author} {\bibfnamefont
  {E.}~\bibnamefont {Lucero}}, \bibinfo {author} {\bibfnamefont
  {R.}~\bibnamefont {Babbush}}, \bibinfo {author} {\bibfnamefont
  {S.}~\bibnamefont {Boixo}}, \bibinfo {author} {\bibfnamefont
  {A.}~\bibnamefont {Megrant}}, \bibinfo {author} {\bibfnamefont
  {Y.}~\bibnamefont {Chen}}, \bibinfo {author} {\bibfnamefont {J.}~\bibnamefont
  {Kelly}}, \bibinfo {author} {\bibfnamefont {V.}~\bibnamefont {Smelyanskiy}},
  \bibinfo {author} {\bibfnamefont {D.~A.}\ \bibnamefont {Abanin}},\ and\
  \bibinfo {author} {\bibfnamefont {P.}~\bibnamefont {Roushan}},\ }\bibfield
  {title} {\bibinfo {title} {Noise-resilient edge modes on a chain of
  superconducting qubits},\ }\href {https://doi.org/10.1126/science.abq5769}
  {\bibfield  {journal} {\bibinfo  {journal} {Science}\ }\textbf {\bibinfo
  {volume} {378}},\ \bibinfo {pages} {785} (\bibinfo {year} {2022})},\ \Eprint
  {https://arxiv.org/abs/https://www.science.org/doi/pdf/10.1126/science.abq5769}
  {https://www.science.org/doi/pdf/10.1126/science.abq5769} \BibitemShut
  {NoStop}%
\bibitem [{\citenamefont {Mi}\ \emph {et~al.}(2023)\citenamefont {Mi},
  \citenamefont {Michailidis}, \citenamefont {Shabani}, \citenamefont {Miao},
  \citenamefont {Klimov}, \citenamefont {Lloyd}, \citenamefont {Rosenberg},
  \citenamefont {Acharya}, \citenamefont {Aleiner}, \citenamefont {Andersen},
  \citenamefont {Ansmann}, \citenamefont {Arute}, \citenamefont {Arya},
  \citenamefont {Asfaw}, \citenamefont {Atalaya}, \citenamefont {Bardin},
  \citenamefont {Bengtsson}, \citenamefont {Bortoli}, \citenamefont {Bourassa},
  \citenamefont {Bovaird}, \citenamefont {Brill}, \citenamefont {Broughton},
  \citenamefont {Buckley}, \citenamefont {Buell}, \citenamefont {Burger},
  \citenamefont {Burkett}, \citenamefont {Bushnell}, \citenamefont {Chen},
  \citenamefont {Chiaro}, \citenamefont {Chik}, \citenamefont {Chou},
  \citenamefont {Cogan}, \citenamefont {Collins}, \citenamefont {Conner},
  \citenamefont {Courtney}, \citenamefont {Crook}, \citenamefont {Curtin},
  \citenamefont {Dau}, \citenamefont {Debroy}, \citenamefont {Barba},
  \citenamefont {Demura}, \citenamefont {Paolo}, \citenamefont {Drozdov},
  \citenamefont {Dunsworth}, \citenamefont {Erickson}, \citenamefont {Faoro},
  \citenamefont {Farhi}, \citenamefont {Fatemi}, \citenamefont {Ferreira},
  \citenamefont {Forati}, \citenamefont {Fowler}, \citenamefont {Foxen},
  \citenamefont {Genois}, \citenamefont {Giang}, \citenamefont {Gidney},
  \citenamefont {Gilboa}, \citenamefont {Giustina}, \citenamefont {Gosula},
  \citenamefont {Gross}, \citenamefont {Habegger}, \citenamefont {Hamilton},
  \citenamefont {Hansen}, \citenamefont {Harrigan}, \citenamefont {Harrington},
  \citenamefont {Heu}, \citenamefont {Hoffmann}, \citenamefont {Hong},
  \citenamefont {Huang}, \citenamefont {Huff}, \citenamefont {Huggins},
  \citenamefont {Ioffe}, \citenamefont {Isakov}, \citenamefont {Iveland},
  \citenamefont {Jeffrey}, \citenamefont {Jiang}, \citenamefont {Jones},
  \citenamefont {Juhas}, \citenamefont {Kafri}, \citenamefont {Kechedzhi},
  \citenamefont {Khattar}, \citenamefont {Khezri}, \citenamefont {Kieferova},
  \citenamefont {Kim}, \citenamefont {Kitaev}, \citenamefont {Klots},
  \citenamefont {Korotkov}, \citenamefont {Kostritsa}, \citenamefont
  {Kreikebaum}, \citenamefont {Landhuis}, \citenamefont {Laptev}, \citenamefont
  {Lau}, \citenamefont {Laws}, \citenamefont {Lee}, \citenamefont {Lee},
  \citenamefont {Lensky}, \citenamefont {Lester}, \citenamefont {Lill},
  \citenamefont {Liu}, \citenamefont {Locharla}, \citenamefont {Malone},
  \citenamefont {Martin}, \citenamefont {McClean}, \citenamefont {McEwen},
  \citenamefont {Mieszala}, \citenamefont {Montazeri}, \citenamefont {Morvan},
  \citenamefont {Movassagh}, \citenamefont {Mruczkiewicz}, \citenamefont
  {Neeley}, \citenamefont {Neill}, \citenamefont {Nersisyan}, \citenamefont
  {Newman}, \citenamefont {Ng}, \citenamefont {Nguyen}, \citenamefont {Nguyen},
  \citenamefont {Niu}, \citenamefont {OBrien}, \citenamefont {Opremcak},
  \citenamefont {Petukhov}, \citenamefont {Potter}, \citenamefont {Pryadko},
  \citenamefont {Quintana}, \citenamefont {Rocque}, \citenamefont {Rubin},
  \citenamefont {Saei}, \citenamefont {Sank}, \citenamefont {Sankaragomathi},
  \citenamefont {Satzinger}, \citenamefont {Schurkus}, \citenamefont
  {Schuster}, \citenamefont {Shearn}, \citenamefont {Shorter}, \citenamefont
  {Shutty}, \citenamefont {Shvarts}, \citenamefont {Skruzny}, \citenamefont
  {Smith}, \citenamefont {Somma}, \citenamefont {Sterling}, \citenamefont
  {Strain}, \citenamefont {Szalay}, \citenamefont {Torres}, \citenamefont
  {Vidal}, \citenamefont {Villalonga}, \citenamefont {Heidweiller},
  \citenamefont {White}, \citenamefont {Woo}, \citenamefont {Xing},
  \citenamefont {Yao}, \citenamefont {Yeh}, \citenamefont {Yoo}, \citenamefont
  {Young}, \citenamefont {Zalcman}, \citenamefont {Zhang}, \citenamefont {Zhu},
  \citenamefont {Zobrist}, \citenamefont {Neven}, \citenamefont {Babbush},
  \citenamefont {Bacon}, \citenamefont {Boixo}, \citenamefont {Hilton},
  \citenamefont {Lucero}, \citenamefont {Megrant}, \citenamefont {Kelly},
  \citenamefont {Chen}, \citenamefont {Roushan}, \citenamefont {Smelyanskiy},\
  and\ \citenamefont {Abanin}}]{mi2023stable}%
  \BibitemOpen
  \bibfield  {author} {\bibinfo {author} {\bibfnamefont {X.}~\bibnamefont
  {Mi}}, \bibinfo {author} {\bibfnamefont {A.~A.}\ \bibnamefont {Michailidis}},
  \bibinfo {author} {\bibfnamefont {S.}~\bibnamefont {Shabani}}, \bibinfo
  {author} {\bibfnamefont {K.~C.}\ \bibnamefont {Miao}}, \bibinfo {author}
  {\bibfnamefont {P.~V.}\ \bibnamefont {Klimov}}, \bibinfo {author}
  {\bibfnamefont {J.}~\bibnamefont {Lloyd}}, \bibinfo {author} {\bibfnamefont
  {E.}~\bibnamefont {Rosenberg}}, \bibinfo {author} {\bibfnamefont
  {R.}~\bibnamefont {Acharya}}, \bibinfo {author} {\bibfnamefont
  {I.}~\bibnamefont {Aleiner}}, \bibinfo {author} {\bibfnamefont {T.~I.}\
  \bibnamefont {Andersen}}, \bibinfo {author} {\bibfnamefont {M.}~\bibnamefont
  {Ansmann}}, \bibinfo {author} {\bibfnamefont {F.}~\bibnamefont {Arute}},
  \bibinfo {author} {\bibfnamefont {K.}~\bibnamefont {Arya}}, \bibinfo {author}
  {\bibfnamefont {A.}~\bibnamefont {Asfaw}}, \bibinfo {author} {\bibfnamefont
  {J.}~\bibnamefont {Atalaya}}, \bibinfo {author} {\bibfnamefont {J.~C.}\
  \bibnamefont {Bardin}}, \bibinfo {author} {\bibfnamefont {A.}~\bibnamefont
  {Bengtsson}}, \bibinfo {author} {\bibfnamefont {G.}~\bibnamefont {Bortoli}},
  \bibinfo {author} {\bibfnamefont {A.}~\bibnamefont {Bourassa}}, \bibinfo
  {author} {\bibfnamefont {J.}~\bibnamefont {Bovaird}}, \bibinfo {author}
  {\bibfnamefont {L.}~\bibnamefont {Brill}}, \bibinfo {author} {\bibfnamefont
  {M.}~\bibnamefont {Broughton}}, \bibinfo {author} {\bibfnamefont {B.~B.}\
  \bibnamefont {Buckley}}, \bibinfo {author} {\bibfnamefont {D.~A.}\
  \bibnamefont {Buell}}, \bibinfo {author} {\bibfnamefont {T.}~\bibnamefont
  {Burger}}, \bibinfo {author} {\bibfnamefont {B.}~\bibnamefont {Burkett}},
  \bibinfo {author} {\bibfnamefont {N.}~\bibnamefont {Bushnell}}, \bibinfo
  {author} {\bibfnamefont {Z.}~\bibnamefont {Chen}}, \bibinfo {author}
  {\bibfnamefont {B.}~\bibnamefont {Chiaro}}, \bibinfo {author} {\bibfnamefont
  {D.}~\bibnamefont {Chik}}, \bibinfo {author} {\bibfnamefont {C.}~\bibnamefont
  {Chou}}, \bibinfo {author} {\bibfnamefont {J.}~\bibnamefont {Cogan}},
  \bibinfo {author} {\bibfnamefont {R.}~\bibnamefont {Collins}}, \bibinfo
  {author} {\bibfnamefont {P.}~\bibnamefont {Conner}}, \bibinfo {author}
  {\bibfnamefont {W.}~\bibnamefont {Courtney}}, \bibinfo {author}
  {\bibfnamefont {A.~L.}\ \bibnamefont {Crook}}, \bibinfo {author}
  {\bibfnamefont {B.}~\bibnamefont {Curtin}}, \bibinfo {author} {\bibfnamefont
  {A.~G.}\ \bibnamefont {Dau}}, \bibinfo {author} {\bibfnamefont {D.~M.}\
  \bibnamefont {Debroy}}, \bibinfo {author} {\bibfnamefont {A.~D.~T.}\
  \bibnamefont {Barba}}, \bibinfo {author} {\bibfnamefont {S.}~\bibnamefont
  {Demura}}, \bibinfo {author} {\bibfnamefont {A.~D.}\ \bibnamefont {Paolo}},
  \bibinfo {author} {\bibfnamefont {I.~K.}\ \bibnamefont {Drozdov}}, \bibinfo
  {author} {\bibfnamefont {A.}~\bibnamefont {Dunsworth}}, \bibinfo {author}
  {\bibfnamefont {C.}~\bibnamefont {Erickson}}, \bibinfo {author}
  {\bibfnamefont {L.}~\bibnamefont {Faoro}}, \bibinfo {author} {\bibfnamefont
  {E.}~\bibnamefont {Farhi}}, \bibinfo {author} {\bibfnamefont
  {R.}~\bibnamefont {Fatemi}}, \bibinfo {author} {\bibfnamefont {V.~S.}\
  \bibnamefont {Ferreira}}, \bibinfo {author} {\bibfnamefont {L.~F. B.~E.}\
  \bibnamefont {Forati}}, \bibinfo {author} {\bibfnamefont {A.~G.}\
  \bibnamefont {Fowler}}, \bibinfo {author} {\bibfnamefont {B.}~\bibnamefont
  {Foxen}}, \bibinfo {author} {\bibfnamefont {E.}~\bibnamefont {Genois}},
  \bibinfo {author} {\bibfnamefont {W.}~\bibnamefont {Giang}}, \bibinfo
  {author} {\bibfnamefont {C.}~\bibnamefont {Gidney}}, \bibinfo {author}
  {\bibfnamefont {D.}~\bibnamefont {Gilboa}}, \bibinfo {author} {\bibfnamefont
  {M.}~\bibnamefont {Giustina}}, \bibinfo {author} {\bibfnamefont
  {R.}~\bibnamefont {Gosula}}, \bibinfo {author} {\bibfnamefont {J.~A.}\
  \bibnamefont {Gross}}, \bibinfo {author} {\bibfnamefont {S.}~\bibnamefont
  {Habegger}}, \bibinfo {author} {\bibfnamefont {M.~C.}\ \bibnamefont
  {Hamilton}}, \bibinfo {author} {\bibfnamefont {M.}~\bibnamefont {Hansen}},
  \bibinfo {author} {\bibfnamefont {M.~P.}\ \bibnamefont {Harrigan}}, \bibinfo
  {author} {\bibfnamefont {S.~D.}\ \bibnamefont {Harrington}}, \bibinfo
  {author} {\bibfnamefont {P.}~\bibnamefont {Heu}}, \bibinfo {author}
  {\bibfnamefont {M.~R.}\ \bibnamefont {Hoffmann}}, \bibinfo {author}
  {\bibfnamefont {S.}~\bibnamefont {Hong}}, \bibinfo {author} {\bibfnamefont
  {T.}~\bibnamefont {Huang}}, \bibinfo {author} {\bibfnamefont
  {A.}~\bibnamefont {Huff}}, \bibinfo {author} {\bibfnamefont {W.~J.}\
  \bibnamefont {Huggins}}, \bibinfo {author} {\bibfnamefont {L.~B.}\
  \bibnamefont {Ioffe}}, \bibinfo {author} {\bibfnamefont {S.~V.}\ \bibnamefont
  {Isakov}}, \bibinfo {author} {\bibfnamefont {J.}~\bibnamefont {Iveland}},
  \bibinfo {author} {\bibfnamefont {E.}~\bibnamefont {Jeffrey}}, \bibinfo
  {author} {\bibfnamefont {Z.}~\bibnamefont {Jiang}}, \bibinfo {author}
  {\bibfnamefont {C.}~\bibnamefont {Jones}}, \bibinfo {author} {\bibfnamefont
  {P.}~\bibnamefont {Juhas}}, \bibinfo {author} {\bibfnamefont
  {D.}~\bibnamefont {Kafri}}, \bibinfo {author} {\bibfnamefont
  {K.}~\bibnamefont {Kechedzhi}}, \bibinfo {author} {\bibfnamefont
  {T.}~\bibnamefont {Khattar}}, \bibinfo {author} {\bibfnamefont
  {M.}~\bibnamefont {Khezri}}, \bibinfo {author} {\bibfnamefont
  {M.}~\bibnamefont {Kieferova}}, \bibinfo {author} {\bibfnamefont
  {S.}~\bibnamefont {Kim}}, \bibinfo {author} {\bibfnamefont {A.}~\bibnamefont
  {Kitaev}}, \bibinfo {author} {\bibfnamefont {A.~R.}\ \bibnamefont {Klots}},
  \bibinfo {author} {\bibfnamefont {A.~N.}\ \bibnamefont {Korotkov}}, \bibinfo
  {author} {\bibfnamefont {F.}~\bibnamefont {Kostritsa}}, \bibinfo {author}
  {\bibfnamefont {J.~M.}\ \bibnamefont {Kreikebaum}}, \bibinfo {author}
  {\bibfnamefont {D.}~\bibnamefont {Landhuis}}, \bibinfo {author}
  {\bibfnamefont {P.}~\bibnamefont {Laptev}}, \bibinfo {author} {\bibfnamefont
  {K.~M.}\ \bibnamefont {Lau}}, \bibinfo {author} {\bibfnamefont
  {L.}~\bibnamefont {Laws}}, \bibinfo {author} {\bibfnamefont {J.}~\bibnamefont
  {Lee}}, \bibinfo {author} {\bibfnamefont {K.~W.}\ \bibnamefont {Lee}},
  \bibinfo {author} {\bibfnamefont {Y.~D.}\ \bibnamefont {Lensky}}, \bibinfo
  {author} {\bibfnamefont {B.~J.}\ \bibnamefont {Lester}}, \bibinfo {author}
  {\bibfnamefont {A.~T.}\ \bibnamefont {Lill}}, \bibinfo {author}
  {\bibfnamefont {W.}~\bibnamefont {Liu}}, \bibinfo {author} {\bibfnamefont
  {A.}~\bibnamefont {Locharla}}, \bibinfo {author} {\bibfnamefont {F.~D.}\
  \bibnamefont {Malone}}, \bibinfo {author} {\bibfnamefont {O.}~\bibnamefont
  {Martin}}, \bibinfo {author} {\bibfnamefont {J.~R.}\ \bibnamefont {McClean}},
  \bibinfo {author} {\bibfnamefont {M.}~\bibnamefont {McEwen}}, \bibinfo
  {author} {\bibfnamefont {A.}~\bibnamefont {Mieszala}}, \bibinfo {author}
  {\bibfnamefont {S.}~\bibnamefont {Montazeri}}, \bibinfo {author}
  {\bibfnamefont {A.}~\bibnamefont {Morvan}}, \bibinfo {author} {\bibfnamefont
  {R.}~\bibnamefont {Movassagh}}, \bibinfo {author} {\bibfnamefont
  {W.}~\bibnamefont {Mruczkiewicz}}, \bibinfo {author} {\bibfnamefont
  {M.}~\bibnamefont {Neeley}}, \bibinfo {author} {\bibfnamefont
  {C.}~\bibnamefont {Neill}}, \bibinfo {author} {\bibfnamefont
  {A.}~\bibnamefont {Nersisyan}}, \bibinfo {author} {\bibfnamefont
  {M.}~\bibnamefont {Newman}}, \bibinfo {author} {\bibfnamefont {J.~H.}\
  \bibnamefont {Ng}}, \bibinfo {author} {\bibfnamefont {A.}~\bibnamefont
  {Nguyen}}, \bibinfo {author} {\bibfnamefont {M.}~\bibnamefont {Nguyen}},
  \bibinfo {author} {\bibfnamefont {M.~Y.}\ \bibnamefont {Niu}}, \bibinfo
  {author} {\bibfnamefont {T.~E.}\ \bibnamefont {OBrien}}, \bibinfo {author}
  {\bibfnamefont {A.}~\bibnamefont {Opremcak}}, \bibinfo {author}
  {\bibfnamefont {A.}~\bibnamefont {Petukhov}}, \bibinfo {author}
  {\bibfnamefont {R.}~\bibnamefont {Potter}}, \bibinfo {author} {\bibfnamefont
  {L.~P.}\ \bibnamefont {Pryadko}}, \bibinfo {author} {\bibfnamefont
  {C.}~\bibnamefont {Quintana}}, \bibinfo {author} {\bibfnamefont
  {C.}~\bibnamefont {Rocque}}, \bibinfo {author} {\bibfnamefont {N.~C.}\
  \bibnamefont {Rubin}}, \bibinfo {author} {\bibfnamefont {N.}~\bibnamefont
  {Saei}}, \bibinfo {author} {\bibfnamefont {D.}~\bibnamefont {Sank}}, \bibinfo
  {author} {\bibfnamefont {K.}~\bibnamefont {Sankaragomathi}}, \bibinfo
  {author} {\bibfnamefont {K.~J.}\ \bibnamefont {Satzinger}}, \bibinfo {author}
  {\bibfnamefont {H.~F.}\ \bibnamefont {Schurkus}}, \bibinfo {author}
  {\bibfnamefont {C.}~\bibnamefont {Schuster}}, \bibinfo {author}
  {\bibfnamefont {M.~J.}\ \bibnamefont {Shearn}}, \bibinfo {author}
  {\bibfnamefont {A.}~\bibnamefont {Shorter}}, \bibinfo {author} {\bibfnamefont
  {N.}~\bibnamefont {Shutty}}, \bibinfo {author} {\bibfnamefont
  {V.}~\bibnamefont {Shvarts}}, \bibinfo {author} {\bibfnamefont
  {J.}~\bibnamefont {Skruzny}}, \bibinfo {author} {\bibfnamefont {W.~C.}\
  \bibnamefont {Smith}}, \bibinfo {author} {\bibfnamefont {R.}~\bibnamefont
  {Somma}}, \bibinfo {author} {\bibfnamefont {G.}~\bibnamefont {Sterling}},
  \bibinfo {author} {\bibfnamefont {D.}~\bibnamefont {Strain}}, \bibinfo
  {author} {\bibfnamefont {M.}~\bibnamefont {Szalay}}, \bibinfo {author}
  {\bibfnamefont {A.}~\bibnamefont {Torres}}, \bibinfo {author} {\bibfnamefont
  {G.}~\bibnamefont {Vidal}}, \bibinfo {author} {\bibfnamefont
  {B.}~\bibnamefont {Villalonga}}, \bibinfo {author} {\bibfnamefont {C.~V.}\
  \bibnamefont {Heidweiller}}, \bibinfo {author} {\bibfnamefont
  {T.}~\bibnamefont {White}}, \bibinfo {author} {\bibfnamefont {B.~W.~K.}\
  \bibnamefont {Woo}}, \bibinfo {author} {\bibfnamefont {C.}~\bibnamefont
  {Xing}}, \bibinfo {author} {\bibfnamefont {Z.~J.}\ \bibnamefont {Yao}},
  \bibinfo {author} {\bibfnamefont {P.}~\bibnamefont {Yeh}}, \bibinfo {author}
  {\bibfnamefont {J.}~\bibnamefont {Yoo}}, \bibinfo {author} {\bibfnamefont
  {G.}~\bibnamefont {Young}}, \bibinfo {author} {\bibfnamefont
  {A.}~\bibnamefont {Zalcman}}, \bibinfo {author} {\bibfnamefont
  {Y.}~\bibnamefont {Zhang}}, \bibinfo {author} {\bibfnamefont
  {N.}~\bibnamefont {Zhu}}, \bibinfo {author} {\bibfnamefont {N.}~\bibnamefont
  {Zobrist}}, \bibinfo {author} {\bibfnamefont {H.}~\bibnamefont {Neven}},
  \bibinfo {author} {\bibfnamefont {R.}~\bibnamefont {Babbush}}, \bibinfo
  {author} {\bibfnamefont {D.}~\bibnamefont {Bacon}}, \bibinfo {author}
  {\bibfnamefont {S.}~\bibnamefont {Boixo}}, \bibinfo {author} {\bibfnamefont
  {J.}~\bibnamefont {Hilton}}, \bibinfo {author} {\bibfnamefont
  {E.}~\bibnamefont {Lucero}}, \bibinfo {author} {\bibfnamefont
  {A.}~\bibnamefont {Megrant}}, \bibinfo {author} {\bibfnamefont
  {J.}~\bibnamefont {Kelly}}, \bibinfo {author} {\bibfnamefont
  {Y.}~\bibnamefont {Chen}}, \bibinfo {author} {\bibfnamefont {P.}~\bibnamefont
  {Roushan}}, \bibinfo {author} {\bibfnamefont {V.}~\bibnamefont
  {Smelyanskiy}},\ and\ \bibinfo {author} {\bibfnamefont {D.~A.}\ \bibnamefont
  {Abanin}},\ }\href@noop {} {\bibinfo {title} {Stable quantum-correlated many
  body states via engineered dissipation}} (\bibinfo {year} {2023}),\ \Eprint
  {https://arxiv.org/abs/2304.13878} {arXiv:2304.13878 [quant-ph]} \BibitemShut
  {NoStop}%
\bibitem [{\citenamefont {Sarma}\ \emph {et~al.}(2025)\citenamefont {Sarma},
  \citenamefont {Väyrynen},\ and\ \citenamefont
  {König}}]{sarma2025designbenchmarksemulatingkondo}%
  \BibitemOpen
  \bibfield  {author} {\bibinfo {author} {\bibfnamefont {S.}~\bibnamefont
  {Sarma}}, \bibinfo {author} {\bibfnamefont {J.~I.}\ \bibnamefont
  {Väyrynen}},\ and\ \bibinfo {author} {\bibfnamefont {E.~J.}\ \bibnamefont
  {König}},\ }\href {https://arxiv.org/abs/2501.08499} {\bibinfo {title}
  {Design and benchmarks for emulating kondo dynamics on a quantum chip}}
  (\bibinfo {year} {2025}),\ \Eprint {https://arxiv.org/abs/2501.08499}
  {arXiv:2501.08499 [cond-mat.str-el]} \BibitemShut {NoStop}%
\bibitem [{\citenamefont {Fr{\"o}ml}\ \emph {et~al.}(2019)\citenamefont
  {Fr{\"o}ml}, \citenamefont {Chiocchetta}, \citenamefont {Kollath},\ and\
  \citenamefont {Diehl}}]{Froml2019}%
  \BibitemOpen
  \bibfield  {author} {\bibinfo {author} {\bibfnamefont {H.}~\bibnamefont
  {Fr{\"o}ml}}, \bibinfo {author} {\bibfnamefont {A.}~\bibnamefont
  {Chiocchetta}}, \bibinfo {author} {\bibfnamefont {C.}~\bibnamefont
  {Kollath}},\ and\ \bibinfo {author} {\bibfnamefont {S.}~\bibnamefont
  {Diehl}},\ }\bibfield  {title} {\bibinfo {title} {Fluctuation-{{Induced
  Quantum Zeno Effect}}},\ }\href
  {https://doi.org/10.1103/PhysRevLett.122.040402} {\bibfield  {journal}
  {\bibinfo  {journal} {Physical Review Letters}\ }\textbf {\bibinfo {volume}
  {122}},\ \bibinfo {pages} {040402} (\bibinfo {year} {2019})}\BibitemShut
  {NoStop}%
\bibitem [{\citenamefont {Damanet}\ \emph {et~al.}(2019)\citenamefont
  {Damanet}, \citenamefont {Mascarenhas}, \citenamefont {Pekker},\ and\
  \citenamefont {Daley}}]{damanet2019controlling}%
  \BibitemOpen
  \bibfield  {author} {\bibinfo {author} {\bibfnamefont {F.~m.~c.}\
  \bibnamefont {Damanet}}, \bibinfo {author} {\bibfnamefont {E.}~\bibnamefont
  {Mascarenhas}}, \bibinfo {author} {\bibfnamefont {D.}~\bibnamefont
  {Pekker}},\ and\ \bibinfo {author} {\bibfnamefont {A.~J.}\ \bibnamefont
  {Daley}},\ }\bibfield  {title} {\bibinfo {title} {Controlling quantum
  transport via dissipation engineering},\ }\href
  {https://doi.org/10.1103/PhysRevLett.123.180402} {\bibfield  {journal}
  {\bibinfo  {journal} {Phys. Rev. Lett.}\ }\textbf {\bibinfo {volume} {123}},\
  \bibinfo {pages} {180402} (\bibinfo {year} {2019})}\BibitemShut {NoStop}%
\bibitem [{\citenamefont {Visuri}\ \emph {et~al.}(2022)\citenamefont {Visuri},
  \citenamefont {Giamarchi},\ and\ \citenamefont
  {Kollath}}]{visuri2022symmetry}%
  \BibitemOpen
  \bibfield  {author} {\bibinfo {author} {\bibfnamefont {A.-M.}\ \bibnamefont
  {Visuri}}, \bibinfo {author} {\bibfnamefont {T.}~\bibnamefont {Giamarchi}},\
  and\ \bibinfo {author} {\bibfnamefont {C.}~\bibnamefont {Kollath}},\
  }\bibfield  {title} {\bibinfo {title} {Symmetry-protected transport through a
  lattice with a local particle loss},\ }\href
  {https://doi.org/10.1103/PhysRevLett.129.056802} {\bibfield  {journal}
  {\bibinfo  {journal} {Phys. Rev. Lett.}\ }\textbf {\bibinfo {volume} {129}},\
  \bibinfo {pages} {056802} (\bibinfo {year} {2022})}\BibitemShut {NoStop}%
\bibitem [{\citenamefont {Visuri}\ \emph {et~al.}(2023)\citenamefont {Visuri},
  \citenamefont {Giamarchi},\ and\ \citenamefont
  {Kollath}}]{visuri2023nonlinear}%
  \BibitemOpen
  \bibfield  {author} {\bibinfo {author} {\bibfnamefont {A.-M.}\ \bibnamefont
  {Visuri}}, \bibinfo {author} {\bibfnamefont {T.}~\bibnamefont {Giamarchi}},\
  and\ \bibinfo {author} {\bibfnamefont {C.}~\bibnamefont {Kollath}},\
  }\bibfield  {title} {\bibinfo {title} {Nonlinear transport in the presence of
  a local dissipation},\ }\href
  {https://doi.org/10.1103/PhysRevResearch.5.013195} {\bibfield  {journal}
  {\bibinfo  {journal} {Phys. Rev. Res.}\ }\textbf {\bibinfo {volume} {5}},\
  \bibinfo {pages} {013195} (\bibinfo {year} {2023})}\BibitemShut {NoStop}%
\bibitem [{\citenamefont {{Visuri}}\ \emph {et~al.}(2023)\citenamefont
  {{Visuri}}, \citenamefont {{Mohan}}, \citenamefont {{Uchino}}, \citenamefont
  {{Huang}}, \citenamefont {{Esslinger}},\ and\ \citenamefont
  {{Giamarchi}}}]{visuri2023dc}%
  \BibitemOpen
  \bibfield  {author} {\bibinfo {author} {\bibfnamefont {A.-M.}\ \bibnamefont
  {{Visuri}}}, \bibinfo {author} {\bibfnamefont {J.}~\bibnamefont {{Mohan}}},
  \bibinfo {author} {\bibfnamefont {S.}~\bibnamefont {{Uchino}}}, \bibinfo
  {author} {\bibfnamefont {M.-Z.}\ \bibnamefont {{Huang}}}, \bibinfo {author}
  {\bibfnamefont {T.}~\bibnamefont {{Esslinger}}},\ and\ \bibinfo {author}
  {\bibfnamefont {T.}~\bibnamefont {{Giamarchi}}},\ }\bibfield  {title}
  {\bibinfo {title} {{DC transport in a dissipative superconducting quantum
  point contact}},\ }\href {https://doi.org/10.48550/arXiv.2304.00928}
  {\bibfield  {journal} {\bibinfo  {journal} {arXiv e-prints}\ ,\ \bibinfo
  {eid} {arXiv:2304.00928}} (\bibinfo {year} {2023})},\ \Eprint
  {https://arxiv.org/abs/2304.00928} {arXiv:2304.00928 [cond-mat.quant-gas]}
  \BibitemShut {NoStop}%
\bibitem [{\citenamefont {Krapivsky}\ \emph {et~al.}(2019)\citenamefont
  {Krapivsky}, \citenamefont {Mallick},\ and\ \citenamefont
  {Sels}}]{krapivsky2019free}%
  \BibitemOpen
  \bibfield  {author} {\bibinfo {author} {\bibfnamefont {P.~L.}\ \bibnamefont
  {Krapivsky}}, \bibinfo {author} {\bibfnamefont {K.}~\bibnamefont {Mallick}},\
  and\ \bibinfo {author} {\bibfnamefont {D.}~\bibnamefont {Sels}},\ }\bibfield
  {title} {\bibinfo {title} {Free fermions with a localized source},\ }\href
  {https://doi.org/10.1088/1742-5468/ab4e8e} {\bibfield  {journal} {\bibinfo
  {journal} {Journal of Statistical Mechanics: Theory and Experiment}\ }\textbf
  {\bibinfo {volume} {2019}},\ \bibinfo {pages} {113108} (\bibinfo {year}
  {2019})}\BibitemShut {NoStop}%
\bibitem [{\citenamefont {Krapivsky}\ \emph {et~al.}(2020)\citenamefont
  {Krapivsky}, \citenamefont {Mallick},\ and\ \citenamefont
  {Sels}}]{krapivsky2020free}%
  \BibitemOpen
  \bibfield  {author} {\bibinfo {author} {\bibfnamefont {P.~L.}\ \bibnamefont
  {Krapivsky}}, \bibinfo {author} {\bibfnamefont {K.}~\bibnamefont {Mallick}},\
  and\ \bibinfo {author} {\bibfnamefont {D.}~\bibnamefont {Sels}},\ }\bibfield
  {title} {\bibinfo {title} {Free bosons with a localized source},\ }\href
  {https://doi.org/10.1088/1742-5468/ab8118} {\bibfield  {journal} {\bibinfo
  {journal} {Journal of Statistical Mechanics: Theory and Experiment}\ }\textbf
  {\bibinfo {volume} {2020}},\ \bibinfo {pages} {063101} (\bibinfo {year}
  {2020})}\BibitemShut {NoStop}%
\bibitem [{\citenamefont {Schiro}\ and\ \citenamefont
  {Scarlatella}(2019)}]{Scarlatella2019}%
  \BibitemOpen
  \bibfield  {author} {\bibinfo {author} {\bibfnamefont {M.}~\bibnamefont
  {Schiro}}\ and\ \bibinfo {author} {\bibfnamefont {O.}~\bibnamefont
  {Scarlatella}},\ }\bibfield  {title} {\bibinfo {title} {Quantum impurity
  models coupled to {{Markovian}} and non-{{Markovian}} baths},\ }\href
  {https://doi.org/10.1063/1.5100157} {\bibfield  {journal} {\bibinfo
  {journal} {The Journal of Chemical Physics}\ }\textbf {\bibinfo {volume}
  {151}},\ \bibinfo {pages} {044102} (\bibinfo {year} {2019})}\BibitemShut
  {NoStop}%
\bibitem [{\citenamefont {Tonielli}\ \emph {et~al.}(2019)\citenamefont
  {Tonielli}, \citenamefont {Fazio}, \citenamefont {Diehl},\ and\ \citenamefont
  {Marino}}]{tonielli2019orthogonality}%
  \BibitemOpen
  \bibfield  {author} {\bibinfo {author} {\bibfnamefont {F.}~\bibnamefont
  {Tonielli}}, \bibinfo {author} {\bibfnamefont {R.}~\bibnamefont {Fazio}},
  \bibinfo {author} {\bibfnamefont {S.}~\bibnamefont {Diehl}},\ and\ \bibinfo
  {author} {\bibfnamefont {J.}~\bibnamefont {Marino}},\ }\bibfield  {title}
  {\bibinfo {title} {Orthogonality catastrophe in dissipative quantum many-body
  systems},\ }\href {https://doi.org/10.1103/PhysRevLett.122.040604} {\bibfield
   {journal} {\bibinfo  {journal} {Phys. Rev. Lett.}\ }\textbf {\bibinfo
  {volume} {122}},\ \bibinfo {pages} {040604} (\bibinfo {year}
  {2019})}\BibitemShut {NoStop}%
\bibitem [{\citenamefont {Dolgirev}\ \emph {et~al.}(2020)\citenamefont
  {Dolgirev}, \citenamefont {Marino}, \citenamefont {Sels},\ and\ \citenamefont
  {Demler}}]{dolgirev2020nongaussian}%
  \BibitemOpen
  \bibfield  {author} {\bibinfo {author} {\bibfnamefont {P.~E.}\ \bibnamefont
  {Dolgirev}}, \bibinfo {author} {\bibfnamefont {J.}~\bibnamefont {Marino}},
  \bibinfo {author} {\bibfnamefont {D.}~\bibnamefont {Sels}},\ and\ \bibinfo
  {author} {\bibfnamefont {E.}~\bibnamefont {Demler}},\ }\bibfield  {title}
  {\bibinfo {title} {Non-gaussian correlations imprinted by local dephasing in
  fermionic wires},\ }\href {https://doi.org/10.1103/PhysRevB.102.100301}
  {\bibfield  {journal} {\bibinfo  {journal} {Phys. Rev. B}\ }\textbf {\bibinfo
  {volume} {102}},\ \bibinfo {pages} {100301} (\bibinfo {year}
  {2020})}\BibitemShut {NoStop}%
\bibitem [{\citenamefont {Ferreira}\ \emph {et~al.}(2024)\citenamefont
  {Ferreira}, \citenamefont {Jin}, \citenamefont {Mannhart}, \citenamefont
  {Giamarchi},\ and\ \citenamefont {Filippone}}]{ferreira2023exact}%
  \BibitemOpen
  \bibfield  {author} {\bibinfo {author} {\bibfnamefont {J.~a.}\ \bibnamefont
  {Ferreira}}, \bibinfo {author} {\bibfnamefont {T.}~\bibnamefont {Jin}},
  \bibinfo {author} {\bibfnamefont {J.}~\bibnamefont {Mannhart}}, \bibinfo
  {author} {\bibfnamefont {T.}~\bibnamefont {Giamarchi}},\ and\ \bibinfo
  {author} {\bibfnamefont {M.}~\bibnamefont {Filippone}},\ }\bibfield  {title}
  {\bibinfo {title} {Transport and nonreciprocity in monitored quantum devices:
  An exact study},\ }\href {https://doi.org/10.1103/PhysRevLett.132.136301}
  {\bibfield  {journal} {\bibinfo  {journal} {Phys. Rev. Lett.}\ }\textbf
  {\bibinfo {volume} {132}},\ \bibinfo {pages} {136301} (\bibinfo {year}
  {2024})}\BibitemShut {NoStop}%
\bibitem [{\citenamefont {Hasegawa}\ \emph {et~al.}(2021)\citenamefont
  {Hasegawa}, \citenamefont {Nakagawa},\ and\ \citenamefont
  {Saito}}]{hasegawa2021kondo}%
  \BibitemOpen
  \bibfield  {author} {\bibinfo {author} {\bibfnamefont {M.}~\bibnamefont
  {Hasegawa}}, \bibinfo {author} {\bibfnamefont {M.}~\bibnamefont {Nakagawa}},\
  and\ \bibinfo {author} {\bibfnamefont {K.}~\bibnamefont {Saito}},\
  }\href@noop {} {\bibinfo {title} {Kondo effect in a quantum dot under
  continuous quantum measurement}} (\bibinfo {year} {2021}),\ \Eprint
  {https://arxiv.org/abs/2111.07771} {arXiv:2111.07771 [cond-mat.mes-hall]}
  \BibitemShut {NoStop}%
\bibitem [{\citenamefont {Vanhoecke}\ and\ \citenamefont
  {Schirò}(2025)}]{vanhoecke2025kondozenocrossoverdynamicsmonitored}%
  \BibitemOpen
  \bibfield  {author} {\bibinfo {author} {\bibfnamefont {M.}~\bibnamefont
  {Vanhoecke}}\ and\ \bibinfo {author} {\bibfnamefont {M.}~\bibnamefont
  {Schirò}},\ }\href {https://arxiv.org/abs/2405.17348} {\bibinfo {title}
  {Kondo-zeno crossover in the dynamics of a monitored quantum dot}} (\bibinfo
  {year} {2025}),\ \Eprint {https://arxiv.org/abs/2405.17348} {arXiv:2405.17348
  [cond-mat.str-el]} \BibitemShut {NoStop}%
\bibitem [{\citenamefont {Stefanini}\ \emph {et~al.}(2024)\citenamefont
  {Stefanini}, \citenamefont {Qu}, \citenamefont {Esslinger}, \citenamefont
  {Gopalakrishnan}, \citenamefont {Demler},\ and\ \citenamefont
  {Marino}}]{stefanini2024dissipative}%
  \BibitemOpen
  \bibfield  {author} {\bibinfo {author} {\bibfnamefont {M.}~\bibnamefont
  {Stefanini}}, \bibinfo {author} {\bibfnamefont {Y.-F.}\ \bibnamefont {Qu}},
  \bibinfo {author} {\bibfnamefont {T.}~\bibnamefont {Esslinger}}, \bibinfo
  {author} {\bibfnamefont {S.}~\bibnamefont {Gopalakrishnan}}, \bibinfo
  {author} {\bibfnamefont {E.}~\bibnamefont {Demler}},\ and\ \bibinfo {author}
  {\bibfnamefont {J.}~\bibnamefont {Marino}},\ }\href@noop {} {\bibinfo {title}
  {Dissipative realization of kondo models}} (\bibinfo {year} {2024}),\ \Eprint
  {https://arxiv.org/abs/2406.03527} {arXiv:2406.03527 [cond-mat.quant-gas]}
  \BibitemShut {NoStop}%
\bibitem [{\citenamefont {Qu}\ \emph {et~al.}(2025)\citenamefont {Qu},
  \citenamefont {Stefanini}, \citenamefont {Shi}, \citenamefont {Esslinger},
  \citenamefont {Gopalakrishnan}, \citenamefont {Marino},\ and\ \citenamefont
  {Demler}}]{qu2025variational}%
  \BibitemOpen
  \bibfield  {author} {\bibinfo {author} {\bibfnamefont {Y.-F.}\ \bibnamefont
  {Qu}}, \bibinfo {author} {\bibfnamefont {M.}~\bibnamefont {Stefanini}},
  \bibinfo {author} {\bibfnamefont {T.}~\bibnamefont {Shi}}, \bibinfo {author}
  {\bibfnamefont {T.}~\bibnamefont {Esslinger}}, \bibinfo {author}
  {\bibfnamefont {S.}~\bibnamefont {Gopalakrishnan}}, \bibinfo {author}
  {\bibfnamefont {J.}~\bibnamefont {Marino}},\ and\ \bibinfo {author}
  {\bibfnamefont {E.}~\bibnamefont {Demler}},\ }\bibfield  {title} {\bibinfo
  {title} {Variational approach to the dynamics of dissipative quantum impurity
  models},\ }\href {https://doi.org/10.1103/PhysRevB.111.155113} {\bibfield
  {journal} {\bibinfo  {journal} {Phys. Rev. B}\ }\textbf {\bibinfo {volume}
  {111}},\ \bibinfo {pages} {155113} (\bibinfo {year} {2025})}\BibitemShut
  {NoStop}%
\bibitem [{\citenamefont {Nakagawa}\ \emph {et~al.}(2018)\citenamefont
  {Nakagawa}, \citenamefont {Kawakami},\ and\ \citenamefont
  {Ueda}}]{Nakagawa2018}%
  \BibitemOpen
  \bibfield  {author} {\bibinfo {author} {\bibfnamefont {M.}~\bibnamefont
  {Nakagawa}}, \bibinfo {author} {\bibfnamefont {N.}~\bibnamefont {Kawakami}},\
  and\ \bibinfo {author} {\bibfnamefont {M.}~\bibnamefont {Ueda}},\ }\bibfield
  {title} {\bibinfo {title} {Non-{{Hermitian Kondo Effect}} in {{Ultracold
  Alkaline}}-{{Earth Atoms}}},\ }\href
  {https://doi.org/10.1103/PhysRevLett.121.203001} {\bibfield  {journal}
  {\bibinfo  {journal} {Physical Review Letters}\ }\textbf {\bibinfo {volume}
  {121}},\ \bibinfo {pages} {203001} (\bibinfo {year} {2018})}\BibitemShut
  {NoStop}%
\bibitem [{\citenamefont {Louren\ifmmode~\mbox{\c{c}}\else \c{c}\fi{}o}\ \emph
  {et~al.}(2018)\citenamefont {Louren\ifmmode~\mbox{\c{c}}\else \c{c}\fi{}o},
  \citenamefont {Eneias},\ and\ \citenamefont {Pereira}}]{lourenco2018kondo}%
  \BibitemOpen
  \bibfield  {author} {\bibinfo {author} {\bibfnamefont {J.~A.~S.}\
  \bibnamefont {Louren\ifmmode~\mbox{\c{c}}\else \c{c}\fi{}o}}, \bibinfo
  {author} {\bibfnamefont {R.~L.}\ \bibnamefont {Eneias}},\ and\ \bibinfo
  {author} {\bibfnamefont {R.~G.}\ \bibnamefont {Pereira}},\ }\bibfield
  {title} {\bibinfo {title} {Kondo effect in a $\mathcal{PT}$-symmetric
  non-hermitian hamiltonian},\ }\href
  {https://doi.org/10.1103/PhysRevB.98.085126} {\bibfield  {journal} {\bibinfo
  {journal} {Phys. Rev. B}\ }\textbf {\bibinfo {volume} {98}},\ \bibinfo
  {pages} {085126} (\bibinfo {year} {2018})}\BibitemShut {NoStop}%
\bibitem [{\citenamefont {Kattel}\ \emph {et~al.}(2024)\citenamefont {Kattel},
  \citenamefont {Zhakenov}, \citenamefont {Pasnoori}, \citenamefont {Azaria},\
  and\ \citenamefont {Andrei}}]{kattel2024dissipation}%
  \BibitemOpen
  \bibfield  {author} {\bibinfo {author} {\bibfnamefont {P.}~\bibnamefont
  {Kattel}}, \bibinfo {author} {\bibfnamefont {A.}~\bibnamefont {Zhakenov}},
  \bibinfo {author} {\bibfnamefont {P.~R.}\ \bibnamefont {Pasnoori}}, \bibinfo
  {author} {\bibfnamefont {P.}~\bibnamefont {Azaria}},\ and\ \bibinfo {author}
  {\bibfnamefont {N.}~\bibnamefont {Andrei}},\ }\href@noop {} {\bibinfo {title}
  {Dissipation driven phase transition in the non-hermitian kondo model}}
  (\bibinfo {year} {2024}),\ \Eprint {https://arxiv.org/abs/2402.09510}
  {arXiv:2402.09510 [cond-mat.str-el]} \BibitemShut {NoStop}%
\bibitem [{\citenamefont {Yamamoto}\ \emph {et~al.}(2025)\citenamefont
  {Yamamoto}, \citenamefont {Nakagawa},\ and\ \citenamefont
  {Kawakami}}]{yamamoto2025correlation}%
  \BibitemOpen
  \bibfield  {author} {\bibinfo {author} {\bibfnamefont {K.}~\bibnamefont
  {Yamamoto}}, \bibinfo {author} {\bibfnamefont {M.}~\bibnamefont {Nakagawa}},\
  and\ \bibinfo {author} {\bibfnamefont {N.}~\bibnamefont {Kawakami}},\
  }\bibfield  {title} {\bibinfo {title} {Correlation versus dissipation in a
  non-hermitian anderson impurity model},\ }\href
  {https://doi.org/10.1103/PhysRevB.111.125157} {\bibfield  {journal} {\bibinfo
   {journal} {Phys. Rev. B}\ }\textbf {\bibinfo {volume} {111}},\ \bibinfo
  {pages} {125157} (\bibinfo {year} {2025})}\BibitemShut {NoStop}%
\bibitem [{\citenamefont {Scarlatella}\ \emph {et~al.}(2021)\citenamefont
  {Scarlatella}, \citenamefont {Clerk}, \citenamefont {Fazio},\ and\
  \citenamefont {Schir\'o}}]{scarlatella2021dynamical}%
  \BibitemOpen
  \bibfield  {author} {\bibinfo {author} {\bibfnamefont {O.}~\bibnamefont
  {Scarlatella}}, \bibinfo {author} {\bibfnamefont {A.~A.}\ \bibnamefont
  {Clerk}}, \bibinfo {author} {\bibfnamefont {R.}~\bibnamefont {Fazio}},\ and\
  \bibinfo {author} {\bibfnamefont {M.}~\bibnamefont {Schir\'o}},\ }\bibfield
  {title} {\bibinfo {title} {Dynamical mean-field theory for markovian open
  quantum many-body systems},\ }\href
  {https://doi.org/10.1103/PhysRevX.11.031018} {\bibfield  {journal} {\bibinfo
  {journal} {Phys. Rev. X}\ }\textbf {\bibinfo {volume} {11}},\ \bibinfo
  {pages} {031018} (\bibinfo {year} {2021})}\BibitemShut {NoStop}%
\bibitem [{\citenamefont {Scarlatella}\ and\ \citenamefont
  {Schirò}(2024)}]{scarlatella2023selfconsistent}%
  \BibitemOpen
  \bibfield  {author} {\bibinfo {author} {\bibfnamefont {O.}~\bibnamefont
  {Scarlatella}}\ and\ \bibinfo {author} {\bibfnamefont {M.}~\bibnamefont
  {Schirò}},\ }\bibfield  {title} {\bibinfo {title} {{Self-consistent
  dynamical maps for open quantum systems}},\ }\href
  {https://doi.org/10.21468/SciPostPhys.16.1.026} {\bibfield  {journal}
  {\bibinfo  {journal} {SciPost Phys.}\ }\textbf {\bibinfo {volume} {16}},\
  \bibinfo {pages} {026} (\bibinfo {year} {2024})}\BibitemShut {NoStop}%
\bibitem [{\citenamefont {Fazio}\ \emph {et~al.}(2025)\citenamefont {Fazio},
  \citenamefont {Keeling}, \citenamefont {Mazza},\ and\ \citenamefont
  {Schirò}}]{fazio2025manybodyopenquantumsystems}%
  \BibitemOpen
  \bibfield  {author} {\bibinfo {author} {\bibfnamefont {R.}~\bibnamefont
  {Fazio}}, \bibinfo {author} {\bibfnamefont {J.}~\bibnamefont {Keeling}},
  \bibinfo {author} {\bibfnamefont {L.}~\bibnamefont {Mazza}},\ and\ \bibinfo
  {author} {\bibfnamefont {M.}~\bibnamefont {Schirò}},\ }\href
  {https://arxiv.org/abs/2409.10300} {\bibinfo {title} {Many-body open quantum
  systems}} (\bibinfo {year} {2025}),\ \Eprint
  {https://arxiv.org/abs/2409.10300} {arXiv:2409.10300 [quant-ph]} \BibitemShut
  {NoStop}%
\bibitem [{\citenamefont {Prosen}(2008)}]{Prosen_2008}%
  \BibitemOpen
  \bibfield  {author} {\bibinfo {author} {\bibfnamefont {T.}~\bibnamefont
  {Prosen}},\ }\bibfield  {title} {\bibinfo {title} {Third quantization: a
  general method to solve master equations for quadratic open fermi systems},\
  }\href {https://doi.org/10.1088/1367-2630/10/4/043026} {\bibfield  {journal}
  {\bibinfo  {journal} {New Journal of Physics}\ }\textbf {\bibinfo {volume}
  {10}},\ \bibinfo {pages} {043026} (\bibinfo {year} {2008})}\BibitemShut
  {NoStop}%
\bibitem [{\citenamefont {Dzhioev}\ and\ \citenamefont
  {Kosov}(2011)}]{dzhioev2011}%
  \BibitemOpen
  \bibfield  {author} {\bibinfo {author} {\bibfnamefont {A.~A.}\ \bibnamefont
  {Dzhioev}}\ and\ \bibinfo {author} {\bibfnamefont {D.~S.}\ \bibnamefont
  {Kosov}},\ }\bibfield  {title} {\bibinfo {title} {{Super-fermion
  representation of quantum kinetic equations for the electron transport
  problem}},\ }\href {https://doi.org/10.1063/1.3548065} {\bibfield  {journal}
  {\bibinfo  {journal} {The Journal of Chemical Physics}\ }\textbf {\bibinfo
  {volume} {134}},\ \bibinfo {pages} {044121} (\bibinfo {year} {2011})},\
  \Eprint
  {https://arxiv.org/abs/https://pubs.aip.org/aip/jcp/article-pdf/doi/10.1063/1.3548065/13788434/044121\_1\_online.pdf}
  {https://pubs.aip.org/aip/jcp/article-pdf/doi/10.1063/1.3548065/13788434/044121\_1\_online.pdf}
  \BibitemShut {NoStop}%
\bibitem [{\citenamefont {Harbola}\ and\ \citenamefont
  {Mukamel}(2008)}]{HARBOLA2008191}%
  \BibitemOpen
  \bibfield  {author} {\bibinfo {author} {\bibfnamefont {U.}~\bibnamefont
  {Harbola}}\ and\ \bibinfo {author} {\bibfnamefont {S.}~\bibnamefont
  {Mukamel}},\ }\bibfield  {title} {\bibinfo {title} {Superoperator
  nonequilibrium green’s function theory of many-body systems; applications
  to charge transfer and transport in open junctions},\ }\href
  {https://doi.org/https://doi.org/10.1016/j.physrep.2008.05.003} {\bibfield
  {journal} {\bibinfo  {journal} {Physics Reports}\ }\textbf {\bibinfo {volume}
  {465}},\ \bibinfo {pages} {191} (\bibinfo {year} {2008})}\BibitemShut
  {NoStop}%
\bibitem [{\citenamefont {Dorda}\ \emph {et~al.}(2014)\citenamefont {Dorda},
  \citenamefont {Nuss}, \citenamefont {von~der Linden},\ and\ \citenamefont
  {Arrigoni}}]{dorda2014auxiliary}%
  \BibitemOpen
  \bibfield  {author} {\bibinfo {author} {\bibfnamefont {A.}~\bibnamefont
  {Dorda}}, \bibinfo {author} {\bibfnamefont {M.}~\bibnamefont {Nuss}},
  \bibinfo {author} {\bibfnamefont {W.}~\bibnamefont {von~der Linden}},\ and\
  \bibinfo {author} {\bibfnamefont {E.}~\bibnamefont {Arrigoni}},\ }\bibfield
  {title} {\bibinfo {title} {Auxiliary master equation approach to
  nonequilibrium correlated impurities},\ }\href
  {https://doi.org/10.1103/PhysRevB.89.165105} {\bibfield  {journal} {\bibinfo
  {journal} {Phys. Rev. B}\ }\textbf {\bibinfo {volume} {89}},\ \bibinfo
  {pages} {165105} (\bibinfo {year} {2014})}\BibitemShut {NoStop}%
\bibitem [{\citenamefont {Arrigoni}\ and\ \citenamefont
  {Dorda}(2018)}]{Arrigoni2018}%
  \BibitemOpen
  \bibfield  {author} {\bibinfo {author} {\bibfnamefont {E.}~\bibnamefont
  {Arrigoni}}\ and\ \bibinfo {author} {\bibfnamefont {A.}~\bibnamefont
  {Dorda}},\ }\bibinfo {title} {Master equations versus keldysh green's
  functions for correlated quantum systems out of equilibrium},\ in\ \href
  {https://doi.org/10.1007/978-3-319-94956-7_4} {\emph {\bibinfo {booktitle}
  {Out-of-Equilibrium Physics of Correlated Electron Systems}}},\ \bibinfo
  {editor} {edited by\ \bibinfo {editor} {\bibfnamefont {R.}~\bibnamefont
  {Citro}}\ and\ \bibinfo {editor} {\bibfnamefont {F.}~\bibnamefont
  {Mancini}}}\ (\bibinfo  {publisher} {Springer International Publishing},\
  \bibinfo {address} {Cham},\ \bibinfo {year} {2018})\ pp.\ \bibinfo {pages}
  {121--188}\BibitemShut {NoStop}%
\bibitem [{\citenamefont {Werner}\ \emph {et~al.}(2023)\citenamefont {Werner},
  \citenamefont {Lotze},\ and\ \citenamefont
  {Arrigoni}}]{werner2023configuration}%
  \BibitemOpen
  \bibfield  {author} {\bibinfo {author} {\bibfnamefont {D.}~\bibnamefont
  {Werner}}, \bibinfo {author} {\bibfnamefont {J.}~\bibnamefont {Lotze}},\ and\
  \bibinfo {author} {\bibfnamefont {E.}~\bibnamefont {Arrigoni}},\ }\bibfield
  {title} {\bibinfo {title} {Configuration interaction based nonequilibrium
  steady state impurity solver},\ }\href
  {https://doi.org/10.1103/PhysRevB.107.075119} {\bibfield  {journal} {\bibinfo
   {journal} {Phys. Rev. B}\ }\textbf {\bibinfo {volume} {107}},\ \bibinfo
  {pages} {075119} (\bibinfo {year} {2023})}\BibitemShut {NoStop}%
\bibitem [{\citenamefont {TAKAHASHI}\ and\ \citenamefont
  {UMEZAWA}(1996)}]{takahashi96}%
  \BibitemOpen
  \bibfield  {author} {\bibinfo {author} {\bibfnamefont {Y.}~\bibnamefont
  {TAKAHASHI}}\ and\ \bibinfo {author} {\bibfnamefont {H.}~\bibnamefont
  {UMEZAWA}},\ }\bibfield  {title} {\bibinfo {title} {Thermo field dynamics},\
  }\href {https://doi.org/10.1142/S0217979296000817} {\bibfield  {journal}
  {\bibinfo  {journal} {International Journal of Modern Physics B}\ }\textbf
  {\bibinfo {volume} {10}},\ \bibinfo {pages} {1755} (\bibinfo {year}
  {1996})},\ \Eprint
  {https://arxiv.org/abs/https://doi.org/10.1142/S0217979296000817}
  {https://doi.org/10.1142/S0217979296000817} \BibitemShut {NoStop}%
\bibitem [{\citenamefont {Ojima}(1981)}]{ojima1981}%
  \BibitemOpen
  \bibfield  {author} {\bibinfo {author} {\bibfnamefont {I.}~\bibnamefont
  {Ojima}},\ }\bibfield  {title} {\bibinfo {title} {Gauge fields at finite
  temperatures—“thermo field dynamics” and the kms condition and their
  extension to gauge theories},\ }\href
  {https://doi.org/https://doi.org/10.1016/0003-4916(81)90058-0} {\bibfield
  {journal} {\bibinfo  {journal} {Annals of Physics}\ }\textbf {\bibinfo
  {volume} {137}},\ \bibinfo {pages} {1} (\bibinfo {year} {1981})}\BibitemShut
  {NoStop}%
\bibitem [{\citenamefont {McDonald}\ and\ \citenamefont
  {Clerk}(2023)}]{mcdonald2023third}%
  \BibitemOpen
  \bibfield  {author} {\bibinfo {author} {\bibfnamefont {A.}~\bibnamefont
  {McDonald}}\ and\ \bibinfo {author} {\bibfnamefont {A.~A.}\ \bibnamefont
  {Clerk}},\ }\bibfield  {title} {\bibinfo {title} {Third quantization of open
  quantum systems: Dissipative symmetries and connections to phase-space and
  keldysh field-theory formulations},\ }\href
  {https://doi.org/10.1103/PhysRevResearch.5.033107} {\bibfield  {journal}
  {\bibinfo  {journal} {Phys. Rev. Res.}\ }\textbf {\bibinfo {volume} {5}},\
  \bibinfo {pages} {033107} (\bibinfo {year} {2023})}\BibitemShut {NoStop}%
\bibitem [{\citenamefont {Vanhoecke}\ and\ \citenamefont
  {Schir\`o}(2024)}]{vanhoecke2024diagrammatic}%
  \BibitemOpen
  \bibfield  {author} {\bibinfo {author} {\bibfnamefont {M.}~\bibnamefont
  {Vanhoecke}}\ and\ \bibinfo {author} {\bibfnamefont {M.}~\bibnamefont
  {Schir\`o}},\ }\bibfield  {title} {\bibinfo {title} {Diagrammatic monte carlo
  for dissipative quantum impurity models},\ }\href
  {https://doi.org/10.1103/PhysRevB.109.125125} {\bibfield  {journal} {\bibinfo
   {journal} {Phys. Rev. B}\ }\textbf {\bibinfo {volume} {109}},\ \bibinfo
  {pages} {125125} (\bibinfo {year} {2024})}\BibitemShut {NoStop}%
\bibitem [{\citenamefont {Bickers}(1987)}]{bickersBickers1987b}%
  \BibitemOpen
  \bibfield  {author} {\bibinfo {author} {\bibfnamefont {N.~E.}\ \bibnamefont
  {Bickers}},\ }\bibfield  {title} {\bibinfo {title} {Review of techniques in
  the large- {{N}} expansion for dilute magnetic alloys},\ }\href
  {https://doi.org/10.1103/RevModPhys.59.845} {\bibfield  {journal} {\bibinfo
  {journal} {Rev. Mod. Phys.}\ }\textbf {\bibinfo {volume} {59}},\ \bibinfo
  {pages} {845} (\bibinfo {year} {1987})}\BibitemShut {NoStop}%
\bibitem [{\citenamefont
  {{M{\"u}ller-Hartmann}}(1984)}]{muller-hartmannMuller-Hartmann1984}%
  \BibitemOpen
  \bibfield  {author} {\bibinfo {author} {\bibfnamefont {E.}~\bibnamefont
  {{M{\"u}ller-Hartmann}}},\ }\bibfield  {title} {\bibinfo {title}
  {Self-consistent perturbation theory of the anderson model: {{Ground}} state
  properties},\ }\href {https://doi.org/10.1007/BF01470417} {\bibfield
  {journal} {\bibinfo  {journal} {Z. Physik B - Condensed Matter}\ }\textbf
  {\bibinfo {volume} {57}},\ \bibinfo {pages} {281} (\bibinfo {year}
  {1984})}\BibitemShut {NoStop}%
\bibitem [{\citenamefont {Meir}\ \emph {et~al.}(1993)\citenamefont {Meir},
  \citenamefont {Wingreen},\ and\ \citenamefont {Lee}}]{meir1993low}%
  \BibitemOpen
  \bibfield  {author} {\bibinfo {author} {\bibfnamefont {Y.}~\bibnamefont
  {Meir}}, \bibinfo {author} {\bibfnamefont {N.~S.}\ \bibnamefont {Wingreen}},\
  and\ \bibinfo {author} {\bibfnamefont {P.~A.}\ \bibnamefont {Lee}},\
  }\bibfield  {title} {\bibinfo {title} {Low-temperature transport through a
  quantum dot: The anderson model out of equilibrium},\ }\href
  {https://doi.org/10.1103/PhysRevLett.70.2601} {\bibfield  {journal} {\bibinfo
   {journal} {Phys. Rev. Lett.}\ }\textbf {\bibinfo {volume} {70}},\ \bibinfo
  {pages} {2601} (\bibinfo {year} {1993})}\BibitemShut {NoStop}%
\bibitem [{\citenamefont {Nordlander}\ \emph {et~al.}(1999)\citenamefont
  {Nordlander}, \citenamefont {Pustilnik}, \citenamefont {Meir}, \citenamefont
  {Wingreen},\ and\ \citenamefont {Langreth}}]{nordlander1999how}%
  \BibitemOpen
  \bibfield  {author} {\bibinfo {author} {\bibfnamefont {P.}~\bibnamefont
  {Nordlander}}, \bibinfo {author} {\bibfnamefont {M.}~\bibnamefont
  {Pustilnik}}, \bibinfo {author} {\bibfnamefont {Y.}~\bibnamefont {Meir}},
  \bibinfo {author} {\bibfnamefont {N.~S.}\ \bibnamefont {Wingreen}},\ and\
  \bibinfo {author} {\bibfnamefont {D.~C.}\ \bibnamefont {Langreth}},\
  }\bibfield  {title} {\bibinfo {title} {How long does it take for the kondo
  effect to develop?},\ }\href {https://doi.org/10.1103/PhysRevLett.83.808}
  {\bibfield  {journal} {\bibinfo  {journal} {Phys. Rev. Lett.}\ }\textbf
  {\bibinfo {volume} {83}},\ \bibinfo {pages} {808} (\bibinfo {year}
  {1999})}\BibitemShut {NoStop}%
\bibitem [{\citenamefont {Eckstein}\ and\ \citenamefont
  {Werner}(2010)}]{ecksteinWerner2010a}%
  \BibitemOpen
  \bibfield  {author} {\bibinfo {author} {\bibfnamefont {M.}~\bibnamefont
  {Eckstein}}\ and\ \bibinfo {author} {\bibfnamefont {P.}~\bibnamefont
  {Werner}},\ }\bibfield  {title} {\bibinfo {title} {Nonequilibrium dynamical
  mean-field calculations based on the noncrossing approximation and its
  generalizations},\ }\href {https://doi.org/10.1103/PhysRevB.82.115115}
  {\bibfield  {journal} {\bibinfo  {journal} {Phys. Rev. B}\ }\textbf {\bibinfo
  {volume} {82}},\ \bibinfo {pages} {115115} (\bibinfo {year}
  {2010})}\BibitemShut {NoStop}%
\bibitem [{\citenamefont {H\"artle}\ \emph {et~al.}(2013)\citenamefont
  {H\"artle}, \citenamefont {Cohen}, \citenamefont {Reichman},\ and\
  \citenamefont {Millis}}]{hartle2013decoherence}%
  \BibitemOpen
  \bibfield  {author} {\bibinfo {author} {\bibfnamefont {R.}~\bibnamefont
  {H\"artle}}, \bibinfo {author} {\bibfnamefont {G.}~\bibnamefont {Cohen}},
  \bibinfo {author} {\bibfnamefont {D.~R.}\ \bibnamefont {Reichman}},\ and\
  \bibinfo {author} {\bibfnamefont {A.~J.}\ \bibnamefont {Millis}},\ }\bibfield
   {title} {\bibinfo {title} {Decoherence and lead-induced interdot coupling in
  nonequilibrium electron transport through interacting quantum dots: A
  hierarchical quantum master equation approach},\ }\href
  {https://doi.org/10.1103/PhysRevB.88.235426} {\bibfield  {journal} {\bibinfo
  {journal} {Phys. Rev. B}\ }\textbf {\bibinfo {volume} {88}},\ \bibinfo
  {pages} {235426} (\bibinfo {year} {2013})}\BibitemShut {NoStop}%
\bibitem [{\citenamefont {Erpenbeck}\ and\ \citenamefont
  {Cohen}(2021)}]{erpenbeck2021resolving}%
  \BibitemOpen
  \bibfield  {author} {\bibinfo {author} {\bibfnamefont {A.}~\bibnamefont
  {Erpenbeck}}\ and\ \bibinfo {author} {\bibfnamefont {G.}~\bibnamefont
  {Cohen}},\ }\bibfield  {title} {\bibinfo {title} {{Resolving the
  nonequilibrium Kondo singlet in energy- and position-space using quantum
  measurements}},\ }\href {https://doi.org/10.21468/SciPostPhys.10.6.142}
  {\bibfield  {journal} {\bibinfo  {journal} {SciPost Phys.}\ }\textbf
  {\bibinfo {volume} {10}},\ \bibinfo {pages} {142} (\bibinfo {year}
  {2021})}\BibitemShut {NoStop}%
\bibitem [{\citenamefont {Dalibard}\ \emph {et~al.}(1992)\citenamefont
  {Dalibard}, \citenamefont {Castin},\ and\ \citenamefont
  {M\o{}lmer}}]{Dalibard1992}%
  \BibitemOpen
  \bibfield  {author} {\bibinfo {author} {\bibfnamefont {J.}~\bibnamefont
  {Dalibard}}, \bibinfo {author} {\bibfnamefont {Y.}~\bibnamefont {Castin}},\
  and\ \bibinfo {author} {\bibfnamefont {K.}~\bibnamefont {M\o{}lmer}},\
  }\bibfield  {title} {\bibinfo {title} {Wave-function approach to dissipative
  processes in quantum optics},\ }\href
  {https://doi.org/10.1103/PhysRevLett.68.580} {\bibfield  {journal} {\bibinfo
  {journal} {Phys. Rev. Lett.}\ }\textbf {\bibinfo {volume} {68}},\ \bibinfo
  {pages} {580} (\bibinfo {year} {1992})}\BibitemShut {NoStop}%
\bibitem [{\citenamefont {Carmichael}(1993)}]{Carmichael_book}%
  \BibitemOpen
  \bibfield  {author} {\bibinfo {author} {\bibfnamefont {H.}~\bibnamefont
  {Carmichael}},\ }\href@noop {} {\emph {\bibinfo {title} {An Open Systems
  Approach to Quantum Optics}}}\ (\bibinfo  {publisher} {Springer},\ \bibinfo
  {address} {Berlin},\ \bibinfo {year} {1993})\BibitemShut {NoStop}%
\bibitem [{\citenamefont {Daley}(2014)}]{Daley2014}%
  \BibitemOpen
  \bibfield  {author} {\bibinfo {author} {\bibfnamefont {A.~J.}\ \bibnamefont
  {Daley}},\ }\bibfield  {title} {\bibinfo {title} {Quantum trajectories and
  open many-body quantum systems},\ }\href
  {https://doi.org/10.1080/00018732.2014.933502} {\bibfield  {journal}
  {\bibinfo  {journal} {Adv. Phys.}\ }\textbf {\bibinfo {volume} {63}},\
  \bibinfo {pages} {77} (\bibinfo {year} {2014})}\BibitemShut {NoStop}%
\bibitem [{\citenamefont {Nakagawa}\ \emph {et~al.}(2020)\citenamefont
  {Nakagawa}, \citenamefont {Tsuji}, \citenamefont {Kawakami},\ and\
  \citenamefont {Ueda}}]{nakagawa2020dynamical}%
  \BibitemOpen
  \bibfield  {author} {\bibinfo {author} {\bibfnamefont {M.}~\bibnamefont
  {Nakagawa}}, \bibinfo {author} {\bibfnamefont {N.}~\bibnamefont {Tsuji}},
  \bibinfo {author} {\bibfnamefont {N.}~\bibnamefont {Kawakami}},\ and\
  \bibinfo {author} {\bibfnamefont {M.}~\bibnamefont {Ueda}},\ }\bibfield
  {title} {\bibinfo {title} {Dynamical sign reversal of magnetic correlations
  in dissipative hubbard models},\ }\href
  {https://doi.org/10.1103/PhysRevLett.124.147203} {\bibfield  {journal}
  {\bibinfo  {journal} {Phys. Rev. Lett.}\ }\textbf {\bibinfo {volume} {124}},\
  \bibinfo {pages} {147203} (\bibinfo {year} {2020})}\BibitemShut {NoStop}%
\bibitem [{\citenamefont {Wauters}\ \emph {et~al.}(2024)\citenamefont
  {Wauters}, \citenamefont {Chung}, \citenamefont {Maffi},\ and\ \citenamefont
  {Burrello}}]{wauters2023simulations}%
  \BibitemOpen
  \bibfield  {author} {\bibinfo {author} {\bibfnamefont {M.~M.}\ \bibnamefont
  {Wauters}}, \bibinfo {author} {\bibfnamefont {C.-M.}\ \bibnamefont {Chung}},
  \bibinfo {author} {\bibfnamefont {L.}~\bibnamefont {Maffi}},\ and\ \bibinfo
  {author} {\bibfnamefont {M.}~\bibnamefont {Burrello}},\ }\bibfield  {title}
  {\bibinfo {title} {Simulations of the dynamics of quantum impurity problems
  with matrix product states},\ }\href
  {https://doi.org/10.1103/PhysRevB.109.115101} {\bibfield  {journal} {\bibinfo
   {journal} {Phys. Rev. B}\ }\textbf {\bibinfo {volume} {109}},\ \bibinfo
  {pages} {115101} (\bibinfo {year} {2024})}\BibitemShut {NoStop}%
\bibitem [{\citenamefont {Kessler}(2012)}]{kessler2012generalized}%
  \BibitemOpen
  \bibfield  {author} {\bibinfo {author} {\bibfnamefont {E.~M.}\ \bibnamefont
  {Kessler}},\ }\bibfield  {title} {\bibinfo {title} {Generalized
  schrieffer-wolff formalism for dissipative systems},\ }\href
  {https://doi.org/10.1103/PhysRevA.86.012126} {\bibfield  {journal} {\bibinfo
  {journal} {Phys. Rev. A}\ }\textbf {\bibinfo {volume} {86}},\ \bibinfo
  {pages} {012126} (\bibinfo {year} {2012})}\BibitemShut {NoStop}%
\bibitem [{\citenamefont {Rosso}\ \emph {et~al.}(2020)\citenamefont {Rosso},
  \citenamefont {Iemini}, \citenamefont {Schirò},\ and\ \citenamefont
  {Mazza}}]{rosso2020dissipative}%
  \BibitemOpen
  \bibfield  {author} {\bibinfo {author} {\bibfnamefont {L.}~\bibnamefont
  {Rosso}}, \bibinfo {author} {\bibfnamefont {F.}~\bibnamefont {Iemini}},
  \bibinfo {author} {\bibfnamefont {M.}~\bibnamefont {Schirò}},\ and\ \bibinfo
  {author} {\bibfnamefont {L.}~\bibnamefont {Mazza}},\ }\bibfield  {title}
  {\bibinfo {title} {{Dissipative flow equations}},\ }\href
  {https://doi.org/10.21468/SciPostPhys.9.6.091} {\bibfield  {journal}
  {\bibinfo  {journal} {SciPost Phys.}\ }\textbf {\bibinfo {volume} {9}},\
  \bibinfo {pages} {091} (\bibinfo {year} {2020})}\BibitemShut {NoStop}%
\end{thebibliography}
\end{document}